\documentclass[twocolumn,twocolappendix]{aastex631}

\usepackage{amsmath,amstext}
\usepackage[T1]{fontenc}
\usepackage{multirow}
\usepackage{xcolor}

\newcommand{\hst}{\textit{HST}}

\newcommand{\jwst}{\textit{JWST}}

\newcommand{\hb}{\hbox{H$\beta$}}
\newcommand{\ha}{\hbox{H$\alpha$}}
\newcommand{\oii}{\hbox{[O \textsc{ii}]}}
\newcommand{\oiii}{\hbox{[O \textsc{iii}]}}
\newcommand{\nii}{\hbox{[N \textsc{ii}]}}
\newcommand{\sii}{\hbox{[S \textsc{ii}]}}

\newcommand{\cigale}{\hbox{\textsc{cigale}}}
\newcommand{\linmix}{\hbox{\textsc{linmix}}}
\newcommand{\grizli}{\hbox{\textsc{grizli}}}

\newcommand{\pysersic}{\hbox{\textsc{pysersic}}}
\newcommand{\izi}{\hbox{\textsc{izi}}}
\newcommand{\oo}{O$_{32}$}
\newcommand{\ebvg}{$\mathrm{E(B-V)_{gas}}$}
\newcommand{\ebvs}{$\mathrm{E(B-V)_{star}}$}
\newcommand{\sm}{$\mathrm{M}_\ast$}
\newcommand{\msun}{$\mathrm{M}_\odot$}

\begin{document}

\title{NGDEEP: The Star Formation and Ionization Properties of Galaxies at 1.7 < $z$ < 3.4}

\correspondingauthor{Lu Shen}
\email{lushen@tamu.edu}

\author[0000-0001-9495-7759]{Lu Shen}
\affiliation{Department of Physics and Astronomy, Texas A\&M University, College Station, TX, 77843-4242 USA}
\affiliation{George P.\ and Cynthia Woods Mitchell Institute for
 Fundamental Physics and Astronomy, Texas A\&M University, College Station, TX, 77843-4242 USA}
 
\author[0000-0001-7503-8482]{Casey Papovich}
\affiliation{Department of Physics and Astronomy, Texas A\&M University, College Station, TX, 77843-4242 USA}
\affiliation{George P.\ and Cynthia Woods Mitchell Institute for
 Fundamental Physics and Astronomy, Texas A\&M University, College Station, TX, 77843-4242 USA}

\author[0000-0002-7547-3385]{Jasleen Matharu}
\affiliation{Cosmic Dawn Center (DAWN), Denmark}
\affiliation{Niels Bohr Institute, University of Copenhagen, Jagtvej 128, DK-2200 Copenhagen N, Denmark}

\author[0000-0003-3382-5941]{Nor Pirzkal}
\affiliation{ESA/AURA Space Telescope Science Institute}

\author[0000-0003-3424-3230]{Weida Hu}
\affiliation{Department of Physics and Astronomy, Texas A\&M University, College Station, TX, 77843-4242 USA}
\affiliation{George P.\ and Cynthia Woods Mitchell Institute for
 Fundamental Physics and Astronomy, Texas A\&M University, College Station, TX, 77843-4242 USA}
 
\author[0000-0002-4153-053X]{Danielle A. Berg}
\affiliation{Department of Astronomy, The University of Texas at Austin, Austin, TX, USA}

\author[0000-0002-9921-9218]{Micaela B. Bagley}
\affiliation{Department of Astronomy, The University of Texas at Austin, Austin, TX, USA}

\author[0000-0001-8534-7502]{Bren E. Backhaus}
\affil{Department of Physics and Astronomy, University of Kansas, Lawrence, KS 66045, USA}

\author[0000-0001-7151-009X]{Nikko J. Cleri}
\affiliation{Department of Astronomy and Astrophysics, The Pennsylvania State University, University Park, PA 16802, USA}
\affiliation{Institute for Computational and Data Sciences, The Pennsylvania State University, University Park, PA 16802, USA}
\affiliation{Institute for Gravitation and the Cosmos, The Pennsylvania State University, University Park, PA 16802, USA}

\author[0000-0001-5414-5131]{Mark Dickinson}
\affiliation{NSF's National Optical-Infrared Astronomy Research Laboratory, 950 N. Cherry Ave., Tucson, AZ 85719, USA}

\author[0000-0001-8519-1130]{Steven L. Finkelstein}
\affiliation{Department of Astronomy, The University of Texas at Austin, Austin, TX, USA}

\author[0000-0001-6145-5090]{Nimish P. Hathi}
\affiliation{Space Telescope Science Institute, 3700 San Martin Drive, Baltimore, MD 21218, USA}

\author[0000-0002-1416-8483]{Marc Huertas-Company}
\affil{Instituto de Astrof\'isica de Canarias, La Laguna, Tenerife, Spain}
\affil{Universidad de la Laguna, La Laguna, Tenerife, Spain}
\affil{Universit\'e Paris-Cit\'e, LERMA - Observatoire de Paris, PSL, Paris, France}

\author[0000-0001-6251-4988]{Taylor A. Hutchison}
\altaffiliation{NASA Postdoctoral Fellow}
\affiliation{Astrophysics Science Division, NASA Goddard Space Flight Center, 8800 Greenbelt Rd, Greenbelt, MD 20771, USA}

\author[0000-0002-7831-8751]{Mauro Giavalisco}
\affiliation{University of Massachusetts Amherst, 710 North Pleasant Street, Amherst, MA 01003-9305, USA}

\author[0000-0001-9440-8872]{Norman A. Grogin}
\affiliation{Space Telescope Science Institute, Baltimore, MD, USA}

\author[0000-0002-6790-5125]{Anne E. Jaskot}
\affiliation{Department of Physics and Astronomy, Williams College, Williamstown, MA 01267, USA}

\author[0000-0003-1187-4240]{Intae Jung}
\affiliation{Space Telescope Science Institute, Baltimore, MD, 21218, USA}

\author[0000-0001-9187-3605]{Jeyhan S. Kartaltepe}
\affiliation{Laboratory for Multiwavelength Astrophysics, School of Physics and Astronomy, Rochester Institute of Technology, 84 Lomb Memorial Drive, Rochester, NY 14623, USA}

\author[0000-0002-6610-2048]{Anton M. Koekemoer}
\affiliation{Space Telescope Science Institute, 3700 San Martin Dr., 
Baltimore, MD 21218, USA}

\author[0000-0003-3130-5643]{Jennifer M. Lotz}
\affiliation{Gemini Observatory/NSF's National Optical-Infrared Astronomy Research Laboratory, 950 N. Cherry Ave., Tucson, AZ 85719, USA}

\author[0000-0003-4528-5639]{Pablo G. P\'erez-Gonz\'alez}
\affiliation{Centro de Astrobiolog\'{\i}a (CAB), CSIC-INTA, Ctra. de Ajalvir km 4, Torrej\'on de Ardoz, E-28850, Madrid, Spain}

\author[0000-0003-2283-2185]{Barry Rothberg}
\affiliation{U.S. Naval Observatory, 3450 Massachusetts Avenue NW, Washington, DC 20392, USA}
\affiliation{Department of Physics and Astronomy, George Mason University, 4400 University Drive, MSN 3F3, Fairfax, VA 22030, USA}

\author[0000-0002-6386-7299]{Raymond C. Simons}
\affiliation{Department of Physics, 196 Auditorium Road, Unit 3046, University of Connecticut, Storrs, CT 06269}

\author[0000-0002-8163-0172]{Brittany N. Vanderhoof}
\affiliation{Space Telescope Science Institute, 3700 San Martin Drive, Baltimore, MD 21218, USA}

\author[0000-0003-3466-035X]{{L. Y. Aaron} {Yung}}
\affiliation{Space Telescope Science Institute, 3700 San Martin Drive, Baltimore, MD 21218, USA}

\begin{abstract}

We use \jwst/NIRISS slitless spectroscopy from the Next Generation Deep Extragalactic Exploratory Public (NGDEEP) Survey to investigate the physical condition of {178} star-forming galaxies at $1.7 < z < 3.4$. 
At these redshifts, the deep NGDEEP NIRISS slitless spectroscopy covers the \oii$\lambda\lambda$3726,3729, \oiii$\lambda\lambda$4959,5007, \hb\ and \ha\ emission features for galaxies with stellar masses $\log(\mathrm{M_\ast/M_\odot}) \gtrsim 7$, nearly a factor of a hundred lower than previous studies.
We focus on the \oii/\oiii\ (\oo) ratio which is primarily sensitive to the ionization state and with a secondary dependence on the gas-phase metallicity of the interstellar medium. 
We find significant ($\gtrsim5\sigma$) correlations between the \oo\ ratio and galaxy properties as \oo\ increases with decreasing stellar mass, decreasing star formation rate (SFR), increasing specific SFR (sSFR$\equiv \mathrm{SFR/M_*}$), and increasing equivalent width (EW) of \ha\ and \hb. 
These trends suggest a tight connection between the ionization parameter and these galaxy properties. 
Galaxies at $z\sim2-3$ exhibit a higher \oo\ than local normal galaxies with the same stellar masses and SFRs, indicating that they have a higher ionization parameter and lower metallicity than local normal galaxies. 
In addition, we observe a {mild} evolutionary trend in the \oo -- EW(\hb) relation from $z\sim0$ {to} $z\gtrsim5$, {where higher redshift galaxies show increased \oo\ and EW, with possibly} higher \oo\ at fixed EW. 
We argue that both the enhanced recent star formation activity and the higher star formation surface density may contribute to the increase in \oo\ and the ionization parameter.

\end{abstract}



\section{Introduction} \label{sec:intro}

Star formation is one of the fundamental processes that drive the evolution of galaxies and determine galaxies' global properties. 
Globally, the cosmic star formation rate (SFR) density peaked between $z\sim1$ and 3 \citep{Madau2014}. 
Understanding the mechanisms behind star formation during this epoch and the factors leading to the subsequent decline is essential for a comprehensive picture of galaxy evolution. 
This requires investigating the physical conditions of star-forming regions within galaxies, particularly the ionized gas in the interstellar medium (ISM), where recently formed stars are embedded.

The physical properties of ISM are typically characterized by gas-phase metallicity and ionization parameters. 
{Understanding the relationship between these ISM properties and galaxy global properties, such as stellar mass and star formation rate (SFR), offers valuable constraints on the formation and evolution of galaxies (e.g., \citealp{Somerville2015}).}
{For example, t}he evolution of stellar mass and gas-phase metallicity relation has been characterized out to $z\sim6$ with a clear trend of decreasing metallicity with redshift at fixed stellar mass \citep{Erb2006, Maiolino2008, Cullen2014, Steidel2014, Sanders2018, Sanders2021, Curti2024}. 
The interpretation of these relations and their evolution is that chemical enrichment depends on the galaxy's star formation history and the interplay between gas infall and outflow. 
%
%

The ionization parameter ($q$, or the dimensionless ionization parameter $U\equiv q/c$) is defined as the ratio between the mean hydrogen ionizing photon flux and the density of hydrogen atoms \citep{Dopita2003, Osterbrock2006}. It is typically estimated from the ratio of emission lines of different ionization stages of the same element, such as the \oiii/\oii\ (\oo) ratio (e.g., \citealp{Kewley2002, Kewley2019}). 
The correlation between the \oo\ ratio and the galaxy global properties such as stellar mass, SFR, and specific SFR (sSFR $\equiv$ SFR/\sm) has been observed up to $z\sim3$ \citep{Nakajima2014, Kewley2015, Sanders2016, Kaasinen2018, Mingozzi2020, Papovich2022}. 
These studies consistently found that star-forming galaxies (SFGs) at $z\sim1-3$ have significantly higher \oo\ ratios than typical SFGs at $z\sim0$ by $\sim$0.6 dex at a fixed stellar mass. 
However, the physical cause of the elevated ionization parameter in high-redshift galaxies remains unclear.

It has been suggested that the evolution of the ionization parameter is correlated with the evolving equivalent width (EW) of H recombination lines and sSFR (\citealp{Kewley2015, Kaasinen2018}). A high sSFR (or EW) indicates a high ratio of young-to-old stars, which corresponds to a relatively increased flux of hydrogen-ionizing photons from young massive stars. 
Some other studies have pointed to the increased electron densities ($n_e$) in high redshift galaxies as a possible factor in increasing the ionization parameter \citep{Davies2021, Reddy2023a, Reddy2023b}. 
This is related to the fact that the ionization parameter is determined by hydrogen gas density, which is approximately the electron density (for an ionized gas). 
In addition, other factors such as low metallicity, a hard ionizing radiation field, and the presence of density-bounded \ion{H}{2} regions could increase the \oo\ ratio \citep{Nakajima2014, Kewley2019}, which could bias the interpretation of \oo\ and the evolutionary trends of ionization parameter.

To gain a deeper understanding of the physical factors driving the elevated ionization parameter in high-redshift galaxies, we investigate the ionization state of SFGs at $1.7 < z < 3.4$, using the deep, \jwst\ NIRISS slitless spectroscopy observations from the Next Generation Deep Extragalactic Exploratory Public (NGDEEP) Survey \citep{Bagley2022, Pirzkal2023, Shen2024}. 
The NIRISS data probe observed-frame near-IR wavelengths ranging from 1.0-2.2 \micron\, corresponding to rest-frame 0.37 - 0.81 \micron\ and 0.23 - 0.50 \micron\ for galaxies at $z=1.7$ and $z=3.4$, respectively. 
This data covers strong emission lines (\oii, \hb, and \oiii) which trace gas ionization parameters and metallicities for galaxies at $z\sim1.7-3.4$. 
Importantly, NGDEEP measures these properties for an unbiased galaxy sample with stellar masses of log(\sm/\msun) $\gtrsim 7$ at these redshifts. 
This enables us to constrain the evolution of ISM properties (i.e.,  ionization and metallicity) and its correlation to galaxy properties in low-mass galaxies ($\mathrm{M_\ast} < 10^{9} \mathrm{M}_\odot$). 
In this paper, we focus on the relationships between the \oo\ ratio and various galaxy properties, comparing them to those measured from local galaxies. We will present results on the evolution with metallicity in a future paper. 

The outline for this paper is as follows. In Section \ref{sec:data} we describe the data sets, methods to derive stellar-population properties using broadband data, and sample selection. 
In Section \ref{sec:resuts}, we present and constrain the relationships between the \oo\ ratio and galaxy properties, including stellar mass, SFR, sSFR, and EW of \ha\ and \hb. 
In Section \ref{sec:discussion}, we discuss the dependence of the \oo\ ratio on the ionization parameter and metallicity, as well as the factors that could influence the ionization parameter. 
Finally, we summarize our findings in Section \ref{sec:summary}. 
Throughout this paper, all magnitudes are presented in the AB system \citep{Oke1983, Fukugita1996}. We adopt a standard $\Lambda$–cold dark matter ($\Lambda$CDM) cosmology with $H_0$ = 70 km s$^{-1}$, $\Omega_{\Lambda,0}$ = 0.70, and $\Omega_\mathrm{M,0}$ = 0.30 \citep{Planck2016}.


\section{Data, Method, and Sample Selection} \label{sec:data}

\subsection{Optical/NIR Imaging and Photometry} \label{sec:imaging}

We utilized a vast array of deep imaging taken with \hst\ and \jwst\ available in the HUDF field. 
This includes \hst\ ACS F435W, F606W, F775W, F814W, and F850LP from the latest reductions processed as part of the Cosmic Assembly Near-IR Deep Extragalactic Legacy Survey (CANDELS, \citealp{Koekemoer2011, Grogin2011}). 
We include \jwst\ NIRCam bands: F090W, F182M, F210M, F277W, F335M, F356W, F430M, F444W, F460M, and F480M from the JADES DR2 \citep{Eisenstein2023, Rieke2023, Williams2023}. 
The \jwst\ NIRISS direct-imaging from NGDEEP is not included in the photometry catalog, because the same filters are covered by the NIRCam data, which are significantly deeper. 

The photometry process is described in detail in \citet{Finkelstein2024}.  Here we summarize the salient steps. 
We matched the PSF of all ACS bands and all NIRCam bands bluer than F277W to that of the NIRCam F277W image. For bands redder than F277W, we calculated correction factors by convolving the F277W image to the larger PSF, and measuring the flux ratio in the native image to that in the convolved image. 
We performed photometry using SExtractor \citep{Bertin1996} in dual image mode, using the inverse-variance-weighted sum of the non-PSF-matched F277W and F356W images as the detection image. 
We ran the SExtractor in hot and cold modes (Finkelstein et al. \textit{in prep.}). To combine these catalogs, we included all objects in the cold-mode catalog, and we included all objects in the hot-mode catalog whose central pixel did not lie in a cold-mode segmentation map, after that cold-mode segmentation map was dilated by a 5$\times$5 kernel. 

We measured flux densities for each galaxy in small Kron apertures. We apply two aperture corrections based on the ratio of the default Kron aperture to our small aperture and an additional aperture correction to measure flux missed on larger scales. 
We derived uncertainties on the flux densities by placing random non-overlapping apertures with diameters from 0.1-1.5\arcsec\ in empty regions of the image. We fitted a four-parameter function to the normalized median absolute deviation (NMAD) in each aperture as a function of the aperture area. The flux uncertainty for each object is calculated from this fit for the area in a given object's aperture. 


\subsection{Grism Data Reduction and Line-flux Measurements} \label{sec:grism}

For the analysis of emission-line maps here, we processed the data using the \grizli\ version 1.9.5 \citep{Brammer2021}. 
\grizli\ performs full end-to-end processing of NIRISS imaging and slitless spectroscopic data sets, including retrieving, pre-processing the raw observations for cosmic rays, flat-fielding, sky subtraction, astrometric corrections, alignment, modeling contamination from overlapping spectra, extracting 1D and 2D spectra, and fitting full continuum+emission-line models.  
For this fitting process, \grizli\ uses a set of templates from the Flexible Stellar Populations Synthesis models (FSPS; \citealp{Conroy2010}) and nebular emission lines. The spectroscopic data is scaled to match the photometry catalog. \grizli\ determines a redshift by minimizing the $\chi^2$ between observed 2D spectra and models.
Then it fits emission-line fluxes using the best-fit redshift and accounts for the continua (correcting for absorption features, e.g., from Balmer lines) using the best-fit stellar population model. For additional details on \grizli\ and its data products, we refer the reader to \cite{Estrada-Carpenter2019, Simons2021, Matharu2021, Papovich2022,Wang2022, Simons2023, Noirot2023, Matharu2023}.

\begin{figure}
    \centering
    \includegraphics[width=\columnwidth]{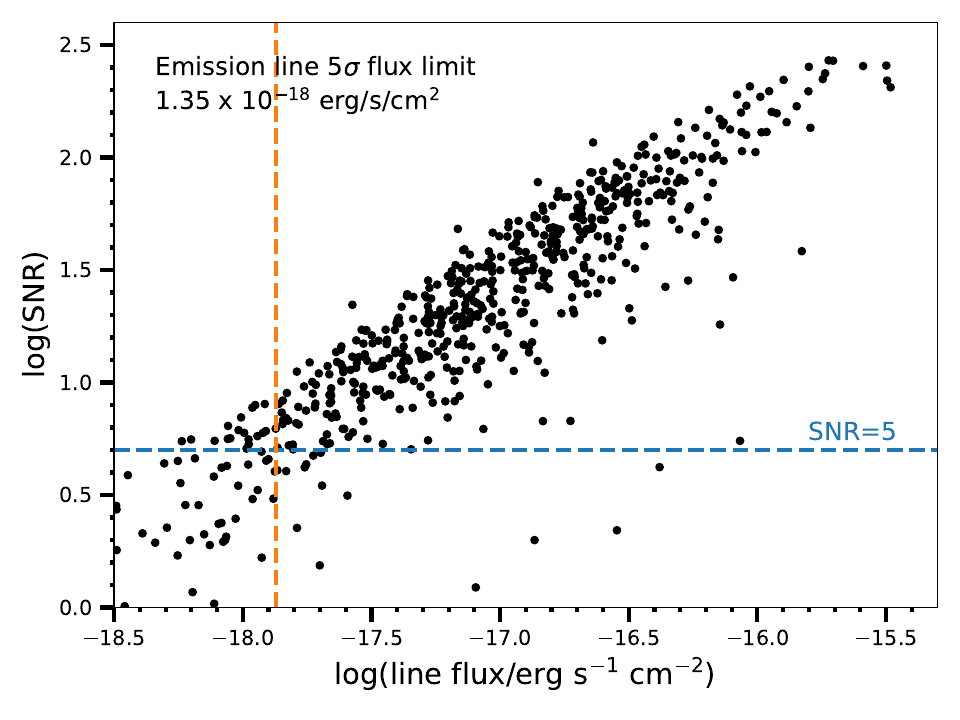}
    \caption{Signal-to-Noise Ratio (SNR) of emission lines as a function of emission line flux for NGDEEP galaxies at $1.7 < z < 3.4$. Emission lines include \oii$\lambda\lambda$3726,3729, \hb, \oiii$\lambda\lambda$4959,5007, and \ha\ across three filters (F115W, F150W, and F200W). }
    \label{fig:lineflux}
\end{figure}

We adopt the \ha, \hb, \oii, and \oiii\ emission line fluxes, equivalent widths, and associated errors measured by \grizli. 
For our selected NGDEEP sample (see Section~\ref{sec:sample}), we measure the average emission line 5$\sigma$ flux limit of $1.35 \times 10^{-18}$~erg/s/cm$^{2}$ as shown in Figure~\ref{fig:lineflux}. We note that due to the low spectral resolution of the NIRISS grism ($\mathrm{R}\sim150$), the \ha\ and \nii\ lines are blended. For the median stellar mass and redshift of the galaxies in our sample, we expect the \nii/(\ha+\nii) to be $\sim0.05$ \citep{Faisst2018}.   
It is worth noting that the flux measurement method employed by \citet[and Vanderhoof in prep.]{Pirzkal2023} adopts more conservative flux uncertainties, which would imply the uncertainties on the line fluxes here are underestimated by a factor of $\approx$1.5. 
We therefore adopt the flux uncertainties from \grizli\ for consistency, and we note that if we increased the errors by a factor of 1.5 it would not otherwise impact any of the findings in this paper.   

%

\subsection{SED Fitting and Estimating Galaxy Properties} \label{sec:sed}

We employed the SED fitting Code Investigating GALaxy Emission (\cigale) \citep{Boquien2019, Yang2020} to estimate properties of the stellar populations of our galaxies using the photometry catalog with flux densities measured in bandpasses spanning the observed wavelength range of 0.43 -- 4.8 $\mu$m (see details in Section \ref{sec:imaging}). 
Note that the NIRISS spectra were not included in the fitting process. 
However, the redshift was fixed to the value determined from the NIRISS spectra as measured by \grizli\ (see Section~\ref{sec:grism}). 

Here we described the modules and parameters adopted in \cigale. We adopted a delayed star formation history (SFH) allowing the $\tau$ and stellar age to vary from 0.05--10~Gyr and 0.02--5~Gyr, respectively. 
In addition to the main star formation history, we allow for a more recent  ``burst'' of star formation.  We model this burst as an exponential SFH with a $\tau$ allowed to vary from 20 to 100~Myr with an age of {5 and} 10 Myr and we allow the mass fraction of this burst ($f_\mathrm{burst}$) to vary from 0 to 0.1. 
We assumed a \citet{Chabrier2003} IMF and the stellar population synthesis models presented by \citet{Bruzual2003} with metallicity ranging from Solar ($Z_\odot$) to sub-Solar values (0.4~$Z_\odot$ and 0.2~$Z_\odot$). 
We include nebular emission using templates of \citet{Inoue2011}. We allow the ionization parameter $\log(U)$ to vary between $-3$ to $-1$, the gas metallicity ($Z_\mathrm{gas}$) to vary between 0.002 and 0.02, and a fixed electron density of 100~cm$^{-3}$. 
The dust attenuation follows the extinction law of \citet{Calzetti2000} for attenuating the stellar continuum and it uses the \citet{Cardelli1989} extinction law with $R_V=3.1$ for attenuating the emission lines. 
We allow the dust attenuation in emission lines from nebular regions \ebvg\ to vary from 0 to 1.1, and a fixed dust attenuation ratio between emission lines and stellar continuum (\ebvs/\ebvg $= 0.44$). 
The details of these parameters and values are listed in Table~\ref{tab:sed}. 

From \cigale\ SED fitting, we adopt the stellar mass (\sm), SFR averaged over 10~Myr as the SED-derived SFR (SFR$_\mathrm{SED}$), the color excess of the stellar continuum \ebvs, and their associated errors. The sSFR is calculated as $\mathrm{sSFR\equiv SFR_{10Myr} / M_*}$. {We discuss the different SFR measurements from SED fitting further in Appendix \ref{sec:app_sfr}. }

\begin{deluxetable}{c|c}
\tablecaption{Parameters used in the SED fitting with \textsc{cigale}. \label{tab:sed}}
\tablewidth{0pt}
\tablehead{
\colhead{Parameter} & \colhead{Values}}
\startdata
\multicolumn{2}{c}{Star formation history (sfhdelayed) }  \\
\hline
\multirow{2}{*}{$\tau$ [Myr]}  & 50, 100, 300, 500, 1000, \\
& 2000, 3000, 5000, 10000 \\
\multirow{2}{*}{Age [Myr]}  & 20, 40, 80, 160, 320, 640, \\
& 1280, 2000, 3000, 5000\\
$\tau$$_{burst}$  & 20, 50, 100 \\
age$_{burst}$  & {5}, 10 \\
f$_{burst}$  & 0.0, 0.01, 0.05, 0.1 \\
\hline
\multicolumn{2}{c}{Simple stellar population \citep{Bruzual2003}} \\
\hline
IMF & \citet{Chabrier2003} \\
Metallicity $Z_*$ &  0.004, 0.008, 0.02 \\ 
\hline
\multicolumn{2}{c}{Nebular emission \citep{Inoue2011}} \\
\hline
log(U) & -3, -2.5, -2.0, -1.5, -1 \\
gas metallicity $Z_{gas}$ &  0.002, 0.005, 0.011, 0.02  \\ 
\hline
\multicolumn{2}{c}{Dust Attenuation \citep{Calzetti2000}}  \\
\hline
\multirow{2}{*}{E(B-V)$_{l}$} & 0, 0.01, 0.05, 0.1, 0.15, \\
& 0.2, 0.3, 0.5, 0.7, 0.9, 1.1 \\
E(B-V)$_{\rm factor}$ & 0.44 \\
\hline
\enddata
\end{deluxetable}

\subsection{Dust reddening correction} \label{sec:dustcorr}

Ideally, we would use the Balmer decrements (e.g., measured from the \ha/\hb\ line ratios) to estimate nebular dust attenuation. However, because not all galaxies have both \hb\ and \ha\ covered {or detected} by NIRISS, we instead rely on the $\mathrm{E(B-V)}$ values derived from SED fitting. 
{Following \citet{Sanders2021}, we calibrate a relation between \ebvg\ based on the Balmer decrement and continuum reddening \ebvs\ derived from SED fitting and SFR using 67 SFGs from the NGDEEP survey at $1.7 < z < 2.3$ that has detections of both \ha\ and \hb\ at SNR $> 5$. 
These best fits are obtained from a Bayesian linear regression code \linmix\ \citep{Kelly2007}, accounting for uncertainties on the two variables. 
The best-fit parameters are the median of the fitted parameters (the slope and intercept) from 10000 random draws from the posterior, and the associated errors are the average between the 16th and 84th percentiles of each parameter. This \linmix\ method is applied throughout the paper to derive fit lines. 
These relations are: 
\begin{align} 
    \label{eq:dust1}
    \mathrm{E(B-V)_{gas, calibrated} - E(B-V)_{stars}} =\qquad\qquad\qquad \nonumber   \\ 
    (0.332 \pm 0.053) \times \log(\mathrm{SFR_{H\alpha}}) +(-0.293\pm 0.045)\\
    \label{eq:dust2}
    \log(\mathrm{SFR_{H\alpha}}) =\qquad\qquad\qquad\qquad\qquad\qquad\qquad \nonumber \\
    (0.879 \pm 0.069) \times \log(\mathrm{SFR_{SED, 10Myrs}}) + (0.063\pm0.060), 
\end{align}
where $\mathrm{E(B-V)_{gas, calibrated}}$ is the calibrated $\mathrm{E(B-V)_{gas}}$, $\mathrm{E(B-V)_{stars}}$ and $\mathrm{SFR_{SED, 10Myrs}}$ are derived from \cigale\ SED fitting (see Section~\ref{sec:sed}), and $\mathrm{SFR_{H\alpha}}$ is calculated from the dust-corrected \ha\ luminosity using \ebvg\ from Balmer decrement and convert to SFR using the \citet{Kennicutt2012} calibration. 
This method recovers the Balmer decrement \ebvg\ with a median offset of 0.04 dex. 
The calibrated \ebvg\ and the \oo\ radio do not vary as a function of stellar mass, though a mild dependence on SFR is observed, with \oo\ potentially being underestimated by 0.04 dex at high SFR ($\log(\mathrm{SFR/M_\odot~yr^{-1}})\sim$1.33). See more details in Appendix \ref{app:dust}. }

{In this paper, for the dust correction of emission lines (i.e., \oiii\, \hb, \oii\, \ha), we adopt the calibrated \ebvg\ calculated from equation \ref{eq:dust1} and \ref{eq:dust2}, or Balmer decrement \ebvg\ if \ha\ and \hb\ detected with SNR $> 5$. 
We adopt the \citet{Cardelli1989} extinction model with $R_V=3.1$. 
No dust correction is applied for galaxies with E(B-V) $< 0$. 
The uncertainties in equation \ref{eq:dust1} and \ref{eq:dust2} are incorporated into the uncertainties of the dust-corrected emission line fluxes. 
In detail, for each galaxy, we generate 10000 mock values of emission line fluxes, \ebvs, SFR, and coefficients in equations \ref{eq:dust1} and \ref{eq:dust2} by randomly sampling from Gaussian distributions, with the measured value as the means and associated uncertainties as the standard deviations. 
For galaxies using Balmer decrement \ebvg, we account for the uncertainties in emission line fluxes and \ebvg\ in the mock sampling process. 
The uncertainties on dust-corrected emission lines are derived from the 16th and 84th percentiles of the mock distributions. }
{As we discuss in Appendix~\ref{app:dust}, applying different E(B-V) for emission lines (a uniform star-to-gas attenuation ratio of 0.44 or 1) does not affect the significance of relations presented in this paper, nor does it substantially impact the slope. However, the intercepts of these relations would be systematically shifted lower by $\sim 0.09$ dex or 0.03 dex, respectively. }
For the dust correction of the continuum, we adopt the \ebvs\ from the SED fits and follow the \citet{Calzetti2000} extinction model with $R_V=3.1$. 

{We note that we found a significant fraction (48\%) of galaxies having \ha/\hb\ ratios lower than the intrinsic \ha/\hb\ value of 2.86 based on the Case B assumption and $T=10^4\ \mathrm{K}$ and $n_\mathrm{e} = 10^2\ \mathrm{cm^{-3}}$. We exclude the possibility that these are caused by low SNR or wavelength-dependent flux calibration. On the other hand, we found significant correlations between the \ha/\hb\ ratio and stellar mass, SFR, and \oo. These correlations suggest that the low Ha/Hb ratios are more likely due to physical conditions, such as a different geometry of dust and gas, and higher temperatures in galaxies with high ionization or low metallicity \citep{Reddy2015, Scarlata2024}. In such environments, the intrinsic \ha/\hb\ of galaxies could be lower than 2.86. See more details and discussions in Appendix \ref{app:dust}. }

\subsection{The \oiii/\oii\ ratio} \label{sec:cal_o32}

Here we focus on the relation between the \oiii/\oii\ line ratio and galaxy properties (i.e., stellar mass and star formation). The \oiii/\oii\ ratio measures the relative amount of emission from double-ionized oxygen to singly ionized oxygen, which traces the ionization of the gas (e.g., \citealp{Kewley2019}).  
We defined the line ratio \oo\ as:
\begin{equation}
    O_{32} \equiv \frac{\oiii\ \lambda\lambda4959, 5007}{\oii\ \lambda\lambda3726, 3729}
\end{equation}
where \oiii\ $\lambda\lambda$4959, 5007 and \oii\ $\lambda\lambda$3726, 3729 are the sum of the emission from both lines in the doublets (as these are unresolved in the NIRISS spectra) and dust corrected as described in section \ref{sec:dustcorr}. For those with \oii\ SNR $<5$, we use the $5\times$ \oii\ flux error as an upper limit on \oii\ and, thus, a lower limit on the \oo\ ratio. 
{The uncertainties in \oo\ ratio include the uncertainties on emission line fluxes, \ebvs\, and uncertainties in converting \ebvs\ to \ebvg\ in equation \ref{eq:dust1} and \ref{eq:dust2} (see Section \ref{sec:dustcorr}). }

\subsection{Sample Selection} \label{sec:sample}

Here we use the NGDEEP NIRISS WFSS spectroscopy to study galaxies with \oii\ $\lambda\lambda$3726, 3729 and \oiii\ $\lambda\lambda$4959, 5007, with the ratio of these two emission lines tracing the ionization parameter. 
Our NIRISS F115W+F150W+F200W data cover both lines for galaxies in the redshift range $1.72 < z < 3.44$. 
Additionally, the data include coverage of \hb\ $\lambda$4862 for all galaxies. 
For galaxies at $1.72 < z < 2.22$, the data also include coverage of \ha\ $\lambda$6564 + \nii\ $\lambda$6548, 6584.

We selected galaxies from NGDEEP for this study using the following criteria and summarized in Table~\ref{tab:sample}
\begin{enumerate}
  \item We visually examined all the \grizli\ extractions with F150W AB magnitude $<28.4$, requiring galaxies to have at least one spectral feature detected (i.e., emission line, absorption feature, and D4000 break) and to be well-fitted with \grizli. {Such spectra were classified as ``good''.} 
  The F150W magnitude cut was chosen because the fraction of galaxies with good spectra is very low ($<$1\%) at magnitudes fainter than 28.4. 
  {We identified and masked contamination if it affects the emission line region of a galaxy. We then re-ran the \grizli\ fitting, excluding the masked regions. We remove objects if the contamination cannot be masked. }
  Visual examination ensures that the derived redshifts are reliable and that the spectra are not affected by contamination from nearby objects. We identified {471} galaxies through this process. 

  \item We required the spectroscopic redshift derived from the NIRISS data to lie in the redshift range of $1.72 < z < 3.44$. This redshift range ensures that \oii\ and \oiii\ are mostly covered by grism. {During this step, we also removed any objects where their \oii\ or \oiii\ emission fell too close to the edge of the NIRISS wavelength range.} A total of {191} galaxies were identified at this stage. 
  
  \item We rejected galaxies identified as AGN by cross-matching to the latest AGN catalog from \citet{Lyu2022} within 1\arcsec. This step removed 9 galaxies identified as AGN. 

  \item We selected SFGs and quiescent galaxies following the UVJ color-color selection criteria from \citet{Whitaker2011}. Four quiescent galaxies are removed in this step. We obtained a total of {178} galaxies following this step.  We denote this as the ``full sample'' and is the main sample in this paper. All of these galaxies have \oiii\ detections at $>5\sigma$ in the combined 1D spectra, because the \oiii\ is the strongest line for the majority of these galaxies, and the redshift identification mostly relies on this emission line.  This does open a selection bias (as we require strong lines for selection), and we will further discuss this effect in Section~\ref{sec:disc-bias}.

 \item Lastly, we require the detection of both \oii\ and \oiii\ emission lines with SNR $>5$ in the combined 1D spectra. This produced a sample of {149} galaxies, which we denote as the  ``final sample''. We also consider and plot those with \oii\ SNR $<5$ (which we call the ``low-SNR sample''), but these galaxies are not included in the calculation of median values or linear relations. Within the final sample, {117} and {75} galaxies have \hb\ and \ha\ SNR $>5$, respectively. These samples will be used in Section \ref{sec:o32-ew}.

\end{enumerate} 


\begin{deluxetable}{l|c}
\tablecaption{Summary of Sample Selection \label{tab:sample}}
\tablewidth{0pt}
\tablehead{
\colhead{Sample Selection Criteria} & \colhead{Galaxies}}
\startdata
F150W magnitude < 28.4 and quality check & {471} \\
$1.72 < z_\mathrm{spec} < 3.44$, spectra covers \oii\ \& \oii & {191} \\
AGN removed & {182} \\
{full sample}: SFGs with UVJ color selection & {178} \\
\hline 
{final sample}: \oiii\ \& \oii\ SNR$>$5 & {149} \\
{low-SNR sample}: & {29} \\
$~~~~~~~$\oiii\ SNR $>$ 5 \& \oii\ SNR$<$5 &  \\
\hline 
final sample with \hb\ SNR$>$5  & {117} \\
final sample with \ha\ SNR$>$5  & {75} \\
\hline
\enddata
\tablecomments{The names of the samples used throughout this paper are denoted in bold.}
\end{deluxetable}

\begin{figure*}
    \centering
    \includegraphics[width=\textwidth]{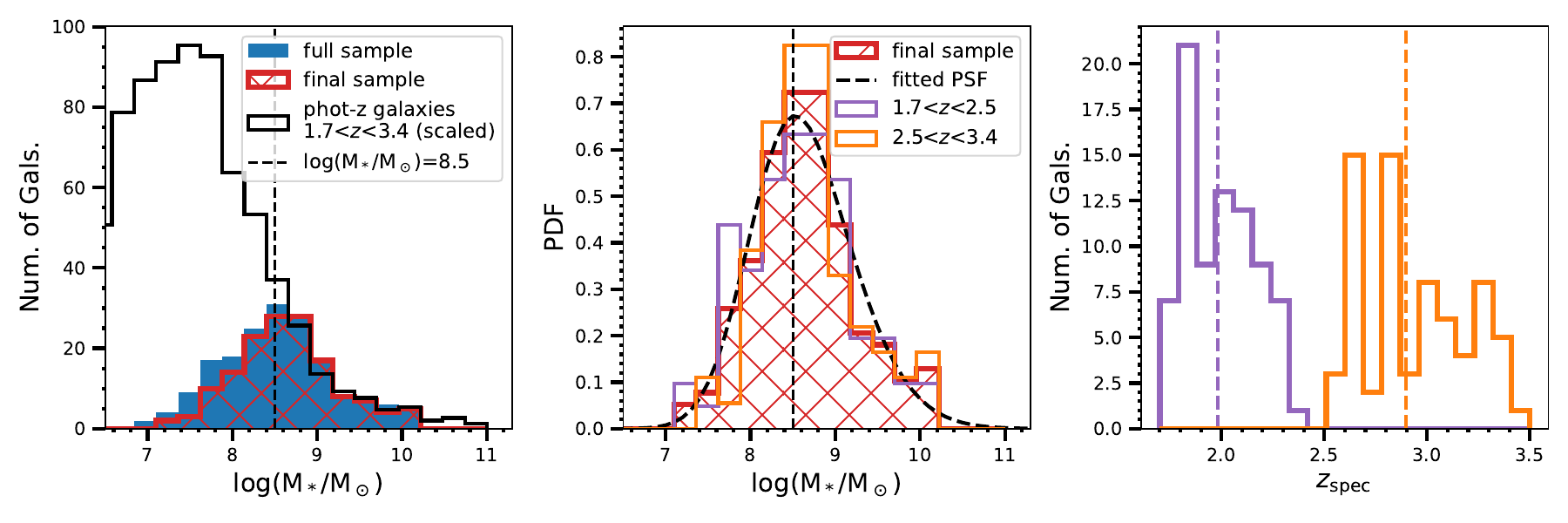}
    \caption{\textit{Left:} The stellar mass histograms for NGDEEP full sample (blue), final sample (red), and ''photometric'' galaxies (black). The latter includes all galaxies with photometry redshift (spectroscopic redshift if available) at $1.7 < z < 3.4$ (grey). The histogram for the photometric galaxies is rescaled to match the massive end of that for NGDEEP galaxies. \textit{Middle:} The probability density function (PDF) of stellar mass for NGDEEP {final sample (red), and the final sample separated by redshift with $1.7 < z < 2.5$ and $2.5 < z < 3.4$ (purple and orange, respectively)}. The fitted skew normal distribution for the NGDEEP final sample is shown in a black dashed line. In both panels, the vertical dashed line marks the mass completeness of $\mathbf{10^{8.50}}$ M$_\odot$. {\textit{Right:} The redshift histograms of the final sample at $1.7 < z < 2.5$ and $2.5 < z < 3.4$ (purple and orange, respectively). Colored vertical lines indicate the median redshifts. }  }
    \label{fig:masshist}
\end{figure*}

In the top panel of Figure~\ref{fig:masshist}, we show the stellar mass histograms for the full sample, the final sample, and for all ``photometric'' galaxies in the dataset (including those with photometric redshifts, or spectroscopic redshift if available) in the range of $1.7 < z < 3.4$. The histogram for the photometric galaxy sample is rescaled to match the massive end of that for NGDEEP galaxies. 
The stellar mass histogram of the full and final samples diverges from that for the photometric galaxies at $\lesssim10^{8.5}$ M$_\odot$. 
The final sample has a similar histogram as the full sample at the massive end, with a small difference present at lower mass. 

The {middle} panel of Figure~\ref{fig:masshist} shows a probability density function (PDF) of stellar mass for the NGDEEP {final} sample and also with the galaxies separated by redshift with $1.7 < z < 2.5$ and $2.5 < z < 3.4$. 
{In the right panel of Figure~\ref{fig:masshist}, we show a redshift histogram of the NGDEEP final sample. 
The median stellar mass and redshift for the final sample are $\log(M_\ast/\mathrm{M_\odot}) = 8.62$ and $z_\mathrm{med}=2.31$, respectively.  
We divide the final sample into two redshift bins at $1.7 < z < 2.5$ (the $z\sim2$ bin) and $2.5 < z < 3.4$ (the $z\sim3$ bin) with 79 and 70 galaxies, respectively. 
The median stellar mass and redshift are $\log(M_\ast/\mathrm{M_\odot}) = 8.59$ and $z_\mathrm{med}=1.98$ for the $z\sim2$ bin, and $\log(M_\ast/\mathrm{M_\odot}) = 8.63$ and $z_\mathrm{med}=2.89$ for the $z\sim3$ bin. }

We adopt the peak of stellar mass PDF as the 50\% mass completeness by fitting a skew-normal distribution with scipy skewnorm package. 
The 50\% mass completeness is $\log(M_\ast/\mathrm{M_\odot}) = 8.50$ for galaxies in the final sample, which is the same when separated in redshift. 
Therefore, throughout the paper, we adopt $\log(M_\ast/\mathrm{M_\odot}) = 8.50$ as our mass completeness. In particular, we will show all galaxies in our samples in subsequent figures, but we only use galaxies with stellar mass above this limit for fitting linear relations.

\subsection{Comparison Galaxy Samples} 

For comparison, we consider both the nearby samples and galaxies at similar redshifts from the literature. 
For $z\sim1-2$, we consider two samples: the CANDELS Ly$\alpha$ Emission At Reionization (CLEAR) survey, which measured \oii, \oiii, and \hb\ emission features for 196 galaxies at $z\sim1.1-2.3$ using HST WFC3 grisms \citep{Papovich2022}; 
and the MOSFIRE Deep Evolution Field (MOSDEF) survey, which obtained spectra for $\sim400$ of galaxies at $z\sim0.8-4.4$ using Keck/MOSFIRE \citep{Kriek2015, Reddy2015}. 
From CLEAR, we selected 158 galaxies with SNR $>3$ for both \oii\ and \oiii\ emission lines. 
{The median and stellar mass and redshift of CLEAR sample is $\log(M_\ast/\mathrm{M_\odot}) = 9.57$ and $z_\mathrm{med}=1.58$.} 
The dust correction is calculated following the \citep{Calzetti2000} extinction curve and using the $A_{V}$ from SED fitting. Stellar mass and SFR are estimated using the SED fitting assuming a \citet{Kroupa2001} IMF. 
{For MOSDEF, we utilized the stacked results from \citet{Sanders2021}. Stacking samples were selected from the MOSDEF SFGs, requiring detection of \oiii$\lambda$5007 with SNR $>3$ and spectral coverage of \oii, \hb\ and \oiii. The stacking sample is 280 (155) at $z\sim2.3$ (3.3) with the median redshift of 2.28 (3.24) and median stellar mass of $\log(M_\ast/\mathrm{M_\odot}) = 9.96~(9.89)$. The \oo\ ratio of MOSDEF stacked spectra is measured with \oiii$\lambda$5007/\oii$\lambda\lambda$3726,3729. To match our \oo\ ratio, we multiply the MOSDEF \oo\ ratio by $(1+1/2.98)$. The dust correction for the MOSDEF stacking results is applied prior to stacking and using the same method as described in Section \ref{sec:dustcorr}). }

For $z\sim0$, we consider four samples: the Sloan Digital Sky Survey (SDSS); the Cosmic Origins Spectrograph Legacy Spectroscopic Survey (CLASSY) treasury \citep{Berg2022}; green pea galaxies \citep{Yang2017a};  and blueberry galaxies \citep{Yang2017b}. 
The latter two samples are named for their compact and distinctive green and blue colors in SDSS false-color $gri$-band images. 
The SDSS sample represents the overall $z\sim0$ population, while the latter three samples are selected to be UV bright (CLASSY) or to have strong emission lines (green peas and blueberry galaxies), which may have properties more similar to those of high redshift galaxies. 
For the SDSS sample, we adopt emission-line measurements and galaxy properties from the MPA-JHU catalog of measurements for SDSS DR8 \citep{Brinchmann2004, Kauffmann2003, Tremonti2004}. We select galaxies with SNR $>$ 3 in all \oii, \hb, \oiii, and \ha\ emission lines, and selected SFGs based on a BPT diagram \citep{Baldwin1981}. Again, to compare these results to those of our NGDEEP samples,  the \oii\ flux is obtained by adding \oii$\lambda$3726 and \oii$\lambda$3729 fluxes, and the \oiii\ flux is obtained by multiplying \oiii$\lambda$5007 flux with (1+1/2.98). 
The dust correction is calculated with \ha/\hb\ assuming an intrinsic ratio of 2.86 and following the \citep{Cardelli1989} extinction curve. We adopted the median estimate of the total stellar mass and SFR PDF as the final stellar mass and SFR. Note that SFRs are derived by combining emission-line measurements and correcting for aperture loss. 

For the CLASSY sample, we adopt measurements for all 45 galaxies from \citep{Berg2022}. The \oo\ ratio of CLASSY sample is measured with \oiii$\lambda$5007/\oii$\lambda\lambda$3726,3729. To match our \oo\ ratio, we multiply the CLASSY \oo\ ratio by (1+1/2.98). 
For green pea galaxies, we adopted all 43 galaxies with all \oii, \hb, \oiii, and \ha\ SNR>3. 
For blueberry galaxies, we adopted all 36 galaxies with all \oii, \hb, and \oiii\ SNR>3. No dust correction is applied for blueberry galaxies because the \ha\ fluxes were probably underestimated due to the poor calibration in the red end of spectra (see more in \citealp{Yang2017b}). 
SFR is calculated from \hb\ for blueberry galaxies and dust-corrected \ha\ for green pea galaxies using the \citet{Kennicutt2012} calibration. 

The SDSS, CLASSY, green pea, and blueberry galaxy catalogs all have \oii\ and \oiii\ measurements, allowing us to compare \oo\ and galaxy properties across samples. 
However, we note that galaxies' properties are derived using different SED fitting methods with different wavelength coverage of photometry, which could result in biases during the comparison to our work. 
In the case of the SFR, our SED-derived SFR values are consistent with \ha-derived SFR, where the latter is used to derive SFRs for the SDSS, green pea, and blueberry galaxies. 
In terms of stellar mass, it is less sensitive to the adopted method and photometry data coverage. 
In addition, most of these comparison catalogs assumed the same \citet{Chabrier2003} IMF and only CLEAR assumed a \citet{Kroupa2001} IMF. 
We expect this difference in IMF to have a negligible impact on the derived galaxy properties. 

\section{Results} \label{sec:resuts}
{In this section, we first compare the SFR -- stellar mass distribution of the NGDEEP final sample to the star formation main sequence (SFMS) at similar redshift and to comparison galaxy samples to provide the context of our sample in Section \ref{sec:SFMS}. 
Then, we present the relations between the \oo\ ratio and galaxies properties, including stellar mass, SFR, sSFR, and Balmer line EW for our sample, along with comparisons to other galaxy samples in Sections \ref{sec:o32-mass}, \ref{sec:o32-sfr}, \ref{sec:o32-ssfr} and \ref{sec:o32-ew}, respectively. }

\subsection{The SFR -- Stellar Mass Relationship} \label{sec:SFMS}

\begin{figure*}
    \centering
    \includegraphics[width=\textwidth]{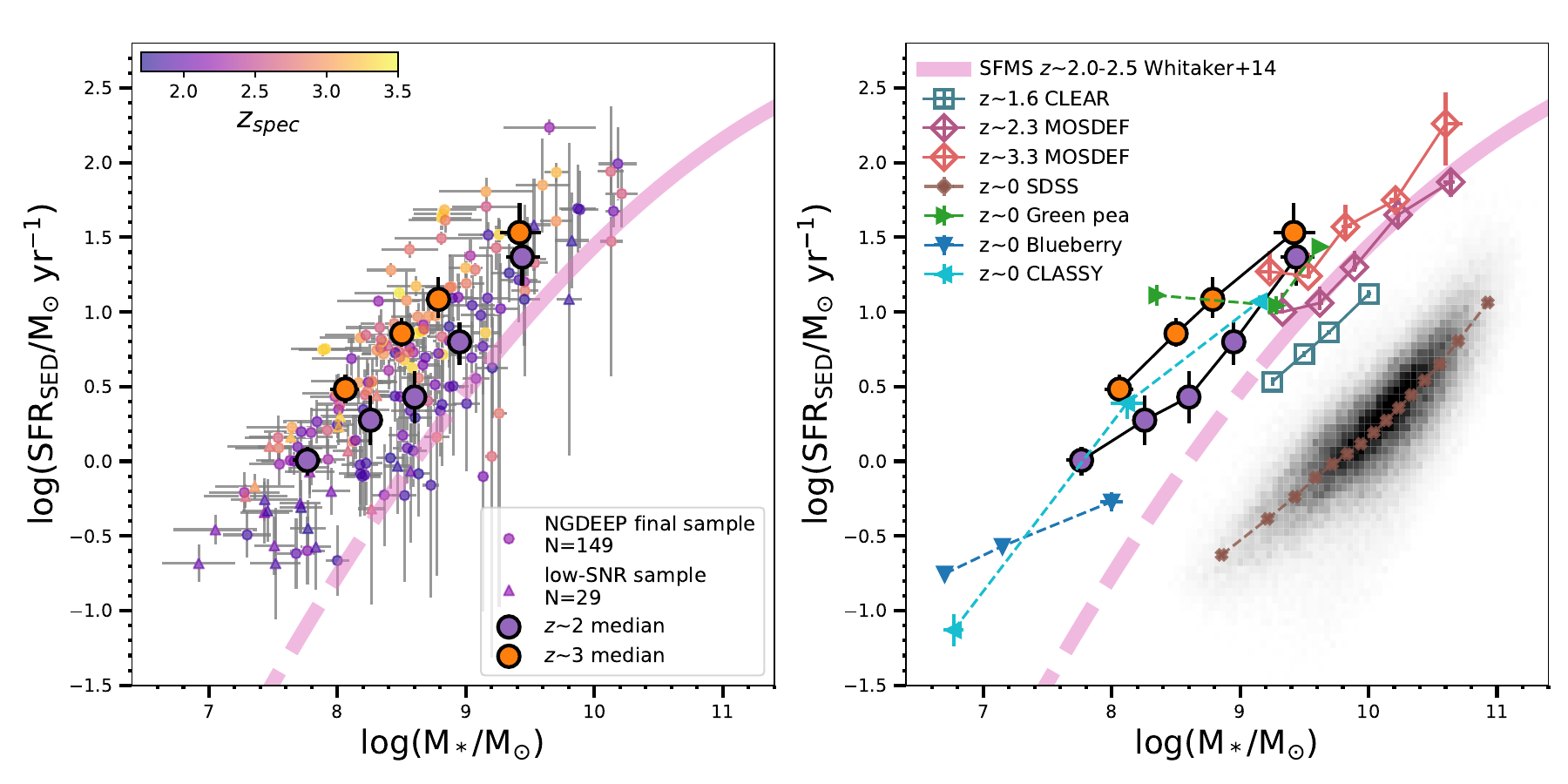}
    \caption{\textit{Left:} The SED-derived SFR versus the stellar mass for galaxies in the NGDEEP final sample (dots) and low-SNR sample (triangle) color-coded by their redshift. The median SFR and stellar mass for galaxies in the final sample at $z\sim2$ and $z\sim3$ are shown (in both panels) in black purple- and orange-filled dots, respectively. For comparison, the SFMS for galaxies at $2.0 < z < 2.5$ from \citet{Whitaker2014} is shown down to their mass completeness in the pink line and extended to the lower stellar mass region in the dashed line. \textit{Right:} The median SFR and stellar mass are shown for galaxies at various redshift: $z\sim1.7$ from CLEAR (\citealp{Papovich2022}, open blue squares), {$z\sim2.3$ and $z\sim3.3$ from the MOSDEF stacked sample (\citealp{Sanders2021}, open magenta and orange diamonds)}, $z\sim0$ from SDSS (brown crosses), $z\sim0$ from CLASSY (\citealp{Berg2022}, cyan triangles), green pea galaxies at $z\sim0$ (\citealp{Yang2017a}, green triangles) and blueberry galaxies at $z\sim0$ (\citealp{Yang2017b}, blue triangles). The SDSS sample is shown in the 2D histograms in black. The same median values from the NGDEEP final sample as \textit{left} are shown in black purple- and orange-filled dots. The NGDEEP \oo\ sample has consistently higher SFR than the SFMS at $z\sim2-2.5$, and has a similar SFR with those of extreme galaxies at $z\sim0$.  } 
    \label{fig:SFR-M}
\end{figure*}

The left panel of Figure~\ref{fig:SFR-M} shows the SED-derived SFRs and stellar masses for galaxies in the NGDEEP final sample and low-SNR sample. 
The figure also shows the median SFRs and stellar masses of galaxies in the final sample divided into redshift bins of $1.7 < z < 2.5$ and $2.5 \le z < 3.4${, with the median redshift of $z\sim2$ and $z\sim3$, respectively. }
The median and associated uncertainties are calculated with a bootstrap method to account for uncertainties on individual measurements and the scatter in each bin following \citet{Shen2023}. 
The binning is chosen to have a similar number of galaxies in each bin, with a minimum of 10 galaxies per bin. 

We color-code individual galaxies by redshift in the left panel of Figure~\ref{fig:SFR-M}. There is a clear trend with redshift such that SFGs have higher SFR with increasing redshift at fixed stellar mass. This is consistent with the redshift evolution of the SFR-\sm\ relation found by previous studies \citep[e.g.,][]{Whitaker2014, Tomczak2016}. 

For comparison, we show the star-formation main-sequence (SFMS) derived from a larger sample of SFGs selected from the UVJ color-color method at $2.0 < z < 2.5$ from \citet{Whitaker2014}. 
We compare our galaxies with this SFMS using galaxies having stellar mass above the mass completeness limits of the \citet{Whitaker2014} samples ($\sim10^{9.2}$ \msun). Our galaxies have a higher SFR than the \citet{Whitaker2014} SFMS with a median difference of $\mathrm{log(SFR/SFR_{MS})} = 0.33$ dex and a scatter from {0.05 to 0.72 dex} from the 16th/84th percentile. 
This bias is similar to other studies of emission-line selected studies of galaxies \citep{Sanders2016, Papovich2022}, such that emission-line samples of SFGs tend to have higher SFRs at fixed stellar mass. We will discuss this effect in our results in Section \ref{sec:disc-bias}. 

In addition, we find that the median SFR values of galaxies with \sm $<10^{8}$ \msun\ at $z < 2.5$ do not decrease with decreasing stellar mass. Instead, galaxies in the low-SNR sample dominate in this low-mass, low-SFR region. Therefore, the flattened median at low mass is mostly likely due to the detection limit, where galaxies with lower SFR have \oii\ (or even \oiii) emission below the detection limit. 

The median stellar mass and SFR of local and $z\sim1-2$ samples are shown in the right panel of Figure~\ref{fig:SFR-M}. The uncertainties on the median are given by $\sigma_\mathrm{MAD}/\sqrt{n-1}$, where $\sigma_\mathrm{MAD}$ is the median absolute deviation and $n$ is the number of the galaxies in each bin. 
The CLEAR sample from \hst\ grism observations is on average more massive ($10^{9.2}$-$10^{10}$ \msun) and has lower SFR values than the NGDEEP final sample, due to the lower redshifts of the CLEAR galaxies. 
The MOSDEF galaxies are also more massive, spanning stellar mass between $10^{9.5}$-$10^{10.5}$ \msun. 
{The median SFR--mass trend of MOSDEF at $z\sim2.3$ generally follow the SFMS at $z\sim2.0-2.5$. MOSDEF galaxies at $z\sim3.3$ have higher SFRs than this SFMS, consistent with the expected evolution of the SFMS. For both redshift samples, the lowest mass bins of MOSDEF galaxies show flattened SFRs, likely due to detection limitations. }
The SFR--mass relations of the extreme local galaxies from the CLASSY and ``green pea'' samples are consistent with that of the NGDEEP final sample, suggesting the properties and physical conditions of these galaxies are comparable with our $z\sim2-3$ galaxies. 
On the other hand, the SDSS sample has significantly lower SFR at the same stellar mass compared to galaxies at intermediate redshift and those extreme local galaxies. 

\subsection{The Mass -- \oo\ Relation} \label{sec:o32-mass}

\begin{figure*}
    \centering
    \includegraphics[width=\textwidth]{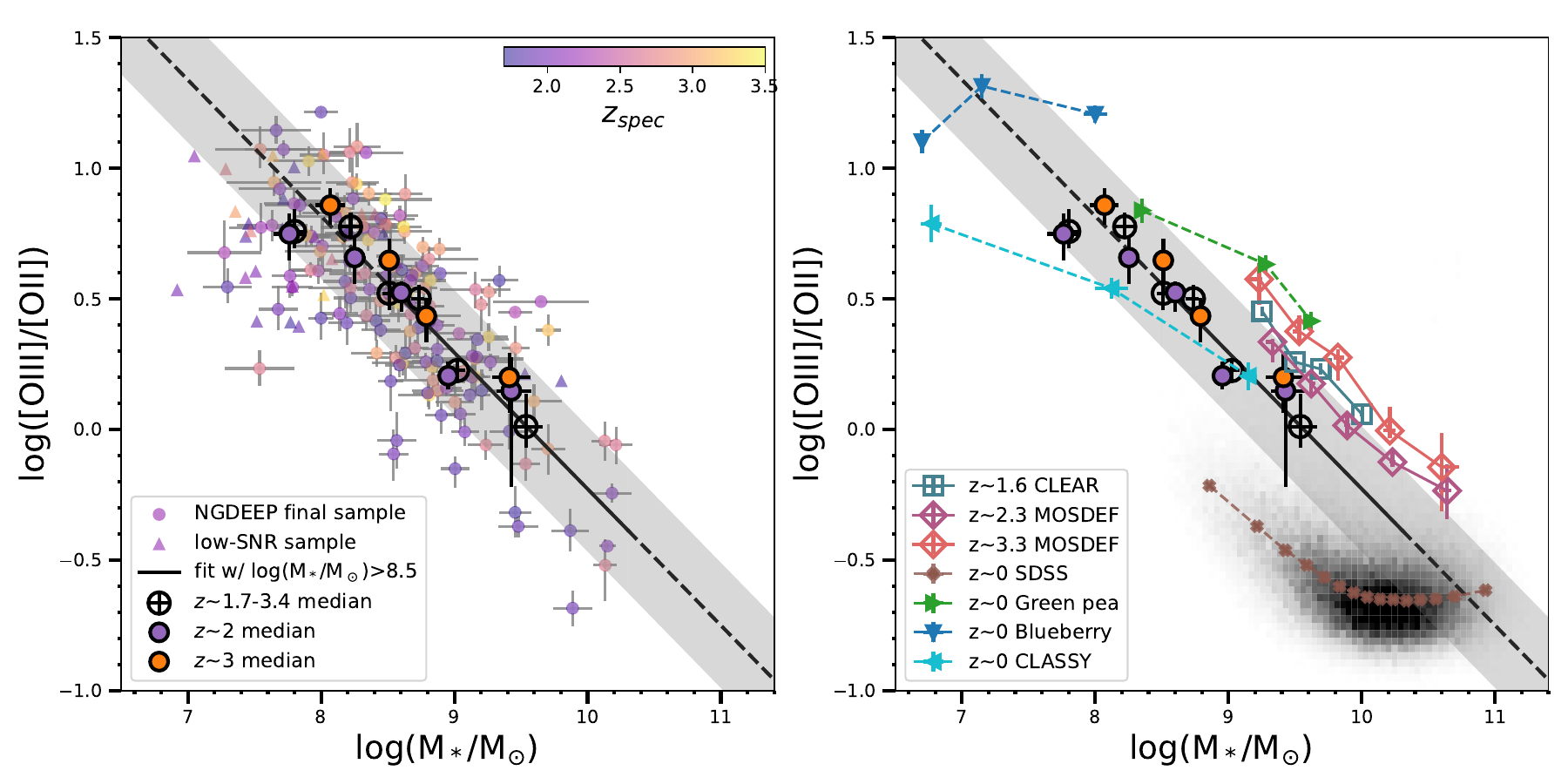}
    \caption{\textit{Left:} The \oo\ ratio versus stellar mass for galaxies in the NGDEEP final sample (dots) and low-SNR sample (triangle), color-coded by redshift. The median \oo\ along the abscissa are shown for galaxies in the final sample (open black circles), for those at $z\sim2$ and $z\sim3$ in purple- and orange-filled dots, respectively. The best fit for the NGDEEP final sample with stellar mass $>10^{8.5}$ M$_\odot$ is shown as the solid black lines, and extended to lower stellar mass as the dashed line. The grey region marks the 1$\sigma$ total scatter. \textit{Right:} The median \oo\ ratios as a function of SFR are shown for CLEAR, MOSDEF, SDSS, CLASSY, green peas, and blueberry galaxies (symbols are the same as in Fig. \ref{fig:SFR-M}). The SDSS sample is shown in the 2D histograms in black. The same median values and fits from the NGDEEP final sample are shown, identical to those in the \textit{left} panel. The NGDEEP sample shows a tight mass--\oo\ relation. }
    \label{fig:o32_mass}
\end{figure*}

Figure~\ref{fig:o32_mass} shows a tight relationship between the \oo\ ratio and the stellar mass of our final sample, such that the \oo\ ratio decreases as stellar mass increases.  
We adopt a Spearman rank correlation test to assess the correlation between stellar mass and \oo\ ratio. 
The test returns a correlation coefficient of $-$0.73 and a p-value\footnote{The p-value quantifies the probability of obtaining the observed data assuming the null hypothesis (i.e., no correlation) is true. the significance of the correlation by giving the probability that the data are uncorrelated (i.e., the null hypothesis). We reject the null hypothesis if the p-value is $\leq 0.05$, which suggests that the observed correlation is statistically significant. } of $\sim10^{-25}$. 
This test confirms the significant correlation between \oo\ and stellar mass.  
The correlation remains significant when separating in redshift. 

We measure the median \oo\ in stellar mass bins for galaxies in the final sample and those in two redshift bins at $z\sim2$ and $z\sim3$. 
{The difference of mass--\oo\ relation at $z\sim2$ and $z\sim3$ is subtle. The median \oo\ at $z\sim3$ is slightly higher than that at $z\sim2$ by 17\% dex.  }

We quantify the mass--\oo\ relation with \linmix\ using galaxies in the final sample with stellar mass above the mass completeness limit ($>10^{8.5}$ \msun). 
{The best-fit parameters are reported in Table~\ref{tab:relations}. }
The median intrinsic scatter measured by \linmix\ is 0.05~dex{, which accounted for measurement uncertainties. 
We measure the total scatter of the mass--\oo\ relation by calculating the sum of the squared deviations of data points from the best-fit relation, which represents the spread of the data. }
The total scatter of the mass--\oo\ relation is 0.24~dex. 
We show the fit extrapolated to lower stellar mass in Figure~\ref{fig:o32_mass}. 
The median values follow the fitted line to the stellar mass bin of $\sim10^{8}$ \msun. 


In comparison, we show the distribution of \oo\ and stellar mass for the local and $z\sim1-3$ samples in the right panel of Figure~\ref{fig:o32_mass}. 
{Both the stacked results from MOSDEF at $z\sim2.3$ and the median from CLEAR at $z\sim1.6$ are higher than our mass--\oo\ relation by a median of $\sim$0.20 dex within the total scatter of our sample. 
The offset is comparable to the total scatter of our relations, though, a larger offset is seen at the massive end $\log(M_*/\mathrm{M_\odot}) > 10$. The stacked spectra from MOSDEF at $z\sim3.3$ are higher than the fitted mass--\oo\ relation by a median of 0.40 dex.}
{The difference between MOSDEF and our results could be due to differences in selection or detection methods. Our NGDEEP sample extends to fainter magnitudes with $m(\mathrm{F150W})\leq28.4$ compared to $m(\mathrm{F160W}) < 25$~mag for MOSDEF \citep{Kriek2015}. This difference is also reflected in the stellar mass range that our NGDEEP sample dominates the lower-mass range ($\log(M_\ast/\mathrm{M_\odot}) < 9$) while MOSDEF galaxies are selected with $\log(M_\ast/\mathrm{M_\odot}) > 9$. This selection difference could account for the observed offset. 
Another issue is possibly related to aperture losses. MOSDEF spectra were obtained with a 0.7\arcsec\ slit, which could exclude the outer regions of galaxies \citep{Kriek2015}. 
These outer regions are likely characterized by lower ionization and lower \oo\ ratios, and the aperture corrections may not fully recover the emission line flux from these regions. In contrast, our observations include the full spatial extent of galaxies, potentially leading to an overall lower ionization. 
Additionally, we cannot rule out the possibility that differences in stellar mass estimates contribute to the discrepancy, as MOSDEF assumes a constant star formation history \citep{Sanders2021}. }
{The comparison between CLEAR and our samples could also affected by the selection method, as CLEAR extends to $m(\mathrm{F105W} < 25$~mag and $9.2 < \log(M_\ast/\mathrm{M_\odot}) < 10.2$. In addition, the emission line detection limit in CLEAR ($3\sigma$ of $2\times10^{-17}~\mathrm{erg~s^{-1}~cm^{-2}}$) is higher than that of NGDEEP and MOSDEF, it is possible that the CLEAR galaxies with detected \oii\ and \oiii\ are biased toward galaxies with elevated excitation properties compared to typical galaxies at $z\sim1.5$. }
{Nevertheless, our NGDEEP sample confirms the existence of tight mass--\oo\ relation, consistent with findings at similar redshift, and shows that it extends to lower masses ($\log(M_\ast/\mathrm{M_\odot}) \sim 8$). }

The mass--\oo\ relation clearly evolved from $z\sim0$ from SDSS to $z\sim2$ from NGDEEP, CLEAR, and MOSDEF. The median \oo\ versus \sm\ values for SDSS show an anti-correlation at low mass ($\sim10^{9} - 10^{10}$\msun) and flattened at high stellar mass. At fixed stellar mass, galaxies have a higher \oo\ at a higher redshift, and the effect is stronger for galaxies of lower stellar mass. The difference of \oo\ is 0.58 dex at \sm\ $\sim10^{9}$ \msun. 
These results suggest that galaxies at $z\sim1-3$ generally have higher ionization parameters compared to local galaxies at the same stellar mass, assuming the same relation between \oo\ and ionization parameters at both redshifts. 
{In contrast, the \oo\ ratio only slightly increases at a fixed stellar mass from $z\sim2$ to $z\sim3$ in NGDEEP and MOSDEF, suggesting a small change in ionization parameter and/or metallicity at fixed stellar mass between these redshifts. 
At $z\sim0$, the median \oo\ appears to flatten at high stellar masses $\sim 10^{10}$\msun. A similar trend is observed in the MOSDEF stacked results. 
We do not see this flattening in our sample, likely because our sample consists mainly of galaxies in the stellar mass range of $10^{8}-10^{9.2}$\msun\ (16th to 84th percentile), and lacks more massive galaxies. }

We then compare to those extreme galaxies at $z\sim0$. 
The median \oo\ ratios of CLASSY galaxies align with our mass--\oo\ relation at stellar mass $\gtrsim 10^{8} \mathrm{M_\odot}$. This consistency indicates that these galaxies have similar ISM physical conditions as galaxies at $z\sim2.5$. 
At low stellar mass $<10^{8} \mathrm{M_\odot}$, the median \oo\ ratio of CLASSY galaxies shows lower \oo\ than our extrapolated line. Due to the limited number of galaxies at these lower stellar masses, we cannot draw any definitive conclusions. 
Green peas have higher \oo\ than our $z\sim2.5$ relation at the same stellar mass, suggesting that these galaxies have higher ionization parameters and/or lower metallicity than $z\sim2.5$ galaxies. 
Blueberry galaxies have similar to higher \oo\ compared to the extended $z\sim2.5$ relation, but they extend to lower stellar mass regions. Also note that no dust correction is applied for blueberry galaxies, which would slightly lower the \oo\, though the correction is expected to be minor for galaxies with low stellar mass.

\subsection{The SFR -- \oo\ Relation} \label{sec:o32-sfr}

\begin{figure*}
    \centering
    \includegraphics[width=\textwidth]{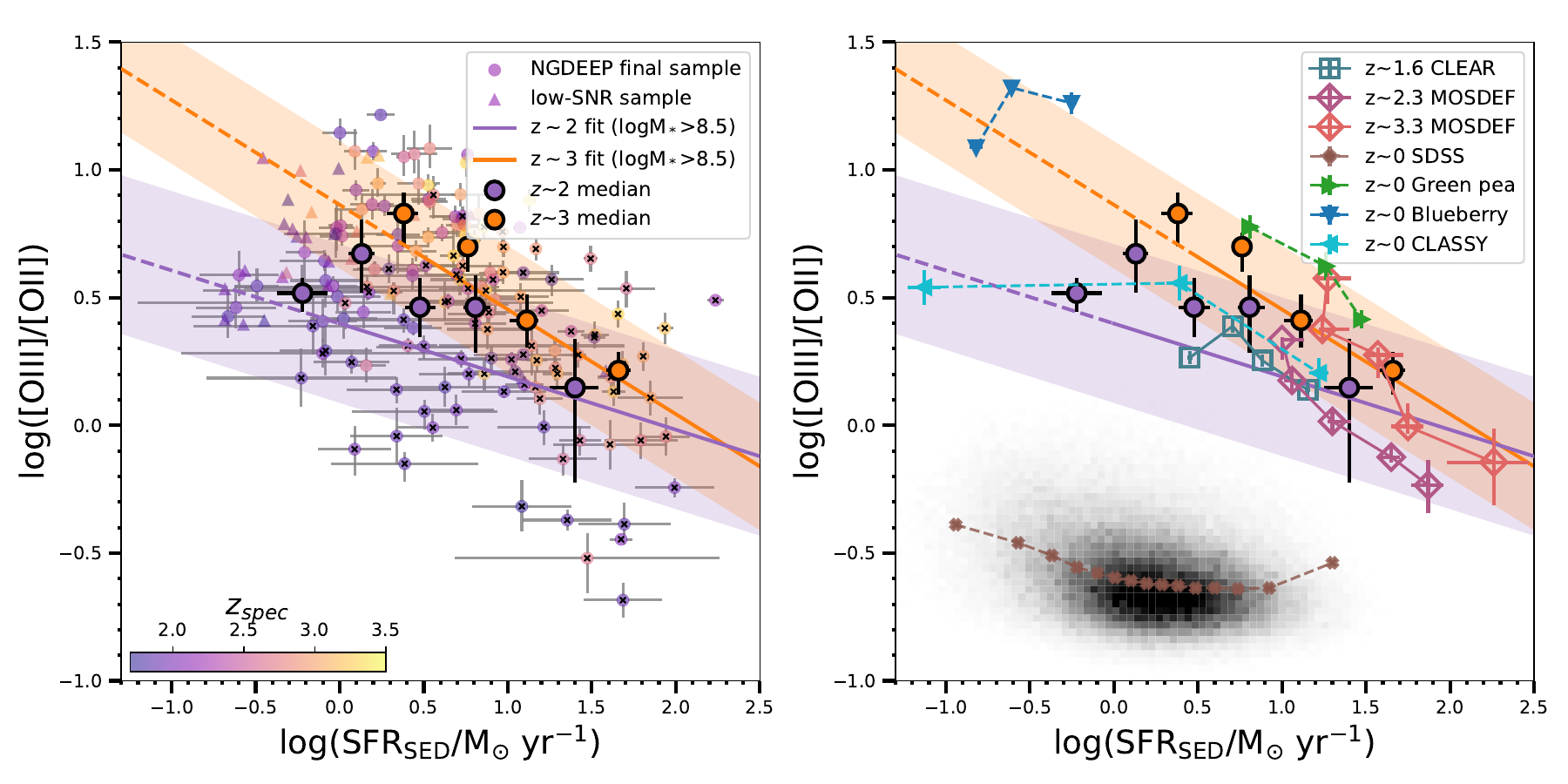}
    \caption{\textit{Left:} The \oo\ versus SFR for galaxies in the NGDEEP final sample (dots) and low-SNR sample (triangle), color-coded by redshift. The median \oo\ values along the abscissa are shown for galaxies at $z\sim2$ and $z\sim3$ in purple- and orange-filled dots, respectively. The best fit for galaxies with stellar mass $>10^{8.5}$ M$_\odot$ at $z\sim2$ and $z\sim3$ are shown as the solid purple and orange lines, and extended to lower SFR as the dashed line. The color-shaded region marks the 1$\sigma$ total scatter for each redshift sample. \textit{Right:} The median \oo\ ratios as a function of SFR are shown for CLEAR, MOSDEF, SDSS, CLASSY, green peas, and blueberry galaxies (symbols are the same as in Fig. \ref{fig:SFR-M}). The SDSS sample is shown in the 2D histograms in black. The same median values and fits from the NGDEEP \oo\ sample are shown, identical to those in the \textit{left} panel. The NGDEEP sample shows a negative SFR--\oo\ relation, with a dependence on redshift.} 
    \label{fig:o32_sfr}
\end{figure*}

In the left panel of Figure \ref{fig:o32_sfr}, we show the \oo\ ratios as a function of SFR for individual galaxies in the NGDEEP final sample and low-SNR sample. 
We find a general negative correlation between \oo\ and SFR, where \oo\ decreases as SFR increases. The Spearman correlation test returns a correlation coefficient of -0.41 and a p-value of $\sim10^{-7}$, confirming the significance of the correlation. 

We show the median values of the final sample in two redshift bins at $z\sim2$ and $z\sim3$ in the left panel of Figure \ref{fig:o32_sfr}. 
The median values reveal a clear dependence on the redshift with galaxies at higher redshift having higher \oo. 
The negative correlation between \oo\ and SFR and its redshift evolution is consistent with the redshift-dependent SFR--\sm\ relation and the tight \oo--\sm\ relation. 

We quantify the SFR-\oo\ relation for galaxies in the final sample with \sm\ $>10^{8.5}$ \msun\ and separated in two redshift bins with \linmix\ \citep{Kelly2007}. 
{The best-fit parameters are reported in Table~\ref{tab:relations}. }
The median intrinsic scatter measured by \linmix\ is 0.10 and 0.05 for $z\sim2$ and $z\sim3$ fits, respectively. 
The total scatter of the SFR--\oo\ relations are 0.31 and 0.25 for $z\sim2$ and $z\sim3$ fits, respectively. 
These fits are slightly lower than the median values, particularly at low SFR. This is because the applied mass cut for the fitting excludes some low SFR galaxies, whereas the median values are computed without any mass cut.

We compare our SFR--\oo\ relations to the local and $z\sim1-2$ samples in the right panel of Figure~\ref{fig:o32_sfr}. 
The MOSDEF at $z\sim2.3$ and CLEAR galaxies are generally consistent with the SFR--\oo\ fit at $z\sim2$, with an average difference of {$\lesssim-0.1$ dex} in \oo. 
{The MOSDEF at $z\sim3.3$ is consistent with the SFR--\oo\ fit at $z\sim3$, with an average difference of {$\sim0.02$ dex} in \oo. }
Similar to the mass--\oo\ relation, the SFR--\oo\ relation evolved from $z\sim0$ from SDSS to $z\sim1-3$ from NGDEEP, CLEAR, and MOSDEF. 
At fixed SFR, galaxies have a higher \oo\ at higher redshift, and the effect is stronger for galaxies with lower SFR. The \oo\ difference between SDSS and $z\sim2$ SFR--\oo\ relation are 1.0 dex and 0.8 dex at SFR= $1$\msun~yr$^{-1}$ and SFR=$10$\msun~yr$^{-1}$, respectively. 
Compared to those extreme galaxies at $z\sim0$, the median \oo\ ratios of CLASSY galaxies lie within our $z\sim2$ and $z\sim3$ relations.  
Green peas and blueberry galaxies on average have higher \oo\ than our $z\sim3$ and the extended $z\sim3$ relation, with their median following the upper 1$\sigma$ total scatter of the $z\sim3$ relation. 

\subsection{The sSFR -- \oo\ Relation} \label{sec:o32-ssfr}

\begin{figure*}
    \centering
    \includegraphics[width=\textwidth]{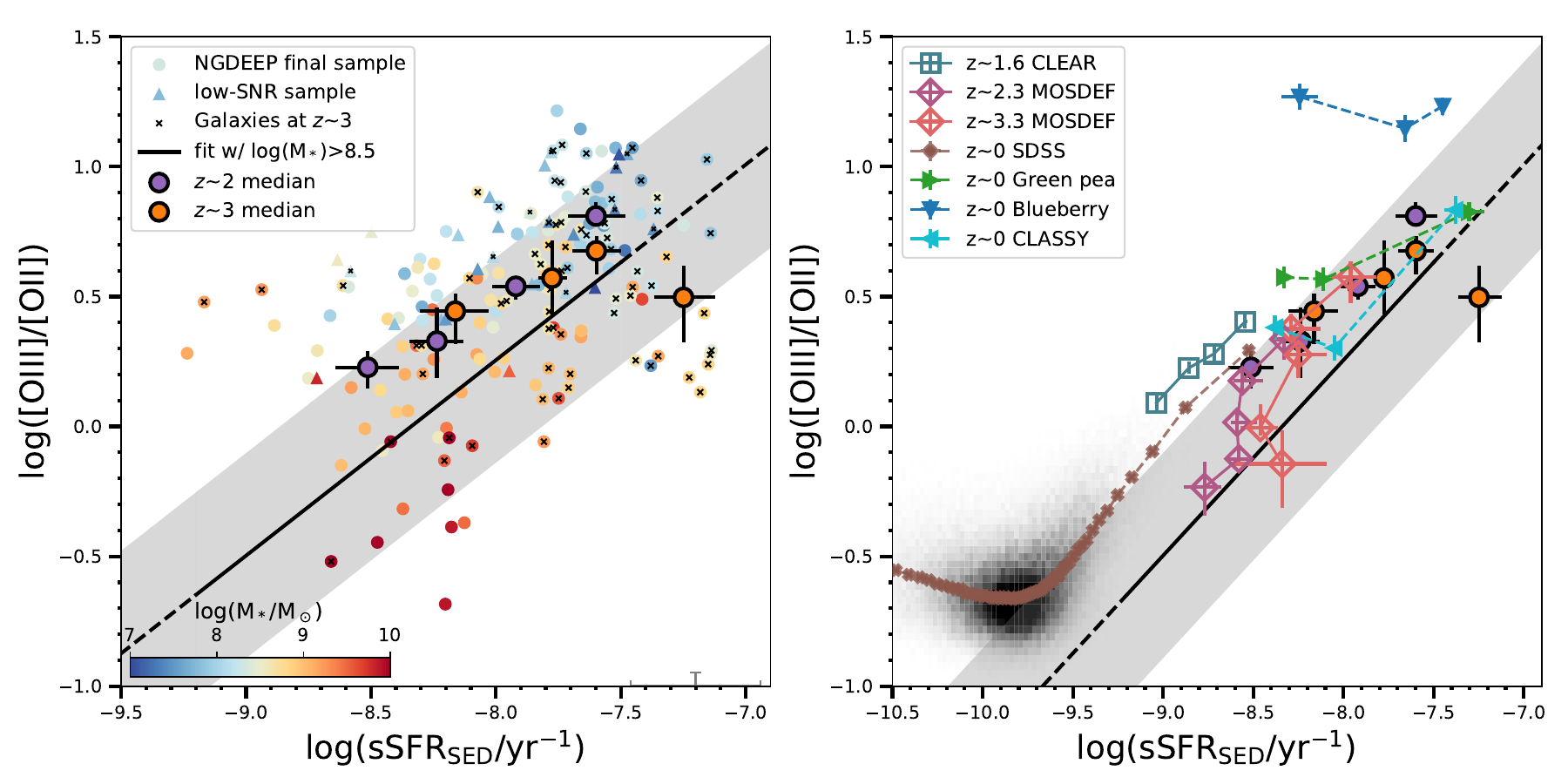}
    \caption{\textit{Left:} The \oo\ versus sSFR for galaxies in the NGDEEP final sample (dots) and low-SNR sample (triangles), color-coded by stellar mass. The median \oo\ values along the abscissa are shown for galaxies at $z\sim2$ and $z\sim3$ in purple- and orange-filled dots, respectively. The best fit for galaxies with stellar mass $>10^{8.5}$ M$_\odot$ is shown as the solid black lines, and extended to lower/higher sSFR as the dashed line. The grey region marks the 1$\sigma$ total scatter. \textit{Left:} The median \oo\ ratios as a function of sSFR are shown for CLEAR, MOSDEF, SDSS, CLASSY, green peas, and blueberry galaxies (symbols are the same as in Fig. \ref{fig:SFR-M}). The SDSS sample is shown in the 2D histograms in black. The same median values and fits from the NGDEEP final sample are shown, identical to those in the \textit{left} panel. Note that the x-axis scale of the right panel is more extended. NGDEEP galaxies show a positive sSFR--\oo\ relation without a redshift-dependence between $z\sim1.7-3.4$, but with a large scatter. }
    \label{fig:o32_ssfr}
\end{figure*}

In the left panel of Figure \ref{fig:o32_ssfr}, we show the \oo\ ratios as a function of sSFR for galaxies in the NGDEEP final sample and low-SNR sample. 
We show the median values for the final sample in two redshift bins. 
We find a significant positive correlation between the \oo\ ratio and sSFR, with \oo\ increasing as sSFR increases. 
This is confirmed by a Spearman test with a correlation coefficient of 0.46 and a p-value of $\sim10^{-9}$. 
When separating in redshift, 
{the Spearman test returns p-values of $\sim10^{-10}$ and 0.08 for galaxies at $z\sim2$ and $z\sim3$, respectively, 
suggesting a less significant correlation for $z\sim3$ galaxies. }


We quantify the sSFR--\oo\ relation for the final sample with \sm\ $>10^{8.5}$ \msun\ with \linmix\ \citep{Kelly2007}, as reported in Table~\ref{tab:relations}. 
The median intrinsic scatter measured by \linmix\ is 0.06 dex. The total scatter of the sSFR--\oo\ relation is 0.40 dex. 

Additionally, we see a diagonal stellar mass gradient as visualized by the color of galaxies with more massive galaxies located in the lower left region, and low-mass galaxies lying in the upper right region. 
Because of such a mass gradient, the fitted line is consistently lower than the median values across the sSFR range due to the mass cut applied in the fitting.

We compare our sSFR--\oo\ relation to the local and $z\sim1-2$ samples in the right panel of Figure~\ref{fig:o32_ssfr}. 
{The MOSDEF at $z\sim2.2$ and $z\sim3.3$ are generally consistent with the sSFR--\oo\ relation within the 1$\sigma$ total scatter, with a median difference of $\sim 0.2$ dex. 
The CLEAR galaxies are higher than the sSFR--\oo\ relation with a median difference of 0.5 dex. }
SDSS galaxies dominate in the low sSFR, {but have higher \oo\ than the extended of our relation.} 
The median difference between SDSS galaxies and the extended sSFR--\oo\ relation is {1.00} dex and a scatter ranging from {0.65 to 1.41} across the 16th and 84th percentiles. 

For those extreme galaxies at $z\sim0$, the median \oo\ ratios of CLASSY and green pea galaxies are located within 1$\sigma$ of our relations, with median offsets of 0.16 dex and 0.33 dex, respectively. 
Blueberry galaxies are located well above our sSFR--\oo\ relation, with a median offset of 0.71 dex. They exhibit higher \oo\ values compared to our sample at the same sSFR. 

\subsection{The EW -- \oo\ Relation} \label{sec:o32-ew}

\begin{figure*}
    \centering
    \includegraphics[width=\textwidth]{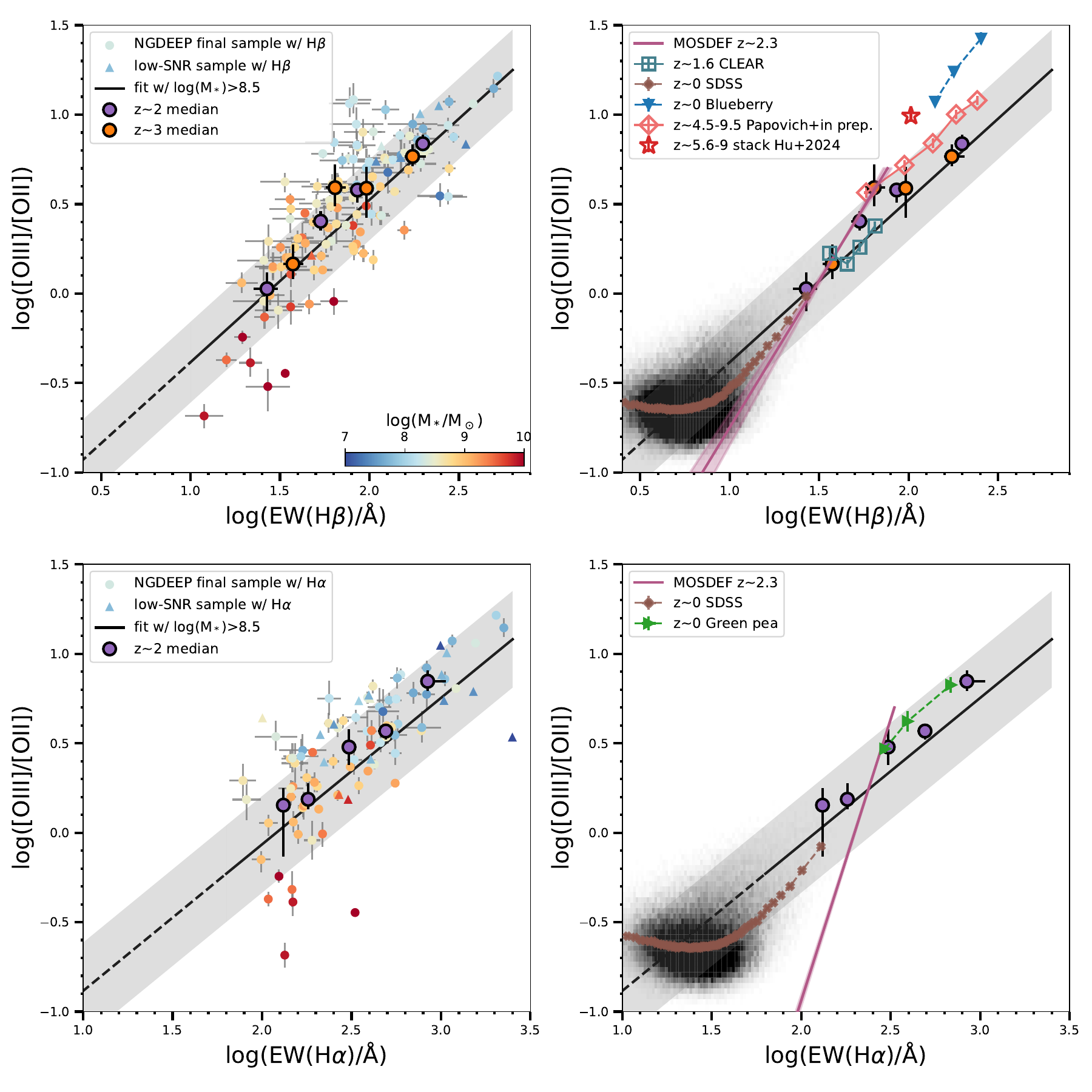}
    \caption{\textit{Lefts:} The relation between the \oo\ ratio and \hb\ EW (\textit{top}) and \ha\ EW (\textit{bottom}) for galaxies in the NGDEEP final sample and low-SNR sample with \hb\ and \ha\ SNR$>$5, respectively. Individual galaxies are color-coded by stellar mass. The median \oo\ values are shown for galaxies at $1.7< z<2.5$ and $2.5 \leq z < 3.4$ as black orange-filled and purple-filled dots, respectively. The best fit for galaxies in the final sample with $\mathrm{M_*>10^{8.4} M_\odot}$ is shown as the solid black line and extended to the low EW region as the dashed black line. \textit{Rights:} Comparison with galaxy samples at different redshift ranges. Median \oo\ ratios are shown for galaxies at $z\sim4.5-9.5$ from CEERS and JADES (Papovich et al. \textit{in prep.}, open magenta diamonds), galaxies at $z\sim5.6-9$ measured from composite spectra (\citealp{Hu2024}, red star), CLEAR galaxies at $z\sim1.1-2.3$ (open orange squares), $z\sim0$ SDSS galaxies (brown crosses), blueberry galaxies (blue triangles), green pea galaxies (green triangles). The SDSS sample is shown in black as the 2D histogram. {The relations from \citet{Reddy2018} using MOSDEF galaxies at $z\sim2.3$ are shown as purple lines.} The same median values and fits from the NGDEEP final sample are shown, identical to those in the \textit{left} panel. The NGDEEP sample shows tight EW--\oo\ relations. A clear evolutionary trend is seen as higher redshift galaxies having higher sSFR and \oo.  }
    \label{fig:o32_ew}
\end{figure*}

We study the distribution of \oo\ with \hb\ and \ha\ EW. 
The EW of \hb\ and \ha\ provides an independent method for determining the sSFR, as it is a ratio of \ha\ (\hb) flux, both of which are SFR indicators, to the underlying continuum flux, which traces the stellar mass. 
In Figure \ref{fig:o32_ew}, we show the \oo\ ratios as a function of EW(\hb) and EW(\ha) for galaxies in the NGDEEP final sample and the low-SNR sample with \hb\ and \ha\ SNR $>5$, respectively. 
We find significant positive correlations between \oo\ and EW(\hb), and \oo\ and EW(\ha). 
These are confirmed by a Spearman test with a correlation coefficient of 0.71 and 0.63 for the two relations, respectively, and p-values of $\sim10^{-11}$ and $\sim10^{-6}$, respectively. 
We quantify the EW--\oo\ relation for the NGDEEP final sample with \sm\ $>10^{8.5}$ \msun\ with \linmix. Best-fitted relations are listed in Table~\ref{tab:relations}.  

For the EW(\hb)--\oo\ relation, we do not find any significant {difference in the median values between $z\sim2$ and $z\sim3$}. 
Similar to sSFR--\oo\ relation, we see a diagonal stellar mass gradient, where, at a fixed EW (or sSFR), more massive galaxies have lower \oo\ ratios. 

In the right panels of Figure~\ref{fig:o32_ew}, we compare our EW--\oo\ relations to the local and $z\sim1.6$ samples that have available EW measurements. 
The CLEAR galaxies show an EW(\hb)--\oo\ relation consistent with ours, with a small median difference of 0.02 dex. 
{We compared with the EW(\hb)--\oo\ and EW(\ha+\nii)--\oo\ relations using MOSDEF galaxies at $z\sim2.3$ from \citet{Reddy2018}. We find that their relations are steeper than our relations, particularly for the EW(\ha). 
Similarly, the sSFR -- \oo\ trends from MOSDEF stacked spectra appear steeper than our median trends.  
The difference may be due to the different stellar mass ranges in MOSDEF and NGDEEP. The more massive galaxies of NGDEEP ($\log(M_*/\mathrm{M_\odot} \gtrsim 9$) tend to follow the relations from the MOSDEF. However, we find a flatter relation in our sample which includes galaxies at lower masses. }

SDSS galaxies mostly occupy the low \oo\ and low EW region, with a tail of high EW and high \oo\ galaxies that tend to align with our $z\sim2-3$ EW--\oo\ relation. 
We measured a relatively small difference between SDSS and our extended EW--\oo\ relations, with a median difference of {0.01 and $-$0.06} for the EW(\hb) and EW(\ha) relations, respectively. These offset values, both for SDSS and CLEAR, are much smaller than those from sSFR--\oo\ comparison, likely due to the different methods used for stellar mass and SFR measurements, while EW are all calculated in a similar way and account for stellar absorption. 
Green pea and blueberry galaxies are located above our EW--\oo\ relation, with median offsets of 0.14 dex and 0.48 dex, respectively. However, these samples were selected to be extreme emission-line galaxies, which likely accounts for their offsets.

Furthermore, we compare galaxies at $z\gtrsim5$ with our EW--\oo\ relation. Firstly, we include 232 galaxies with \oii, \oiii, and \hb\ emission line fluxes SNR $>3$ measured using the \jwst\ {NIRSpec} PRISM data taken as part of CEERS and JADES surveys (Papovich et al. \textit{in prep.}). 
The median dust-corrected \oo\ ratios as a function of EW(\hb) are shown in the top right panel of Figure~\ref{fig:o32_ew}. 
Secondly, we include the \oo\ ratio and EW(\hb) obtained based on the composite spectrum of 63 galaxies at $5.6 < z < 9$ using the \jwst\ medium resolution grating spectra (M-Grating) from CEERS and JADES survey \citep{Hu2024}. Note that this EW(\hb) might be underestimated, due to some systematic background issues shown in the composite spectrum (see Figure 2 in \citealp{Hu2024} and more discussion there). 
These $z\gtrsim5$ galaxies dominate in the high EW \hb\ region. 
The median \oo\ of $z\gtrsim5$ galaxies lie {slightly} above our $z\sim2-3$ relations with a median offset of 0.22 dex. 
We find an {mild} evolutionary trend in \oo\ and EW(\hb) from $z\sim0$ to $z\gtrsim5$ that as higher redshift galaxies have higher EW(\hb) and higher \oo. 
These $z\gtrsim5$ galaxies seem to occupy a region similar to green pea galaxies and blueberry galaxies, suggesting they have similar star formation and ISM properties (i.e., ionization parameter and metallicity). 
{We acknowledge potential biases towards emission line galaxies of these $z\gtrsim5$ galaxies, which will need future surveys with deeper observations to test. See more discussion in Section \ref{sec:disc-bias}. }


\begin{deluxetable*}{lcccccc}
\tablecaption{Summary of correlations between \oo\ and galaxies properties \label{tab:relations}}
\tablewidth{0pt}
\tablehead{
\colhead{Properties} & \colhead{redshift range$^{a}$} & \colhead{$c_1$$^{b}$} & \colhead{$c_0$$^{b}$} & \colhead{intrinsic scatter$^{c}$} & \colhead{total scatter$^{d}$} & \colhead{p-value$^{e}$} }
\startdata
M$_\ast$ & $1.7<z<3.4$ (<$z$>=2.3) & -0.52$\pm$0.06 & 4.97$\pm$0.59 & 0.05$\pm$0.01 & 0.24 & $10^{-25}$\\
SFR  & $1.7<z<3.4$ (<$z$>=2.3) & -0.21$\pm$0.08 & 0.50$\pm$0.09 & 0.09$\pm$0.01 & 0.30 & $10^{-7}$\\ 
SFR  & $1.7<z<2.5$ (<$z$>=2.0) & -0.21$\pm$0.11 & 0.40$\pm$0.11 & 0.10$\pm$0.02 & 0.31 & $10^{-3}$\\ 
SFR & $2.5<z<3.4$ (<$z$>=2.9) & -0.41$\pm$0.10 & 0.86$\pm$0.14 & 0.05$\pm$0.01 & 0.25 & $10^{-10}$\\ 
sSFR & $1.7<z<3.4$ (<$z$>=2.3) & 0.74$\pm$0.22 & 6.18$\pm$1.76 & 0.06$\pm$0.02 & 0.47 & $10^{-9}$\\ 
EW(\hb) & $1.7<z<3.4$ (<$z$>=2.3) & 0.90$\pm$0.11 & -1.29$\pm$0.20 & 0.05$\pm$ 0.01 & 0.23 & $10^{-11}$\\ 
EW(\ha) & $1.7<z<3.4$ (<$z$>=2.3) & 0.82$\pm$0.19 & -1.69$\pm$0.44 & 0.08$\pm$0.02 & 0.27 & $10^{-6}$\\ 
\enddata
\tablecomments{$^{a}$ Only the SFR--\oo\ relation shows a significant redshift dependence, so we separate the SFR--\oo\ relation into two redshift bins. The EW(\ha)--\oo\ relation is restricted to galaxies at $1.7<z<2.3$ as our NIRISS data cover \ha\ emission line up to $z=2.3$.  \\
$^{b}$ $\log$\oo $= c_0 + c_1\times x$, where $x$ is the logarithm of the galaxy property in the first column of each row. \\
$^{c}$ The median intrinsic scatter is measured from \linmix {, and the uncertainty is derived from the average difference between 16th and 84th percentiles}, in a unit of dex. \\
$^{d}$ The total scatter {is measured by summing the squared deviations between data and the best-fit relation}, in a unit of dex. \\
$^{e}$ The p-value is measured from the Spearmen rank correlation test. }
\end{deluxetable*}

\section{Discussion} \label{sec:discussion}

We compared the \oo\ ratio with galaxy properties including stellar mass, SFR, sSFR, and EW of \ha\ and \hb\ for NGDEEP galaxies at $z\sim1.7-3.4$ in the final sample (i.e., those galaxies with \oii\ and \oiii\ $>$5$\sigma$ detections). 
The \oo\ ratio increases with (1) decreasing stellar mass, (2) decreasing SFR, (3) increasing sSFR, and (4) increasing EW of \ha\ and \hb.  
We quantify the significance and scatter of these relations and summarize them in Table~\ref{tab:relations}.
{These correlations are qualitatively consistent with results from previous studies at both low and high redshifts (e.g., \citealp{Nakajima2014, Kewley2015, Sanders2016, Bian2016, Kaasinen2018, Kashino2019, Sanders2020, Sanders2021, Papovich2022}). }

{We compare relations from MOSDEF at similar redshift $z\sim2-3$ but derived from more massive galaxies. 
Galaxies at low mass ($\log(M_\ast/\mathrm{M_\odot}) < 9$) generally follow the same trend as more massive galaxies in the mass-\oo\ relation. 
However, we find a flattening in SFR--, sSFR-- and EW--\oo\ relations at a low SFR, high sSFR, and high EW ($\log (\mathrm{sSFR/yr^{-1}}) \gtrsim -8$). 
These regions are dominated by low-mass galaxies ($\log(M_\ast/\mathrm{M_\odot}) < 9$), suggesting a non-linear correlation between \oo\ and galaxies properties. 
These imply either a non-linear correlation at low masses between ionization (or metallicity) and galaxy properties at low masses or a non-linear correlation between \oo\ and ionization (or metallicity). We further explore this in Section \ref{sec:disc-o32}.  }


The NGDEEP final-sample galaxies at $z\sim2-3$ have on average higher \oo\ compared to local normal star-forming galaxies from SDSS at the same stellar mass and SFR. This is consistent with other findings for galaxies at $1 \lesssim z \lesssim 3$ \citep{Sanders2016, Papovich2022, Sanders2021}.  
We interpret this as being driven by higher sSFR values in $z\sim2-3$ galaxies, a manifestation of ``downsizing''. 
On the other hand, $z\sim2-3$ galaxies have similar to lower \oo\ than those of extreme local galaxies from low-redshift samples such as CLASSY, and the SDSS Green pea, and blueberry galaxies at fixed stellar mass, SFR, sSFR, and EW. 
Therefore these local samples are more typical of the high-\oo\ tail of the high-redshift samples than of the typical galaxy in the NGDEEP final sample.

Our NGDEEP sample spans a wide range of \oo, stellar mass, and sSFR (or EW), extending to lower stellar masses and higher sSFR than previously studied at these redshifts. 
This helps bridge the gap between the local galaxies and galaxies at $z>5$.  
Compared to the higher redshift galaxies, we find an evolutionary trend in EW(\hb) -- \oo\ from $z\sim0$ to $z\gtrsim5$.  
Higher redshift galaxies have {slightly} higher \oo\ at fixed EW(\hb) (see Figure~\ref{fig:o32_ew}).
However, the evolutionary trend in the EW(\hb)--\oo\ relation is mild, {as the offset in the \oo\ ratios for SDSS and galaxies $z\gtrsim5$, compared to NGDEEP galaxies, fall within the total scatter of NGDEEP sample. }
This contrasts with the strong evolution in \oo\ at fixed stellar mass and SFR. 
This merits future study to understand how this connects to the physical properties of the galaxies as a function of redshift, which we will explore in a future paper.


{The mass--\oo\ and EW(\hb)--\oo\ relations have the smallest intrinsic and total scatters among properties studies in this paper. On the other hand, the SFR--\oo\ relation has a larger scatter and shows redshift dependence, suggesting that it is more likely a byproduct of the correlation between SFR--mass and mass--\oo\ relations. } 
The \oo\ ratio is sensitive to the ionization parameter. \citet{Papovich2022} found a tight correlation between \oo\ and ionization parameter derived by modeling the \oii, \oiii, and \hb\ of the CLEAR galaxies with the MAPPINGS V photoionization models \citep{Sutherland1993, Dopita1996}. They also found a secondary effect that some of the galaxies with high \oo\ ratio shift below the \oo--ionization relation, likely due to the effect of metallicity \citep{Strom2018}. 
This metallicity dependence has been previously seen in photoionization models \cite{Kewley2013}, and the existence of anti-correlation between \oo\ and metallicity up to $z\sim9$ (e.g., \citealp{Maiolino2008, Jones2015, Sanders2024}). 
Because the \oo\ ratio has a dependence on metallicity, the evolution in the relation between \oo\ and galaxy properties would be a consequence of, or largely driven by, the evolution of the mass--metallicity relation (e.g., \citep{Erb2006, Maiolino2008, Zahid2011, Zahid2014, Sanders2018, Curti2020, Sanders2021}). 

In the remainder of this section, we consider potential biases in our results, and we explore the implications of some of the trends further.  As mentioned above (Section~\ref{sec:sample}), all galaxies in the NGDEEP full sample have \oiii\ detections above $5\sigma$, which potentially introduces selection bias. We investigate this further and discuss any potential impacts on our results in Section~\ref{sec:disc-bias}. 
We explore the dependence of the \oo\ ratio on ionization parameters and metallicity in Section \ref{sec:disc-o32}. 
Finally, we discuss the effect on ionization parameters that drives the changes in the \oo\ ratio in Section~\ref{sec:disc-q}. 

\subsection{Testing for Selection Bias} \label{sec:disc-bias}

\begin{figure}
    \centering
    \includegraphics[width=\columnwidth]{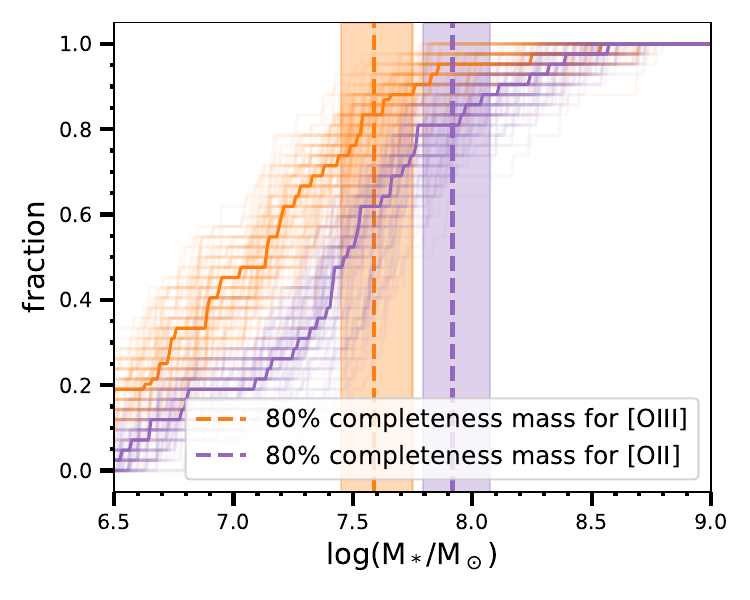}
    \caption{{Median fraction of galaxies with \oiii\ or \oii\ detection above our 5$\sigma$ emission line flux limit from down-scaling galaxies as orange and purple solid lines. The fraction distributions of 100 bootstrapping iterations are shown as faint lines. Our empirically derived 80\% mass-completeness limits are indicated by vertical dashed lines, and the 16th to 80th percentiles are shown as shaded regions. }}
    \label{fig:oiii_flux}
\end{figure} 


During the sample selection (see Section~\ref{sec:sample}), we first choose galaxies with reliable redshift from the NIRISS grism data. Because \oiii\ is the strongest emission line in our sample, we primarily rely on its detection for the redshift determination. However, this approach may introduce a selection bias as it could limit our detection of galaxies with fainter \oiii\ emission.  
{In addition, our analysis also requires \oii\ to be detected, which may lead to missing galaxies with low stellar mass. This is evident in Figure~\ref{fig:o32_mass}, where low-SNR sample (i.e., \oii\ SNR $<5$) dominated in low mass region. }
{To evaluate the completeness of our sample given the requirement for \oiii\ or \oii\ detection, we adapted and modified the mass-completeness limit method from \citet{Tomczak2014}, bootstrapping a completeness limit from the data.
We select galaxies that have \oiii\ (or \oii) detected $3\times$ above our 5$\sigma$ emission line flux limit ($1.35\times10^{-18} erg/s/cm^2$, see Figure~\ref{fig:lineflux}) and stellar mass $\log(M_\ast/\mathrm{M_\odot}) > 9$ (0.5 dex above the mass completeness limit from Section~\ref{sec:sample}).  This restricts the sample of galaxies to be sufficient above the detection limits that they should be reasonably complete. 
We obtain 42 galaxies that meet these criteria, with all of them having \oiii\ and \oii\ above 15$\sigma$ emission line flux limit. 
We then randomly select galaxies from this sample with replacement and scale down their masses and emission line fluxes assuming a mass-to-light ratio of 1.54, derived from a fit between stellar mass and F150W flux. 
From this scaled-down sample, we then calculate the fraction of galaxies with \oiii\ (\oii) above our 5$\sigma$ emission line flux limit. We then repeat this process by lowering the stellar mass of the selected sample. 
In this way, we compute the fraction of galaxies recovered as a function of stellar mass.  
From 100 bootstrapping iterations, we take the median stellar mass that encompasses 80\% of the galaxies as the mass-completeness limit. 
From this technique, we measure an 80\% mass-completeness limit of $\log(M_\ast/\mathrm{M_\odot}) =7.59_{-0.14}^{+0.16}$ for \oiii\ and $\log(M_\ast/\mathrm{M_\odot}) =7.91_{-0.13}^{+0.16}$ for \oii.
These limits are lower than the mass completeness derived in Section \ref{sec:sample} (Figure ~\ref{fig:masshist}), minimizing the selection bias introduced by the requirement of \oiii\ and \oii\ detection.  }

With this said, our sample selection also requires that galaxies have rest-frame $UVJ$ colors that classify them as ``star-forming'' (see Section~\ref{sec:sample} and Table~\ref{tab:sample}).  Therefore our sample, by construction, is devoid of quiescent galaxies at these stellar masses and redshifts, which likely contain objects with low or absent nebular emission. Our results therefore apply only to the star-forming population. 
Our sample is biased to higher SFRs at fixed stellar mass compared to measurements of the SFMS at the same redshifts (see Section~\ref{sec:SFMS} and Figure \ref{fig:SFR-M}). 
This bias is likely related to the fact that our sample is based on emission-line selection, while most measurements of the SFMS used photometric bands with color-selection of star-forming galaxies (e.g., \citealp{Whitaker2014, Tomczak2016}). 
We also note that the SFR values are calculated in different methods that we adopt SFR from SED fitting and well-matched with \ha-derived SFR, while, \citet{Whitaker2014} adopted SFR calculated from modeling the broadband imaging form the rest-frame UV to mid/far-IR. 

{In Section \ref{sec:o32-ew}, we compared our EW--\oo\ relation with those derived from $z\sim5$ samples. These high redshift samples are from CEERS and JADES spectroscopic samples which could be biased towards emission line galaxies and affected by their pre-selection methods.  Nevertheless, we see only mild evolution toward higher \oo\ values at fixed EW from $z\sim 2$ to $z >4.5$. Any bias would likely act to bring these closer in line. Future studies with deeper spectroscopy, such as \citet{Dickinson2024} will test for bias and better measure this evolution. }

\subsection{The Dependence of \oo\ on Ionization Parameter and Metallicity} \label{sec:disc-o32}

\begin{figure}
    \centering
    \includegraphics[width=\columnwidth]{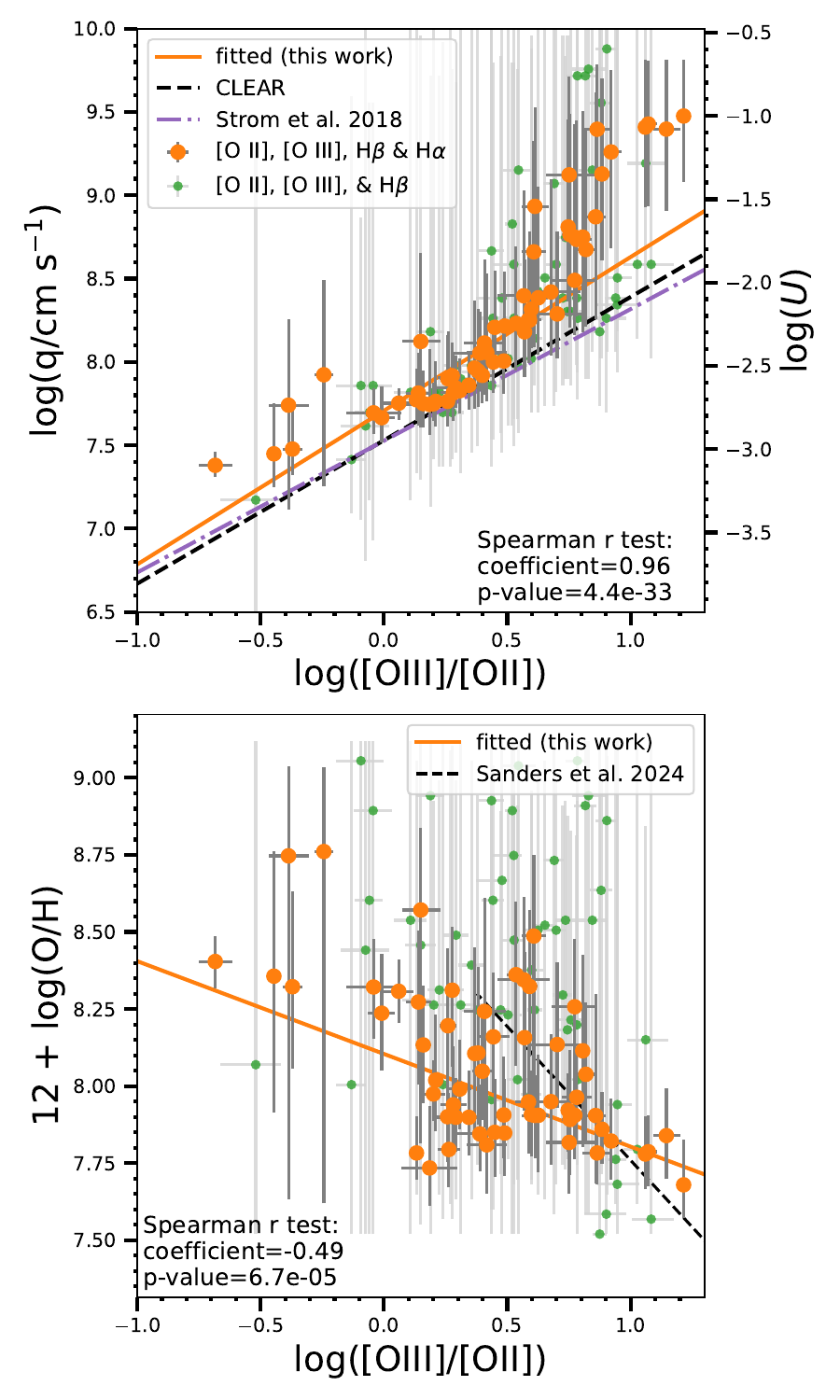}
    \caption{The distribution of ionization parameter ($q$, \textit{top}) and metallicity (12+log(O/H), \textit{bottom}) as function of \oo. Galaxies with all four emission lines $>5\sigma$ and with three lines (\oii, \oiii, and \hb) $>5\sigma$ are shown in orange and green, respectively. The orange solid lines show a linear fit to the NGDEEP galaxies with all four emission line detection. The \textit{top} panel shows the relations from CLEAR galaxies \citep{Papovich2022} (black dashed line) and a sample of $z\sim2.3$ SFGs from \citet{Strom2018} (purple dashed line). The \textit{bottom} panel shows the relation derived from a sample of $z\sim2-9$ galaxies from \citet{Sanders2024} (black dashed line). While \oo\ correlates with both the ionization parameter and metallicity, the relationship between \oo\ and the ionization parameter is more pronounced. }
    \label{fig:q_o32}
\end{figure}

To investigate the \oo\ ratio dependence on ionization parameter and metallicity, we applied the Python version of code ``Inferring the gas-phase metallicity (Z) and Ionization parameter'' (\izi, \citealp{Blanc2015}) developed by \citet{Mingozzi2020} to our dataset. \izi\ is a Bayesian code that computes posterior likelihoods for the gas-phase metallicity and the ionization parameter by comparing the measured emission line fluxes to predictions from the photoionization models. Based on the original \izi, the new Python version adds the option of using a more efficient Markov-chain Monte-Carlo (MCMC) algorithm and simultaneously fits the dust attenuation following the \citet{Calzetti2000} extinction law. 
However, the dust-attenuation modeling requires coverage of both \hb\ and \ha, while \oii, \oiii, and \hb\ are the minimum needed to constrain metallicity and the ionization parameter. 
For the 60 galaxies in the NGDEEP final sample with all four emission lines (\oii, \oiii, \hb, and \ha+\nii) detected with SNR $>5$, we use the MCMC algorithm. For the 57 galaxies in the NGDEEP final sample only with \oii, \oiii, and \hb\ SNR $>5$, we use the original \izi\ algorithm and the dust-corrected fluxes following Section~\ref{sec:dustcorr}. 

For the photoionization models, we adopt the MAPPINGS V grid models \citep{Kewley2019, Kewley2019b}. 
We generated isobaric models with the ISM pressure of $\log (P/k) [\mathrm{K~cm^{-3}}] = 7$, a grid of ionization parameters $\log(q)~[\mathrm{cm~s^{-1}}]$ from 6 to 10 in intervals of 0.25 {(corresponding to $\log(U)$ from -4.5 to -0.5)}, and a grid of metallicities ($Z$) with [0.05, 0.2, 0.4, 1.0] $Z_\odot$ \citep{Jung2024}. 
The choice of ISM pressure is motivated by the combination of median electron density, $n_e \sim 250-300~cm^{-3}$ for galaxies at $z\sim2-3$ \citep{Sanders2016, Strom2017} and the nebular temperature of $\sim$10,000-20,000K \citep{Sanders2020}. 
We also tested models with $\log (P/k) = 6.5$ and $\log (P/k) = 7.5$. Adopting these models does not alter our conclusions. 

We show the ionization parameter as a function of \oo\ for our NGDEEP final sample in the top panel of Figure~\ref{fig:q_o32}. The ionization parameter increases as \oo\ increases. The Spearman correlation test returns a correlation coefficient of 0.96 and a p-value of $\sim10^{-33}$ with all four emission lines, and a p-value of $\sim10^{-42}$ when including galaxies with three emission lines. 
We fit a linear relation between $\log$(\oo) and $\log(q)$ for galaxies with all four emission lines using \linmix:
\begin{equation}
    \mathrm{log}~q = (7.71 \pm 0.04) + (0.91 \pm 0.10) \times \mathrm{log(O_{32})}
\end{equation}
where the ionization parameter, $q$, is in the units of $\mathrm{cm~s}^{-1}$. The total scatter of the \oo--$q$ relation is 0.32 dex. 

\begin{figure*}
    \centering
    \includegraphics[width=0.45\textwidth]{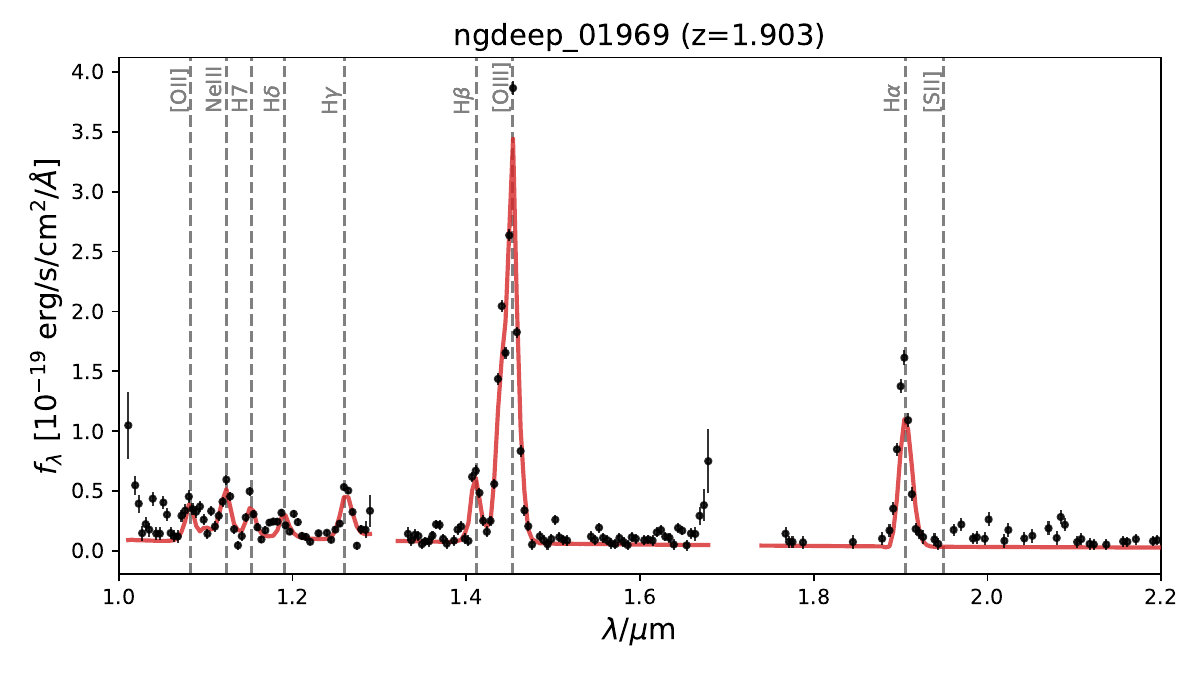}
    \includegraphics[width=0.45\textwidth]{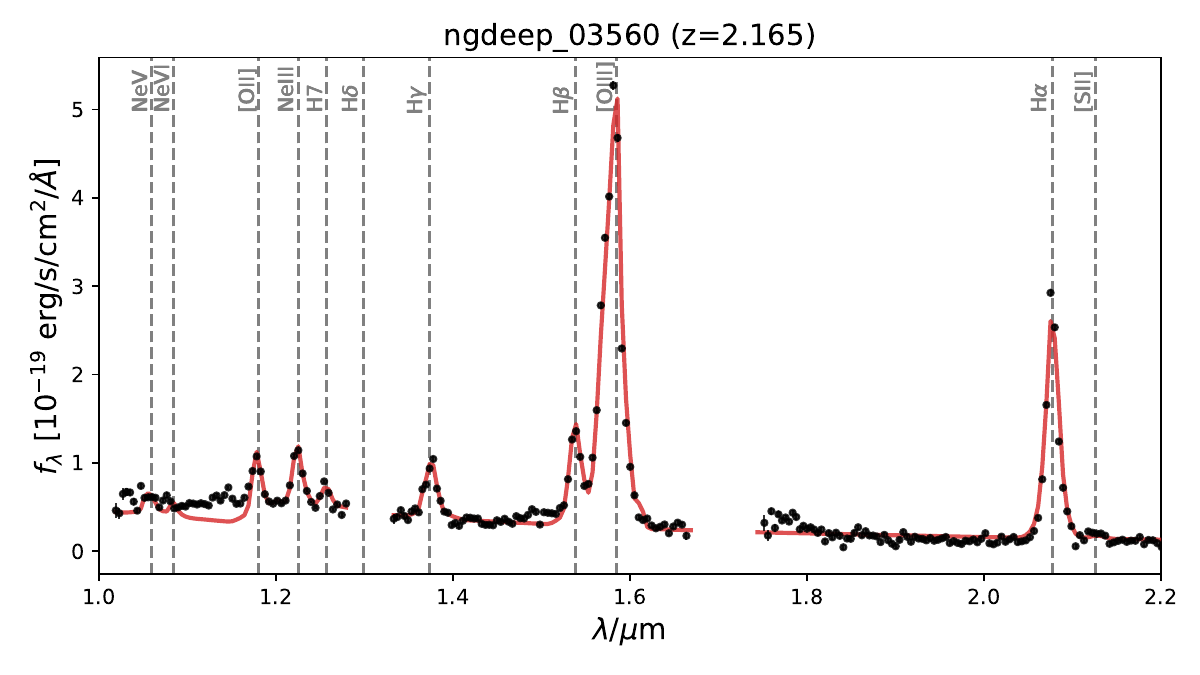}
    \includegraphics[width=0.45\textwidth]{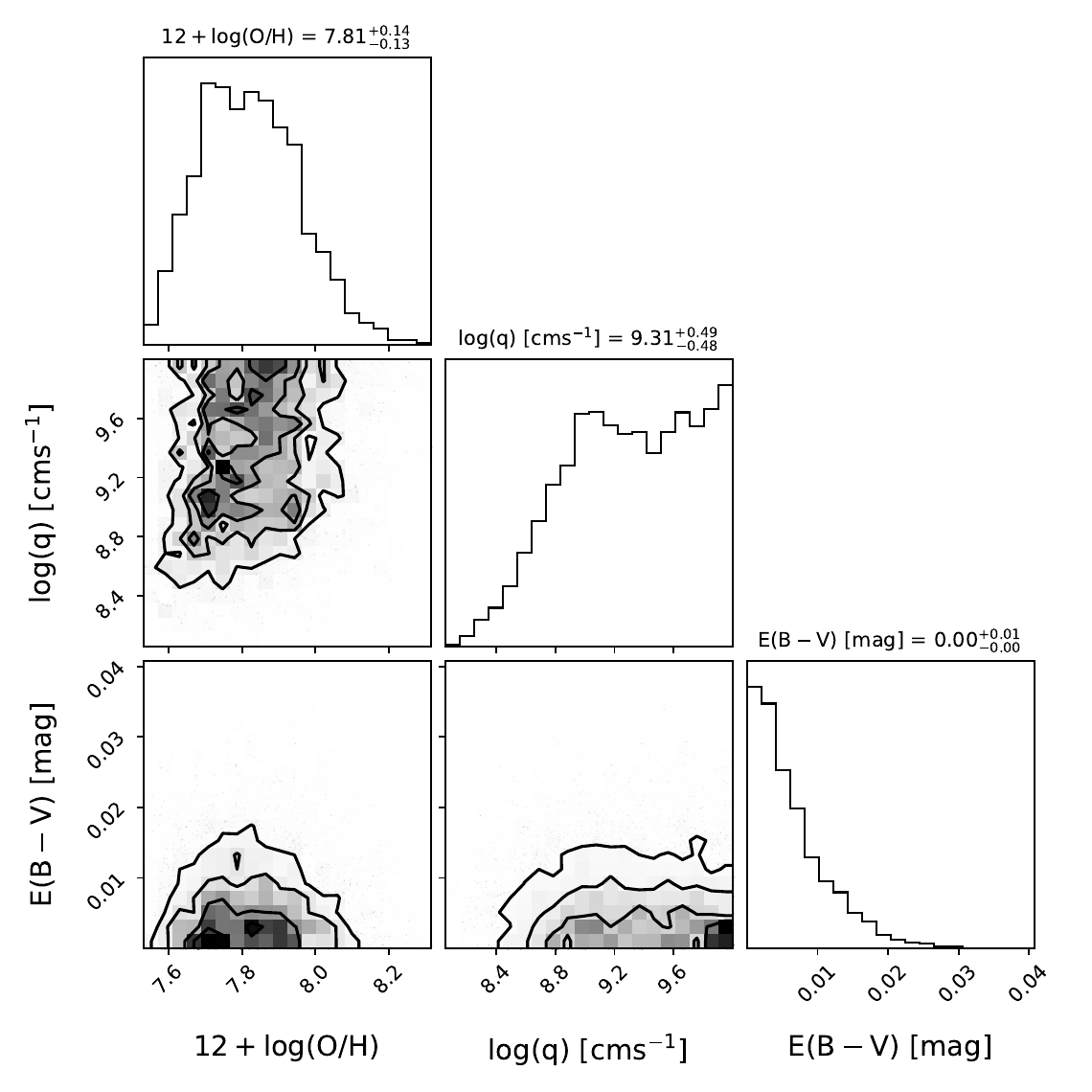}
    \includegraphics[width=0.45\textwidth]{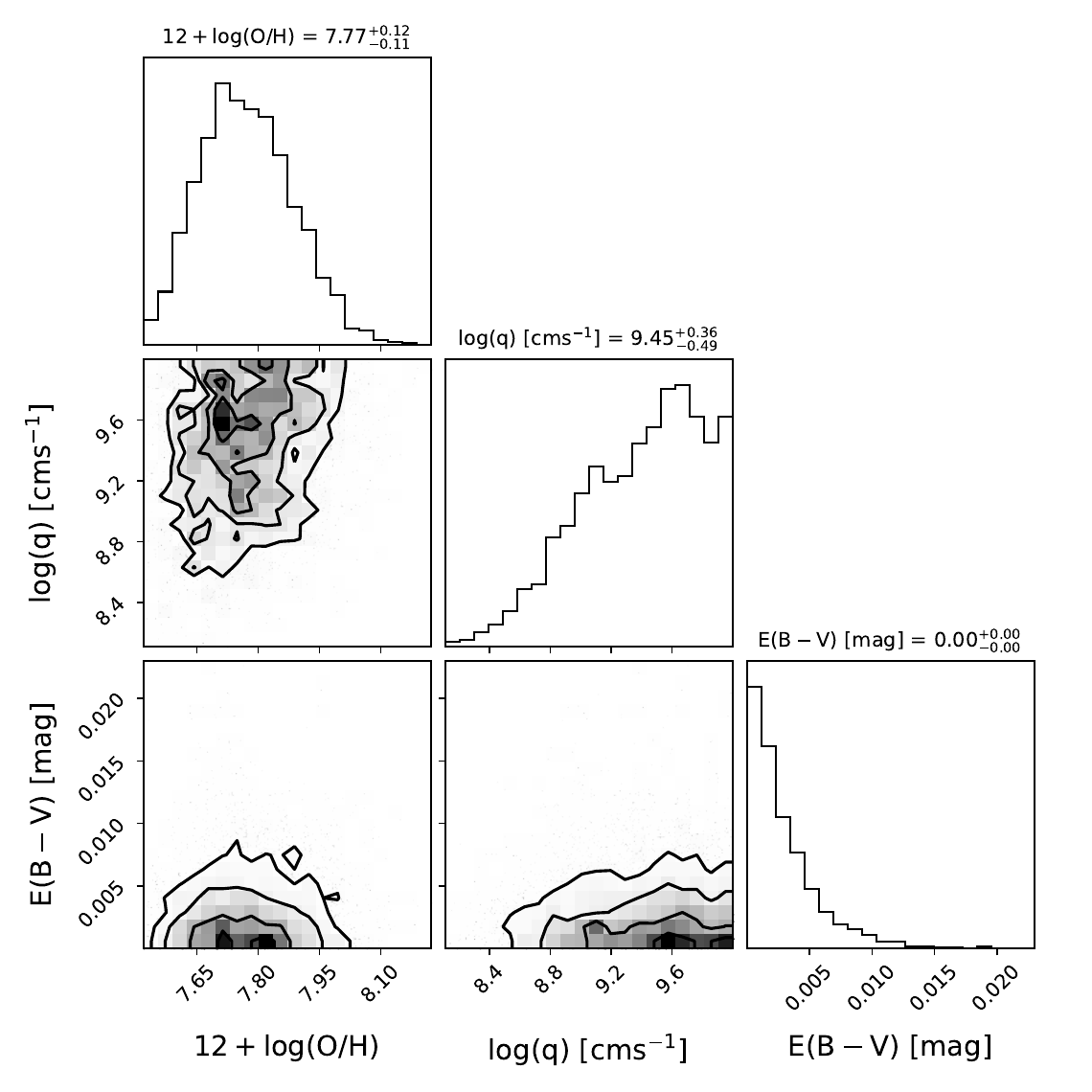}
    \caption{\textit{Top:} 1D spectrum of two galaxies with log(\oo)>0.75 and \izi-derived $E(B-V)\simeq0$. Each panel shows the NIRISS grism data (black dots) along with the best-fit spectrum  (red line). Important emission lines are marked by vertical grey dashed lines. \textit{Bottom:} The corner plot of $12+\log(\mathrm{O/H})$, $\log(q)$, and $E(B-V)$ from \izi\ for the same galaxies as \textit{top} panel. The median and associated uncertainties are displayed on the top of the histograms. }
    \label{fig:1dspec_izi}
\end{figure*}

{The large scatter is primarily driven by high \oo\  ($\log(\mathrm{O_{32}})>0.8$ or $\log(q)>8.5$), where} galaxies have higher $\log(q)$ values than predicted by the linear fit. 
{It is known that $\log(O_{32})$ versus $\log(q)$ is sub-linear at the high \oo\ where the slope of log(\oo)--log(q) relation decreases at high $q$ \citep{Kewley2002, Sanders2016}, or in the log(q)--log(\oo) plot (as shown in Figure~\ref{fig:q_o32}), the slope increases at high \oo. }
{Meanwhile, it is interesting that among galaxies with all four emission lines detected, 10 galaxies with log(\oo) $>$ 0.8, 8 of them have no dust attenuation with $\mathrm{E(B-V)_{gas}}=0$, and the remaining 2 have low dust attenuation with $\mathrm{E(B-V)_{gas}}\sim0.09$. 
These galaxies also appear to have low metallicity, as shown in the bottom panel of Figure~\ref{fig:q_o32}. 
The high ionization and low metallicity of these galaxies suggest that they may have a high electron temperature in these galaxies (e.g., \citealp{Yates2020}). 
Because the intrinsic ratio of Balmer decrement is temperature dependent, with 2.86 for $10^4$ K and decreasing to 2.79 for $1.5\times10^4$ K, a lower intrinsic ratio would be appropriate for these galaxies. 
However, in \izi\ (and as adopted in this paper), the intrinsic ratio of \ha/\hb\ is assumed to be 2.86. This assumption may force} \izi\ to increase the ionization parameter to model the emission line ratios to account for the low $\mathrm{E(B-V)}$ in these galaxies. 
We show the 1D spectra of two examples of these galaxies and their PDFs from \izi\ in Figure~\ref{fig:1dspec_izi}.
%


This linear fit is slightly steeper than those obtained from other studies at these redshifts \citep{Strom2018,Papovich2022}. 
This difference mostly appears in the high \oo\ region, where the majority of galaxies from \citet{Papovich2022} and \citet{Strom2018} do not extend to these high \oo\ ratios. 
Therefore, our results suggest the linear relation between \oo\ and the ionization parameter breaks down at high values of this ratio ($\log(\mathrm{O_{32}})>0.8$). 
%


We show the metallicity as a function of the \oo\ ratio for our NGDEEP final sample in the bottom panel of Figure~\ref{fig:q_o32}. The metallicity decreases with \oo\ increases. The Spearman correlation test returns a correlation coefficient of $-0.49$ and a p-value of $\sim10^{-5}$ with all four emission lines and a p-value of $\sim10^{-4}$ when including galaxies with three emission lines. 

We quantified the (anti-)correlation between log(\oo) and 12+log(O/H) using galaxies with all four emission lines (\oii, \oiii, \hb, and \ha+\nii) using \linmix:
\begin{equation}
    \mathrm{12+log(O/H)}  = (8.10 \pm 0.04) + (-0.30 \pm 0.06) \times \mathrm{log(O_{32})},
\end{equation}
where the 1$\sigma$ total scatter is 0.22 dex. 
Our fitted linear relation is flatter than that from \citet{Sanders2024}, who derived their relation using metallicity from the direct $T_e$ method from a sample of galaxies at $z\sim2-9$. 
However, our data points are scattered around their fit at high \oo\ region or low metallicity regions. Our sample extends to low \oo\ region where \citet{Sanders2024} did not cover, while this region predominantly drives our fit. 
Overall, our results suggest that the \oo\ ratio is a strong tracer of gas ionization parameter, with a secondary dependence on metallicity.

The tight mass--\oo\ relation with a scatter of 0.24 dex is difficult to explain by the scatter of \oo--metallicity calibrations ($\sim$0.29 dex from \citealt{Sanders2024} and 0.22 dex from our data) and the scatter of mass--metallicity relation ($\sim$0.14 dex from \citealt{Sanders2021}). 
This scatter is also revealed from the distribution of $\mathrm{O_{32}}$ and $\mathrm{R_{23}}$ $\equiv$(\oiii+\oii/\hb) which is sensitive to the gas-phase metallicity \citep{Nakajima2014, Sanders2016, Papovich2022}. 
Although, we cannot rule out the possible effect of the metallicity on the mass--, SFR-- and sSFR--\oo\ relations, the ionization parameter may play a more important role in these \oo\ -- galaxy properties relations.


\subsection{The Relation between the \oiii/\oii\ Ratio, Ionization Production, and SFR}
\label{sec:disc-q}


As the \oo\ ratio primarily depends on the ionization parameter, in this section, we focus on the possible effect on the ionization parameter that drives the changes in the \oo\ ratio. 
The ionization parameter is, by definition, the ratio of the hydrogen ionizing photon flux and the density of hydrogen atoms. 
For a radiation-bounded nebula, the average ionization parameter can be written as: 
\begin{equation} \label{eq:U}
    q = (\frac{3Q n_\mathrm{H} \epsilon \alpha_B^2}{4\pi})^{1/3}
\end{equation}
where $Q$ is the ionizing photon production rate, $\alpha_B$ is the case-B hydrogen recombination coefficient, $n_{H}$ is the number density of hydrogen, and $\epsilon$ is the volume filling factor of the gas. Here, we assume no change in $\alpha_B$ and $\epsilon$ for galaxies with different SFR and \sm\ and across redshift (see more discussion in \citealp{Papovich2022}).
Thus, the ionization parameter in this case depends only on $Q$ and $n_\mathrm{H}$. 

\subsubsection{Ionizing photon production rate} \label{sec:Q}

The ionizing photon production rate depends on the properties of massive stars, their evolution, and the IMF. 
We estimate the ionizing photon production rate $Q$ following \citet{Leitherer1995} and assuming an ionizing photon escape fraction of zero:
\begin{equation}
    Q = 7.35 \times 10^{11}  L(\mathrm{H}\alpha)\  \mathrm{[s^{-1}]}
\end{equation}
where the $L(\mathrm{H}\alpha)$ is the dust-corrected \ha\ luminosity in units of erg~s$^{-1}$. 
{$L(\mathrm{H}\beta)$ is converted to $L(\mathrm{H}\alpha)$ assuming the intrinsic ratio of \ha/\hb\ $=2.86$ (as we discussed in Section \ref{sec:disc-o32}, this may not be valid for all galaxies in our sample. }
We also adopt the ionizing production efficiency $\xi_{ion}$ defined as the ionizing photon production rate per unit UV luminosity, as follows: 
\begin{equation}
    \xi_{ion} = \frac{Q}{L(UV)} \ \mathrm{[s^{-1}/erg~ s^{-1}~ Hz^{-1}]}
\end{equation}
where $L(\mathrm{UV})$ is the intrinsic UV luminosity at rest-frame 1500~\AA. We adopt the rest-frame FUV luminosity from \cigale\ SED fits for each galaxy in our sample and convert it to intrinsic UV luminosity by correcting the dust with \ebvs\ from SED fitting and following the \citet{Calzetti2000} law (see more details in Section~\ref{sec:dustcorr}).

\begin{figure}
    \centering
    \includegraphics[width=\columnwidth]{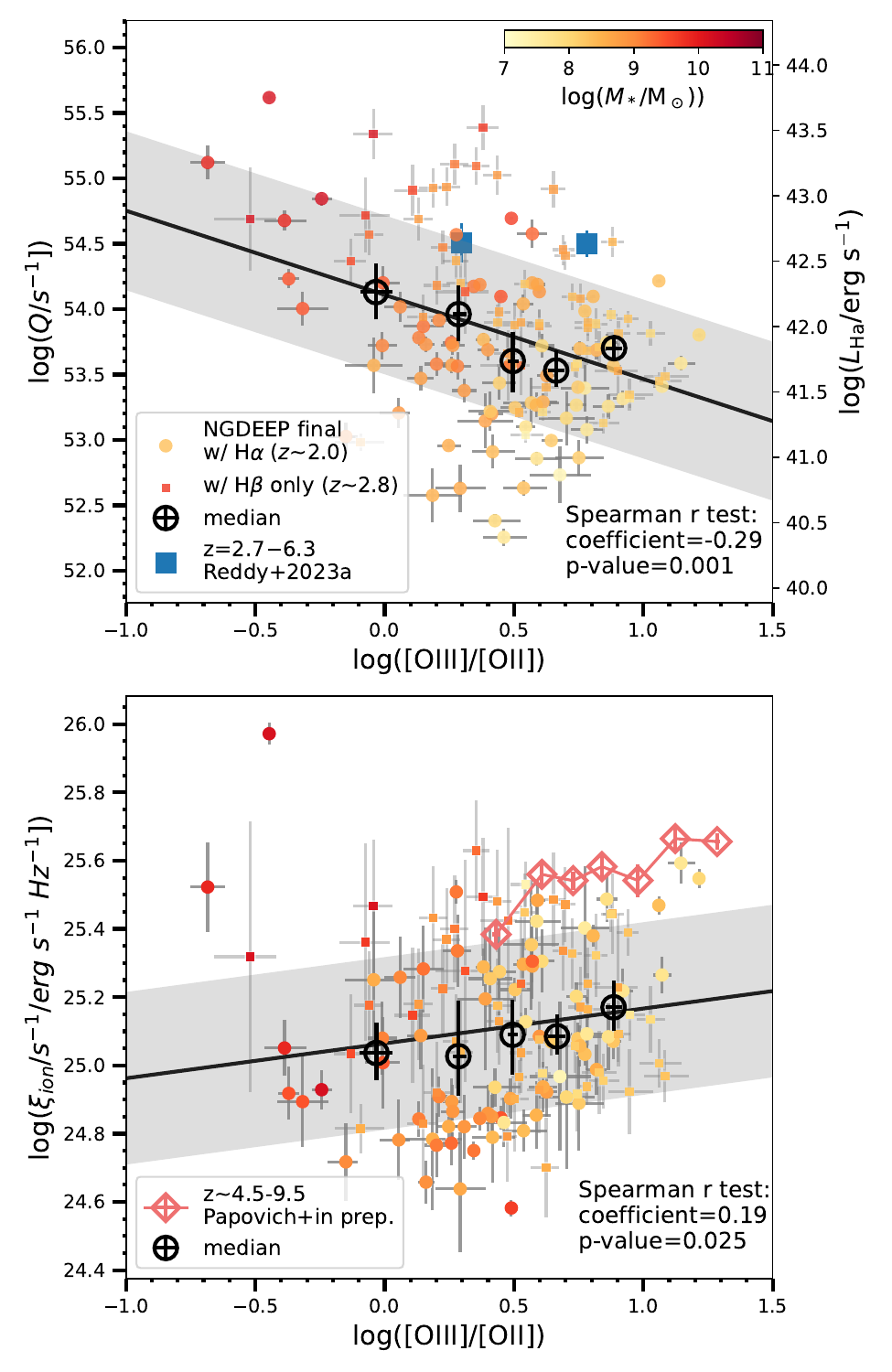}
    \caption{The distribution of \oo\ and ionizing photon production rate $Q$ (\textit{top}) and ionizing production efficiency $\xi_{ion}$ (\textit{bottom}) for the NGDEEP final sample with \ha\ {or \hb\ SNR $> 5$ detections}, color-coded by stellar mass. The median values and best fits for the NGDEEP \oo\ galaxies {with \ha\ or \hb\ detection} (without mass cut) are shown as black dots and solid black lines, and 1$\sigma$ total scatter is shown as the grey regions. {Results from the Spearman rank test are shown as well. } The median values of galaxies at $z=2.7 - 6.3$ from \citet{Reddy2023b} are shown as blue squares in the top panel. The median values of galaxies at $z\sim4.5-9.5$ from CEERS and JADES (Papovich et al. \textit{in prep.}) are shown in open magenta diamonds. Both $Q$ and $\xi_{ion}$ show correlations with \oo. }
    \label{fig:q_xion}
\end{figure}

In the top panel of Figure~\ref{fig:q_xion}, we show the distribution of the $Q$ and $\xi_{ion}$ values as a function of \oo\ for the NGDEEP final sample with \ha\ or \hb\ SNR $> 5$ detection. 
{For galaxies with both \ha\ and \hb\ detections,  we used values from \ha, which typically has higher SNR. Additionally, the \ha\ and \hb\ derived values are consistent for these galaxies, as demonstrated by \ha- and \hb- derived SFR in Figure~\ref{fig:sfr_sfr}. }
{We find a negative correlation between \oo\ and $Q$ with a Spearman rank correlation coefficient of $-$0.29 and a p-value of 0.001 when including galaxies with \ha\ or \hb. The significance of this correlation decreases, with the p-value reduced to 0.04, when only including galaxies with \ha\ detections. }
Because $Q$ depends only on \ha\ {and \hb\ }luminosity, which traces the SFR (also see Figure~\ref{fig:sfr_sfr}), $Q$ versus \oo\ is similar to the SFR versus \oo\ as shown in Figure~\ref{fig:o32_sfr}.
{However, as discussed in Section \ref{sec:discussion}, the SFR--\oo\ is largely a byproduct of the mass--\oo\ and SFR--mass relations. 
This mostly rules out the possibility that the ionizing photon production rate plays a dominant role in driving the \oo\ ratio.  }
{We compare results from }\citet{Reddy2023b}, who studied a sample of 48 SFGs at $z=2.7 - 6.3$ with \oo\ spanning a range of 0.5 $-$ 100. They found no correlation between the ionizing photon production rate between low- and high-\oo\ galaxies with mean \oo\ values of 1.98 and 6.06, respectively, as shown in Figure~\ref{fig:q_xion}. 
Their mean $Q$ values are higher than the trend derived from our sample at fixed \oo, which could be due to their sample containing galaxies at higher redshift. 

In the bottom panel of Figure~\ref{fig:q_xion}, we find a mild positive correlation between \oo\ and $\xi_{ion}$ such that $\xi_{ion}$ increases with increasing \oo. 
The Spearman rank correlation test confirms the significance of this relation, with a correlation coefficient of {$0.19$} and a p-value of {0.03} when considering galaxies with \ha\ or \hb\ detection. The p-value decreases to 0.003 when only galaxies with \ha\ detection are included. 
The scatter is also non-negligible, 0.26 dex. 
This could be related to the star formation history of galaxies. The $\xi_{ion}$ values are calculated using the \ha/UV ratio, which is commonly used as a tracer for recent bursty star-formation activity (e.g., \citealp{Emami2019, Guo2016, Faisst2019}). 
An elevated \ha\ emission with respect to the UV continuum is a sign of bursty star-formation activity in the past $\sim$10 Myr.  
Similarly, the EW(\ha) and sSFR are sensitive to recent star-formation activity. The difference between EW \ha\ and \ha/UV ratio is their denominator: the stellar continuum flux and UV continuum flux, respectively. The UV continuum flux traces SF activity in the recent 100-200 Myr. 
Thus, the two measurements are comparable for a galaxy with an age less than $\sim$100-200 Myr or a galaxy with a constant star formation history. 
Therefore, it is not surprising that a correlation between $\xi_{ion}$ and sSFR exists for high redshift galaxies up to $z\sim7$ \citep{Faisst2019, Emami2020, Endsley2021, Emami2020, Prieto-Lyon2023} and for extreme galaxies at lower redshift $z\sim1$ \citep{Tang2019, Izotov2021}. 
In Section~\ref{sec:o32-ssfr}, we found \oo\ is correlated with sSFR and EW \ha. Here we find that \oo\ is correlated with ionizing production efficiency, confirming the connection between ionization parameters and recent star-formation activity. 

In addition, we compare galaxies at $z\sim4.5-9.5$ from Papovich et al.\ (\textit{in prep.}) with our \oo--$\xi_{ion}$ relation. The median $\xi_{ion}$ of $z>5$ galaxies generally follows our relation but is offset to slightly higher $\xi_{ion}$ at fixed \oo. 
The higher $\xi_{ion}$ could be due to a recent starburst which increases the number of massive young stars and boosts the production of ionizing photons. This is consistent with the comparison of EW(\hb)--\oo\ relation where the $z>5$ galaxies tend to have higher EW(\hb) than our NGDEEP galaxies at $z\sim2-3$ (see Figure~\ref{fig:o32_ew}). 
It is unclear what causes the shift toward higher $\xi_{ion}$ with increasing redshift, though we expect it could be due to low metallicity as metal-poor stars are hotter and more efficient at generating ionizing radiation.

One galaxy (NGDEEP\_02938) is significantly offset from the fitted relation,
with very high ionizing production efficiency, $\xi_\mathrm{ion}$, and relatively low \oo\ ratio.  We conclude this offset is likely linked to the fact that it has the highest $E(B-V)$ in our sample, with $E(B-V)$ = 1.13 from Balmer decrement and \ebvs = 0.48 from SED fitting. These values correspond to an attenuation of $A_V \gtrsim $1.5, and an optical depth of $\tau \gtrsim 1.6$, indicating the galaxy is optically thick. 
Due to the optical thickness, the UV luminosity of this galaxy could be underestimated as the intrinsic UV emission may be higher than what is observed. This underestimation could lead to an overestimation of the $\xi_\mathrm{ion}$.




\subsubsection{SFR surface density} \label{sec:gasdensity}

\begin{figure}
    \centering
    \includegraphics[width=\columnwidth]{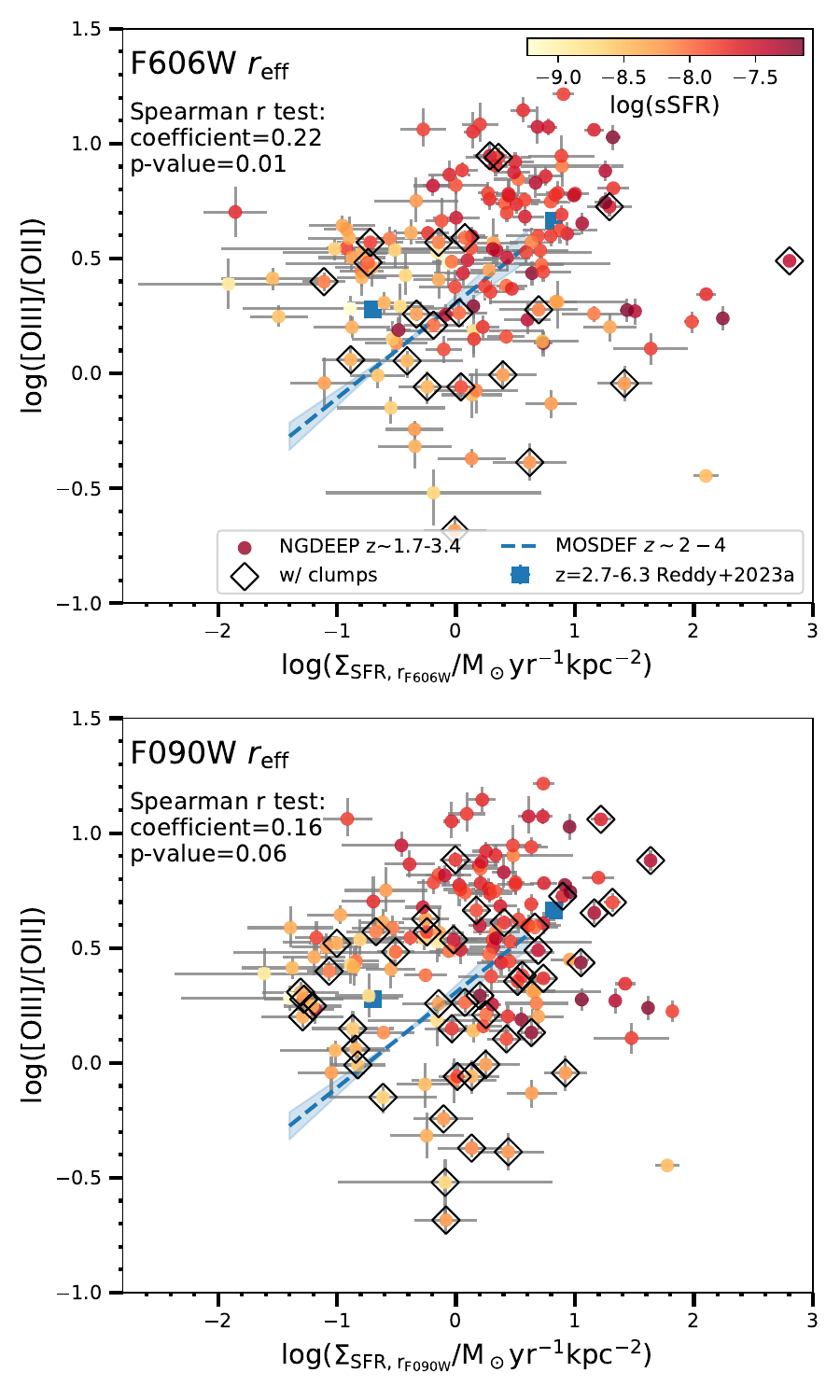}
    \caption{The distribution of the \oo\ ratio and $\Sigma_\mathrm{SFR}$ values for the NGDEEP final sample using effective radius measured from the HST F060W (\textit{top}) and JWST F090W images (\textit{bottom}). The data points are color-coded by sSFR. For galaxies identified with more than one component, we use the effective radius of the brightest clump. Such galaxies are marked with open black diamonds. In both panels, the median values of galaxies at $z=2.7-6.3$ from \citet{Reddy2023b} are shown as blue squares, and the fit for MOSDEF galaxies at $z\sim2-4$ \citep{Reddy2023a} are shown as a blue dotted line. The SFR surface density may contribute to affecting \oo, but with a considerable scatter and sensitivity to the bandpass used for size measurements. }
    \label{fig:SigmaSFR}
\end{figure}

As mentioned in the introduction, the hydrogen density is approximately the electron density of an ionized gas. 
Some studies have suggested that the increased electron densities in high-redshift galaxies could be a major factor contributing to the increase in the ionization parameter \citep{Davies2021, Reddy2023a, Reddy2023b}. 
It is important to note that the electron densities from these studies are derived from \sii\ $\lambda\lambda$6716, 6731 or \oii\ $\lambda\lambda$3726, 3729, which measures the density in the low-ionization zone. This may differ from the density in higher ionization zones and the average electron density \citep{Berg2021}. 

However, we do not have measurements of the electron density. 
Some studies have found a significant correlation between electron density and SFR \textit{surface density} for galaxies at $z\sim2$ \citep{Shimakawa2015, Reddy2023a}. 
These are related because the $n_e$ of \ion{H}{2} regions is governed by the molecular gas density and the balance between stellar feedback and ambient pressure (e.g., \citealp{Davies2021}), and the molecular gas density is related to the SFR \textit{surface density} via the Kennicutt--Schmidt relation (e.g., \citealp{Kennicutt2007, delosReyes2019, Bacchini2019, Pessa2021}). 
Here, we investigate whether the SFR surface density is correlated with \oo, and whether it affects the ionization parameter.

We derive the SFR surface density as: 
\begin{equation}
    \Sigma_\mathrm{SFR} = \frac{\mathrm{SFR}}{2\pi r_\mathrm{eff}^2 }\ \mathrm{[M_\odot\ yr^{-1}\ kpc^{-2}]}, 
\end{equation}
where $r_\mathrm{eff}$ is the effective radius of galaxies in units of kpc. 
For each galaxy, we estimate $r_\mathrm{eff}$ using two different bandpasses.  We used both the \hst\ F606W image and \jwst/NIRCAM F090W images, and measured the effective radii with the \pysersic\ code \citep{Pasha2023}. 
Because the SFR is derived from SED fitting which mostly relies on far-UV luminosity, we chose the \hst\ F606W image as this corresponds to rest-frame wavelengths of $\simeq$ 1400--2250~\AA. However, even though the F090W covers a less ideal set of rest-frame wavelengths 2050--3200~\AA, it provides better spatial resolution and deeper sensitivity, which is particularly important for measuring the size of small, low-mass galaxies.  
For the F606W sizes, we run \pysersic\ with a $2\arcsec\times2\arcsec$ cutout centered on the source with the image in a 60 mas scale, a PSF adopted from \cite{Finkelstein2022}, and priors from \textsc{SEP} \citep{Bertin1996, Barbary2016} with a detection threshold of 5$\times$ the RMS noise and a \textsc{minarea} of 5 pixels. 
We perform the \pysersic\ on F090W with the same setup, but with the image in a 30 mas scale and the PSF from \textsc{webbpsf}. 
For galaxies identified with more than one component in the segmentation, we use the effective radius of the brightest clump. 

Figure~\ref{fig:SigmaSFR} shows the distribution of \oo\ and $\Sigma_\mathrm{SFR}$ based on effective radius measured in F606W and F090W on the top and bottom panels, respectively. 
We find a significant correlation between \oo\ and $\Sigma_\mathrm{SFR}$ when using F606W $r_\mathrm{eff}$, with a Spearman correlation coefficient of 0.22 and a p-value of 0.01. 
No {significant} correlation is found between \oo\ and $\Sigma_\mathrm{SFR}$ when using F090W $r_\mathrm{eff}$ with a p-value of 0.06. 
As a comparison, we show the fitted relation for MOSDEF galaxies at $z\sim 1.9-3.7$ derived by \citet{Reddy2023a} and the median values of galaxies at $z=2.7 - 6.3$ from \citet{Reddy2023b}. 
Our galaxies are scattered around the fitted relation from MOSDEF galaxies with a relatively large scatter of $\sim$0.45 dex in both panels. Several potential factors could lead to the difference. 
Firstly, 18\% and 31\% of our galaxies show multiple components in the F606W and F090W images, respectively. As mentioned above, for galaxies with multiple detected clumps, we use only the effective radius of the brightest clump, which could underestimate the size of these galaxies and lead to an increase in the $\Sigma_\mathrm{SFR}$. 
This effect is likely more pronounced in F090W, as more galaxies are spatially resolved into multiple components in this band. 
In addition, dust could bias our UV size measurements by artificially increasing the apparent size in UV when dust obscuration dominates in the central region of the galaxy (see further discussion in \citealp{Shen2023}). 
This effect is likely negligible in low-mass, high-sSFR galaxies in our sample as they have lower dust content and small $\mathrm{E(B-V)}$. 
Another possible reason is that different data is used for size measurements. MOSDEF used the size measured from F160W, where light is dominated by an older stellar population. Meanwhile, the PSF of F160W is larger than F606W and F090W, which could bias the size of compact galaxies. 
Overall, we argue that the SFR surface density may influence \oo, but with a considerable scatter and sensitivity to the bandpass used for size measurements. 

Instead of using SFR from SED fitting and UV size, we can directly use \ha\ emission line map to obtain a $\Sigma_\mathrm{SFR}$ map. 
This method is more straightforward and \ha\ is less affected by dust compared with UV. 
In fact, we see evidence of a positive relation between \oo\ and $\Sigma_\mathrm{SFR}$ from the \oo\ ratio maps and \ha-derived $\Sigma_\mathrm{SFR}$ maps. We will fully explore this in the next paper (Shen et al. in prep).

\subsubsection{Density-bounded nebulae} \label{sec:density-bounded}

In equation \ref{eq:U}, we assumed a radiation-bounded nebula, where all ionizing photons are absorbed and converted to nebular emission. 
Alternatively, in the case of ``density-bounded'' nebulae, ionizing photons can escape.  This can elevate the \oo\ ratio as the low-ionization region (where most of the \oii\ originates) is not fully formed (see \citealt{Nakajima2014}). 
Studies of low-redshift galaxies find that the \oo\ ratio correlates with the escape fraction H-ionizing photons, $f_\mathrm{esc}$ \citep{Chisholm2018, Izotov2018}. 
However, more recent studies suggest that a high \oo\ ratio may be a necessary but not sufficient condition for the large escape fraction \citep{Nakajima2020, Choustikov2024}. 
Future direct measurements of the escape fraction for our sample would allow us to test this.  This study may be achievable using existing data from UVUDF \citep{Teplitz2013, Rafelski2015}, which covers rest-frame far-UV continuum at wavelengths $<$912\AA.   
We will explore this in future work. 

\section{Summary} \label{sec:summary}

We study the relation between \oo\ ratios and galaxy properties for 178 SFGs at $1.7 < z < 3.4$ using NIRISS slitless spectroscopy observations from the NGDEEP survey. 
The \oo\ ratio primarily traces the ionization parameter of the ISM, with a secondary dependence on metallicity.  
Our sample is selected based on the deep NIRISS spectra, primarily relying on the \oiii\ emission lines.  All galaxies lie above the star-formation main sequence and span a stellar mass range of $10^{7}-10^{10.2}$ \msun. 
Galaxy properties are estimated with ancillary deep HST and JWST imaging from CANDELS, JADES, and JEMS. 
Our results are summarized as follows: 
\begin{itemize}
    \item We find significant correlations between the \oo\ ratios and galaxies properties such that the \oo\ ratio increases with decreasing stellar mass, decreasing SFR, increasing sSFR, and increasing EW of \ha\ and \hb. 

    \item We find that the \oo\ ratio is primarily driven by ionization parameters with a secondary dependence on metallicity, consistent with previous observational results and photoionization models. 
    This suggests that the ionization parameter has a similar dependence on galaxy properties as is found for \oo. 
    
    \item Compared to local normal galaxies, galaxies at $z\sim2-3$ have a higher \oo\ ratio at fixed stellar mass and SFR, suggesting higher ionization parameter and/or low metallicity in  $z\sim2-3$ galaxies. 

    \item We find that $z\sim2-3$ galaxies have comparable or lower \oo\ to that measured for extreme galaxies at $z\sim0$ at the fixed stellar mass, SFR, sSFR and \ha\ and \hb\ EW, indicating these extreme galaxies at $z\sim0$ have similar to higher ionization parameter and similar to lower metallicity. 

    \item Our NGDEEP sample spans a wide range of \oo\ and stellar mass, which helps bridge the gap between the local and the $z>5$ universe. We find an evolutionary trend in \oo\ and EW(\hb) from $z\sim0$ to $z\gtrsim5${, where higher redshift galaxies show increased \oo\ and EW, with possibly} higher \oo\ at fixed EW. 

    \item We investigate the possible physical mechanisms behind the relation between the ionization parameter and recent star formation activity. We find that the \oo\ ratio is significantly correlated with the ionizing production efficiency. 
    Meanwhile, the SFR surface density may contribute to affecting \oo, but with a considerable scatter and sensitivity to the bandpass used for size measurements. We argue that both the enhanced recent star formation activity and SFR surface density could be the main contributors to the increase in \oo\ and the ionization parameter.

\end{itemize}

This paper demonstrates the capability of \jwst/NIRISS slitless spectroscopy data to measure the nebular emission line for low-mass galaxies $\log \mathrm{(M_\ast/M_\odot)} \sim 10^{7} - 10^{9}$ at $z\sim2-3$. This dataset enables us to constrain star formation, metallicity, ionization parameter, and dust content of entire galaxies, as well as their spatially resolved profiles. 
With the extended mass range of galaxies, this dataset can bridge the gap between the local and $z>5$ universe, enabling us to constrain the evolutionary trends of key galaxies and ISM properties.

\begin{acknowledgments}
We acknowledge the hard work of our colleagues in the NGDEEP collaboration and everyone involved in the \jwst\ mission.  
We wish to thank the anonymous referee for a thorough and constructive report that improved the quality and clarity of this work. 
This work benefited from support from the George P. and Cynthia Woods Mitchell Institute for Fundamental Physics and Astronomy at Texas A\&M University. 
CP thanks Marsha and Ralph Schilling for generous support of this research. 
This work is based on observations made with the NASA/ESA/CSA {\it JWST}. The data were obtained from the Mikulski Archive for Space Telescopes at the Space Telescope Science Institute, which is operated by the Association of Universities for Research in Astronomy, Inc., under NASA contract NAS 5-03127 for \jwst. These observations are associated with program \#2079. 
Some/all the {\it JWST} data presented in this paper were obtained from the Mikulski Archive for Space Telescopes (MAST) at the Space Telescope Science Institute. The specific observations analyzed can be accessed via 
\cite{https://doi.org/10.17909/3s7h-8k54}, \cite{https://doi.org/10.17909/fsc4-dt61} and \cite{https://doi.org/10.17909/8tdj-8n28}. 
JM is grateful to the Cosmic Dawn Center for the DAWN Fellowship. 

\end{acknowledgments}

%

\vspace{5mm}
\facilities{HST (ACS, WFC3), JWST (NIRCAM, NIRISS)}


\software{The python packages \textsc{astropy} \citep{Astropy2013, Astropy2018}, \textsc{matplotlib} \citep{Hunter2007}, \textsc{numpy} \citep{vanderWalt2011}, and \textsc{scipy} \citep{2020SciPy-NMeth} were extensively used. } 



\appendix
\restartappendixnumbering

\section{Model Estimated Star Formation Rates} \label{sec:app_sfr}
\begin{figure}
    \centering
    \includegraphics[width=\columnwidth]{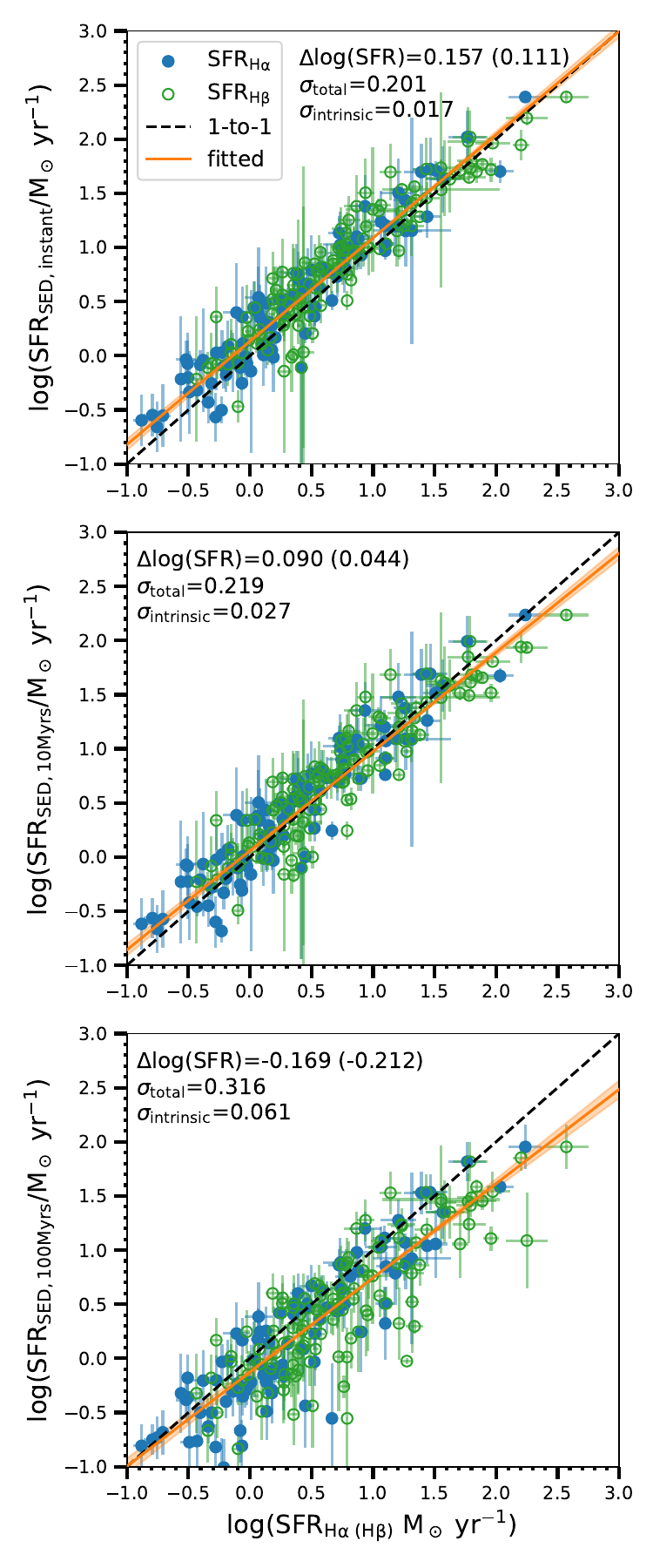}
    \caption{Comparison of the SED-derived instantaneous SFR (\textit{top}), and the SFR averaged over 10 Myr (\textit{middle}) and average over 100 Myr (\textit{bottom}) and \ha- and \hb-derived SFR for NGDEEP full sample with \ha\ and \hb\ SNR > 5, respectively. In each panel, the solid orange line shows the best-fitted line using \linmix\ package \citep{Kelly2007} using SFR$_\mathrm{H\alpha}$ and SFR$_\mathrm{H\beta}$ (if SFR$_\mathrm{H\alpha}$ is not available). The black dashed line is the 1-to-1 relation. The median offset between SFR$_\mathrm{SED}$ and SFR$_\mathrm{H\alpha}$ (SFR$_\mathrm{H\beta}$) are presented in each panel.  } 
    \label{fig:sfr_sfr}
\end{figure}

CIGALE calculates several different SFRs, including an instantaneous SFR ($\mathrm{SFR_{instant}}$), and the SFR averaged over 10 Myr ($\mathrm{SFR_{10Myrs}}$) and 100 Myr ($\mathrm{SFR_{100Myrs}}$). These SFRs are influenced by the assumed parameterization and the fitted star formation history.  
To better understand these estimated SFRs, we compare the \ha-derived and \hb-derived SFRs ($\mathrm{SFR_{H\alpha}}$ and $\mathrm{SFR_{H\beta}}$) with these SED-derived SFRs in Figure~\ref{fig:sfr_sfr} for NGDEEP SFGs with \ha\ and \hb\ emission line flux detected with SNRs $>5$. 
The $\mathrm{SFR_{H\alpha}}$ and $\mathrm{SFR_{H\beta}}$ are calculated from the dust-corrected \ha\ and \hb\ luminosity using the \citet{Kennicutt2012} calibration. 
We assume an intrinsic ratio of \ha/\hb\ = 2.86 (based on the Case B assumption for $T=10^4$~K and $n_e = 10^2$~cm$^{-3}$) when converting the \hb\ luminosity to an SFR. 
{The dust correction applied to \ha\ and \hb\ is based on \ebvs\ and SFR derived from SED fitting and follows the \citep{Cardelli1989} extinction model with $R_V = 3.1$ (as described in Section \ref{sec:dustcorr} and see Appendix \ref{app:dust}). }

As shown in Figure~\ref{fig:sfr_sfr}, the instantaneous SFRs and $\mathrm{SFR_{10Myrs}}$ from CIGALE correlate well with the SFRs derived from the \ha\ and \hb\ lines. The median difference between the SED-derived SFRs and $\mathrm{SFR_{H\alpha}}$ are {0.15} and {0.09} dex, for the instantaneous and 10 Myr-averaged SFRs, respectively. 
However, the $\mathrm{SFR_{100Myrs}}$ is on average lower than $\mathrm{SFR_{H\alpha}}$ by {-0.17} dex (a factor of order 1.5). 
We found similar trends when comparing SED-derived SFRs and $\mathrm{SFR_{H\beta}}$ with differences of {0.10, 0.03, and -0.23} dex for $\mathrm{SFR_{instant}}$, $\mathrm{SFR_{10Myrs}}$, and $\mathrm{SFR_{100Myrs}}$, respectively. 

We adopt the \linmix\ to fit the relation between SED-derived and $\mathrm{SFR_{H\alpha}}$ {and $\mathrm{SFR_{H\beta}}$ (if $\mathrm{SFR_{H\alpha}}$ is not avaliable)}. 
The best-fitted lines for $\mathrm{SFR_{instant}}$ and $\mathrm{SFR_{10Myrs}}$ align {close to} the 1-to-1 relation. 
The 1$\sigma$ {total scatter and intrinsic scatter} are similar ($\sim$0.2 dex and $\sim$0.02-0.03 dex respectively) in both cases. 
In contrast, the best-fitted line for $\mathrm{SFR_{100Myrs}}$ is offset from the 1-to-1 relation and has a larger total {and intrinsic scatter (0.32 dex and 0.06 dex, respectively)}.
This is consistent with the fact that the \ha\ emission is sensitive to ionization from O-type stars with lifetimes of $\sim$ 5 Myr. 
In the paper, we use $\mathrm{SFR_{10Myrs}}$ as the SED-derived SFR, which shows the smallest offset compared to $\mathrm{SFR_{H\alpha}}$ and $\mathrm{SFR_{H\beta}}$.

\section{Dust reddening correction} \label{app:dust}

Ideally, we would use the Balmer decrements (e.g., measured from the \ha/\hb\ line ratios) to estimate nebular dust attenuation. 
However, because not all galaxies have both \hb\ and \ha\ covered by NIRISS, we instead rely on the $\mathrm{E(B-V)}$ values derived from SED fitting. 
{We consider two methods here: (1) applying a uniform star-to-gas attenuation ratio to convert the SED-derived \ebvs\ to \ebvg\ for nebular emission; and (2) adopting an SFR-dependent star-to-gas attenuation ratio, following the method from \citet{Reddy2015, Sanders2021}. }
{We calibrated and tested these two methods using 65 SFGs from the NGDEEP survey at $1.7<z<2.3$ that has detections of both \ha\ and \hb\ at SNR $> 5$. The median redshift of this sample is 2.0. }

\begin{figure}
    \centering
    \includegraphics[width=\columnwidth]{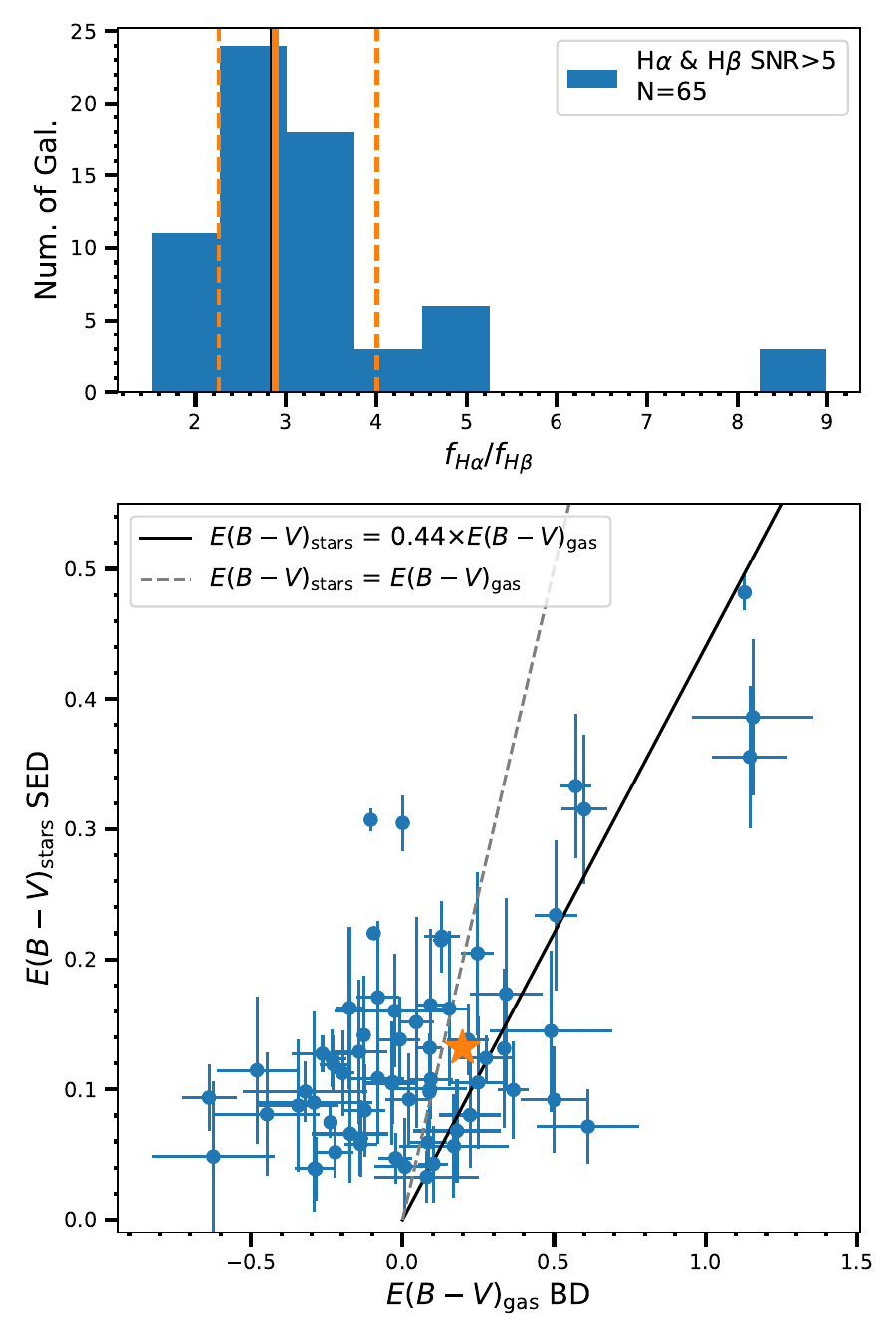}
    \caption{\textit{Top:} Histogram of \ha/\hb\ emission line ratios for the NGDEEP final sample with \ha\ and \hb\ SNR $> 5$. {The vertical orange solid line indicates the median \ha/\hb\ ratio, while the dashed lines represent the 16th and 84th percentiles. } The vertical black line marks the theoretical minimum value of 2.86 in the absence of dust for Case B recombination.  \textit{Bottom:} Comparison of the color excesses derived for the stellar continuum from SED fitting and that for the ionized gas calculated from \ha/\hb. The \ebvg\ is computed assuming the \citet{Cardelli1989} extinction curve with Rv = 3.1. The median stellar-to-gas attenuation ratio for galaxies with \ha/\hb\ $> 2.86$ is marked by a {orange} star. The black solid line shows the relation \ebvs=0.44$\times$\ebvg\ from \citet{Calzetti2000}. The grey dashed line shows the 1-to-1 relation between \ebvs\ and \ebvg. }
    \label{fig:ebv}
\end{figure}

In the top panel of Figure~\ref{fig:ebv}, we show the histogram of the \ha/\hb\ line ratios. There are {48\% of the 65} galaxies having \ha/\hb\ ratios lower than the Case B value of 2.86 (based on the Case B assumption for $T=10^4$~K and $n_e = 10^2$~cm$^{-3}$). {The median \ha/\hb\ ratios of our sample is 2.88, with a 16th and 84th percentile range of 2.26 to 4.01. A significant fraction of galaxies with \ha/\hb\ ratios below 2.86 was also reported in \citet{Pirzkal2023} which used NGDEEP epoch 1 data and adopted a different data reduction and extraction method. } 
{The low \ha/\hb\ ratios are unlikely to be caused by low SNR or wavelength-dependent flux calibration, as no correlation is observed between these ratios and the SNR of \ha\ or \hb\ emission lines or the observed wavelength.  }
{On the other hand, we find that the \ha/\hb\ ratio is significantly correlated with stellar mass, SFR, and \oo\, as \ha/\hb increases with increasing mass, SFR, and with decreasing \oo. The correlation between the \ha/\hb\ ratios and SFR is consistent with the finding from \citep{Reddy2015}. Furthermore, \citep{Scarlata2024} also shows that galaxies with high \oo\ are more likely to have low \ha/\hb\ ratios. These results support that the low Ha/Hb ratios are more likely due to physical conditions, such as a different geometry of dust and gas, and higher temperatures in galaxies with high ionization or low metallicity. In such environments, the intrinsic \ha/\hb\ of galaxies could be lower than 2.86. This topic is beyond the primary scope of this paper, so we have not included this analysis in the current work. However, we plan to explore this in greater detail in a future study. }

We compared the SED-derived \ebvs\ and \ha/\hb-derived \ebvg\ and shown in the bottom panel of Figure~\ref{fig:ebv}. For those with \ha/\hb\ $> 2.86$, the SED-derived \ebvs\ and \ha/\hb-derived \ebvg\ generally follow the \ebvs$=0.44\times$\ebvg. We measure a median star-to-gas attenuation ratio of {0.67}. 
For those with \ha/\hb\ $< 2.86$, they tend to have low SED-derived \ebvs\ with the median \ebvs\ of {0.10}, suggesting that the dust attenuation is relatively low in these galaxies. 
{Here, we test on a star-to-gas attenuation ratio of 0.44 and 1. 
However, if \ebvg=0 is assumed for galaxies with \ha/\hb\ $< 2.86$, applying this method will overestimate dust in these galaxies. }

\begin{figure*}
    \centering
    \includegraphics[width=0.8\textwidth]{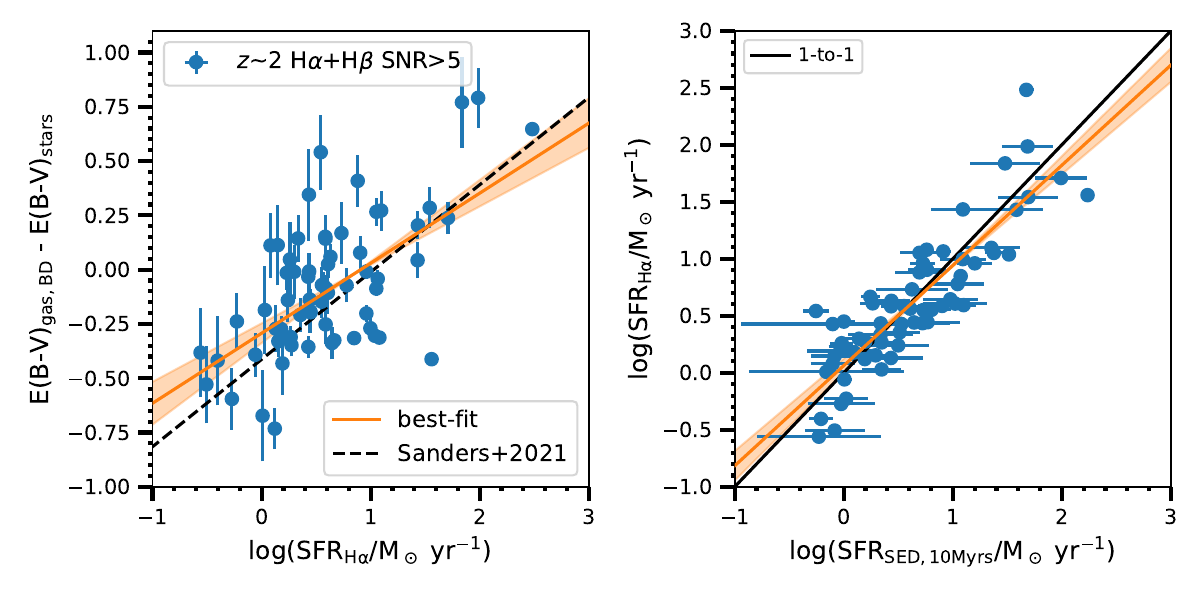}
    \caption{{\textit{Left:} The difference between nebular reddening \ebvg\ derived using the Balmer decrement (BD) and stellar reddening \ebvs\ from SED fitting as a function of SFR$_\mathrm{H\alpha}$ for SFGs at $z \sim2$ with detections of both \ha\ and \hb\ SNR $> 5$. The orange solid line shows the best fit, with the shaded region indicating the $1\sigma$ uncertainties. The dashed black line shows the relation from \citet{Sanders2021}. \textit{Right:} SFR$_\mathrm{H\alpha}$ vs. SFR$_\mathrm{SED, 10Myrs}$. The solid black line shows the 1-to-1 relation. }}
    \label{fig:dustcorr2}
\end{figure*}

{For the second method, it assumes that the difference between \ebvg\ and \ebvs\ is a function of SFR$_\mathrm{H\alpha}$, as found in \citet{Reddy2015} (and references therein). Following \citet{Sanders2021}, we quantified relations between \ebvg-\ebvs\ and \ha-derived SFR, and between the \ha-derived SFR and SED-derived SFR as shown in \ref{fig:dustcorr2}. 
The best-fit linear relations from \linmix\ are:
\begin{align} 
    \label{eq_app:dust1}
    \mathrm{E(B-V)_{gas, calibrated} - E(B-V)_{stars}} =\qquad\qquad\qquad \nonumber   \\ 
    (0.332 \pm 0.053) \times \log(\mathrm{SFR_{H\alpha}}) +(-0.293\pm 0.045)\\
    \label{eq_app:dust2}
    \log(\mathrm{SFR_{H\alpha}}) =\qquad\qquad\qquad\qquad\qquad\qquad\qquad \nonumber \\
    (0.879 \pm 0.069) \times \log(\mathrm{SFR_{SED, 10Myrs}}) + (0.063\pm0.060), 
\end{align}
, where $\mathrm{SFR_{SED, 10Myrs}}$ is the SFR average over 10 Myrs derived from \cigale\ SED fitting, and $\mathrm{SFR_{H\alpha}}$ is calculated from the dust-corrected \ha\ luminosity using \ebvg\ from Balmer decrement and convert to SFR using the \citet{Kennicutt2012} calibration. 
The total scatter of $\mathrm{E(B-V)_{gas} - E(B-V)_{stars}}$--$\log(\mathrm{SFR_{H\alpha}})$ and $\log(\mathrm{SFR_{H\alpha}})$--$\log(\mathrm{SFR_{SED}})$ relations are 0.24 and 0.30 dex, respectively. The intrinsic scatters are 0.05 and 0.08 dex, respectively. }

\begin{figure*}
    \centering
    \includegraphics[width=0.8\textwidth]{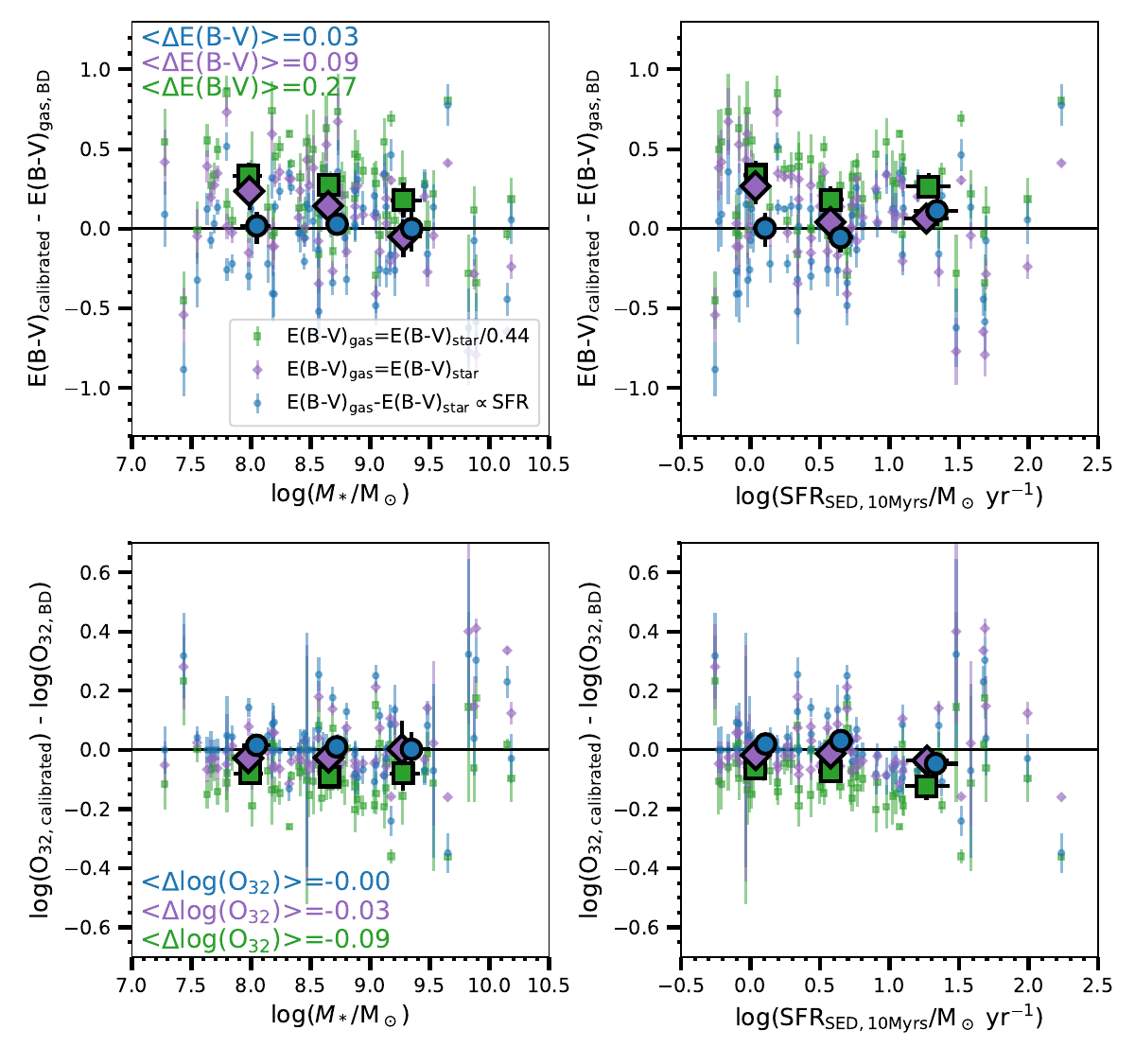}
    \caption{{\textit{Tops:} Comparison of calibrated E(B-V) using a uniform star-to-gas attenuation ratio of 0.44 (green squares) or 1 (purple diamonds) or an SFR-dependent star-to-gas attenuation ratio (blue dots) to \ebvg\ from the Balmer decrement as a function of stellar mass (left) and SFR (right). \textit{Bottoms:} Comparison of \oo\ using calibrated E(B-V) based on \ebvs\ and \ebvg from the Balmer decrement as a function of stellar mass and SFR. The median values are shown as larger markers. The median values using an SFR-dependent star-to-gas attenuation ratio (blue circles) are shifted to the right for clarity.  } }
    \label{fig:dustcorr}
\end{figure*}

{In the top panels of Figure \ref{fig:dustcorr}, we compare the calibrated E(B-V) values derived using these two methods to \ebvg\ obtained from the Balmer decrement, as a function of stellar mass and SFR. 
The median offset between the calibrated E(B-V) and \ebvg\ is smaller when using an SFR-dependent star-to-gas attenuation ratio, with a median offset of 0.03 dex. This offset does not vary across stellar mass but shows a mild dependence on SFR, with higher calibrated E(B-V) at higher SFRs. 
In the case of using a uniform star-to-gas attenuation ratio of 0.44 and 1, the median offsets are 0.09 dex and 0.27 dex, respectively, indicating an overestimation of dust correction for these galaxies. This overestimation is particularly pronounced in low stellar masses and low SFRs. }

{As we focus on the relations between the \oo\ ratio and various galaxy properties, we compare the offset between \oo\ using the calibrated E(B-V) values derived using these two methods to those using \ebvg\ from the Balmer decrement. 
We adopt the same \citet{Cardelli1989} extinction model with $R_V=3.1$. 
For galaxies with E(B-V) $< 0$, no dust correction is applied.  
The \oo\ offsets are shown as a function of stellar mass in the bottom panels of Figure \ref{fig:dustcorr}. 
The median offset is 0 when the SFR-dependent attenuation ratio, and the median offset increases to -0.03 and -0.09 when using a uniform star-to-gas attenuation ratio of 1 and 0.04, respectively.  
Similar to the E(B-V) comparison, the offsets from method 2 do not vary across stellar mass but show a mild dependence on SFR, suggesting that \oo\ might be underestimated at high SFR by $\sim$0.04. 
The offsets from method 1 using a ratio of 1 show dependence on both stellar mass and SFR. 
This comparison shows that applying different reddening corrections could impact the \oo\ ratio up to $\sim$0.1 dex. The effect may depend on stellar mass and SFR, potentially influencing the derived relations. }

{In this paper, for the dust correction of emission lines (i.e., \oiii\, \hb, \oii\, \ha), we adopt the second method using equation \ref{eq:dust1} and \ref{eq:dust2}, which provides consistent E(B-V) and \oo\ compared to those using the Balmer decrement and consistent SFR from dust-corrected $H\alpha$ and $H\beta$ (see Figure~\ref{fig:sfr_sfr}). }
{We note that while the significances and slopes of the \oo--galaxy property relations presented in this paper remain unchanged if a uniform star-to-gas attenuation ratio of 0.44 or 1 (first method) is adopted, the intercept of these relations would be systematically shifted lower by $\sim$0.09 dex or $\sim$0.03 dex. }




\begin{thebibliography}{}
\expandafter\ifx\csname natexlab\endcsname\relax\def\natexlab#1{#1}\fi
\providecommand{\url}[1]{\href{#1}{#1}}
\providecommand{\dodoi}[1]{doi:~\href{http://doi.org/#1}{\nolinkurl{#1}}}
\providecommand{\doeprint}[1]{\href{http://ascl.net/#1}{\nolinkurl{http://ascl.net/#1}}}
\providecommand{\doarXiv}[1]{\href{https://arxiv.org/abs/#1}{\nolinkurl{https://arxiv.org/abs/#1}}}

\bibitem[{{Astropy Collaboration} {et~al.}(2013){Astropy Collaboration},
  {Robitaille}, {Tollerud}, {Greenfield}, {Droettboom}, {Bray}, {Aldcroft},
  {Davis}, {Ginsburg}, {Price-Whelan}, {Kerzendorf}, {Conley}, {Crighton},
  {Barbary}, {Muna}, {Ferguson}, {Grollier}, {Parikh}, {Nair}, {Unther},
  {Deil}, {Woillez}, {Conseil}, {Kramer}, {Turner}, {Singer}, {Fox}, {Weaver},
  {Zabalza}, {Edwards}, {Azalee Bostroem}, {Burke}, {Casey}, {Crawford},
  {Dencheva}, {Ely}, {Jenness}, {Labrie}, {Lim}, {Pierfederici}, {Pontzen},
  {Ptak}, {Refsdal}, {Servillat}, \& {Streicher}}]{Astropy2013}
{Astropy Collaboration}, {Robitaille}, T.~P., {Tollerud}, E.~J., {et~al.} 2013,
  \aap, 558, A33, \dodoi{10.1051/0004-6361/201322068}

\bibitem[{{Astropy Collaboration} {et~al.}(2018){Astropy Collaboration},
  {Price-Whelan}, {Sip{\H{o}}cz}, {G{\"u}nther}, {Lim}, {Crawford}, {Conseil},
  {Shupe}, {Craig}, {Dencheva}, {Ginsburg}, {VanderPlas}, {Bradley},
  {P{\'e}rez-Su{\'a}rez}, {de Val-Borro}, {Aldcroft}, {Cruz}, {Robitaille},
  {Tollerud}, {Ardelean}, {Babej}, {Bach}, {Bachetti}, {Bakanov}, {Bamford},
  {Barentsen}, {Barmby}, {Baumbach}, {Berry}, {Biscani}, {Boquien}, {Bostroem},
  {Bouma}, {Brammer}, {Bray}, {Breytenbach}, {Buddelmeijer}, {Burke},
  {Calderone}, {Cano Rodr{\'\i}guez}, {Cara}, {Cardoso}, {Cheedella}, {Copin},
  {Corrales}, {Crichton}, {D'Avella}, {Deil}, {Depagne}, {Dietrich}, {Donath},
  {Droettboom}, {Earl}, {Erben}, {Fabbro}, {Ferreira}, {Finethy}, {Fox},
  {Garrison}, {Gibbons}, {Goldstein}, {Gommers}, {Greco}, {Greenfield},
  {Groener}, {Grollier}, {Hagen}, {Hirst}, {Homeier}, {Horton}, {Hosseinzadeh},
  {Hu}, {Hunkeler}, {Ivezi{\'c}}, {Jain}, {Jenness}, {Kanarek}, {Kendrew},
  {Kern}, {Kerzendorf}, {Khvalko}, {King}, {Kirkby}, {Kulkarni}, {Kumar},
  {Lee}, {Lenz}, {Littlefair}, {Ma}, {Macleod}, {Mastropietro}, {McCully},
  {Montagnac}, {Morris}, {Mueller}, {Mumford}, {Muna}, {Murphy}, {Nelson},
  {Nguyen}, {Ninan}, {N{\"o}the}, {Ogaz}, {Oh}, {Parejko}, {Parley}, {Pascual},
  {Patil}, {Patil}, {Plunkett}, {Prochaska}, {Rastogi}, {Reddy Janga},
  {Sabater}, {Sakurikar}, {Seifert}, {Sherbert}, {Sherwood-Taylor}, {Shih},
  {Sick}, {Silbiger}, {Singanamalla}, {Singer}, {Sladen}, {Sooley},
  {Sornarajah}, {Streicher}, {Teuben}, {Thomas}, {Tremblay}, {Turner},
  {Terr{\'o}n}, {van Kerkwijk}, {de la Vega}, {Watkins}, {Weaver}, {Whitmore},
  {Woillez}, {Zabalza}, \& {Astropy Contributors}}]{Astropy2018}
{Astropy Collaboration}, {Price-Whelan}, A.~M., {Sip{\H{o}}cz}, B.~M., {et~al.}
  2018, \aj, 156, 123, \dodoi{10.3847/1538-3881/aabc4f}

\bibitem[{{Bacchini} {et~al.}(2019){Bacchini}, {Fraternali}, {Iorio}, \&
  {Pezzulli}}]{Bacchini2019}
{Bacchini}, C., {Fraternali}, F., {Iorio}, G., \& {Pezzulli}, G. 2019, \aap,
  622, A64, \dodoi{10.1051/0004-6361/201834382}

\bibitem[{{Bagley} {et~al.}(2022){Bagley}, {Finkelstein}, {Koekemoer},
  {Ferguson}, {Arrabal Haro}, {Dickinson}, {Kartaltepe}, {Papovich},
  {P{\'e}rez-Gonz{\'a}lez}, {Pirzkal}, {Somerville}, {Willmer}, {Yang}, {Yung},
  {Fontana}, {Grazian}, {Grogin}, {Hirschmann}, {Kewley}, {Kirkpatrick},
  {Kocevski}, {Lotz}, {Medrano}, {Morales}, {Pentericci}, {Ravindranath},
  {Trump}, {Wilkins}, {Calabr{\`o}}, {Cooper}, {Costantin}, {de la Vega},
  {Hutchison}, {Lucas}, {McGrath}, {Wang}, \& {Wuyts}}]{Bagley2022}
{Bagley}, M.~B., {Finkelstein}, S.~L., {Koekemoer}, A.~M., {et~al.} 2022, arXiv
  e-prints, arXiv:2211.02495.
\newblock \doarXiv{2211.02495}

\bibitem[{{Baldwin} {et~al.}(1981){Baldwin}, {Phillips}, \&
  {Terlevich}}]{Baldwin1981}
{Baldwin}, J.~A., {Phillips}, M.~M., \& {Terlevich}, R. 1981, \pasp, 93, 5,
  \dodoi{10.1086/130766}

\bibitem[{{Barbary} {et~al.}(2016){Barbary}, {Boone}, {McCully}, {Craig},
  {Deil}, \& {Rose}}]{Barbary2016}
{Barbary}, K., {Boone}, K., {McCully}, C., {et~al.} 2016, {kbarbary/sep:
  v1.0.0}, v1.0.0,  Zenodo, \dodoi{10.5281/zenodo.159035}

\bibitem[{{Berg} {et~al.}(2021){Berg}, {Chisholm}, {Erb}, {Skillman}, {Pogge},
  \& {Olivier}}]{Berg2021}
{Berg}, D.~A., {Chisholm}, J., {Erb}, D.~K., {et~al.} 2021, \apj, 922, 170,
  \dodoi{10.3847/1538-4357/ac141b}

\bibitem[{{Berg} {et~al.}(2022){Berg}, {James}, {King}, {McDonald}, {Chen},
  {Chisholm}, {Heckman}, {Martin}, {Stark}, {Aloisi}, {Amor{\'\i}n},
  {Arellano-C{\'o}rdova}, {Bayliss}, {Bordoloi}, {Brinchmann}, {Charlot},
  {Chevallard}, {Clark}, {Erb}, {Feltre}, {Gronke}, {Hayes}, {Henry},
  {Hernandez}, {Jaskot}, {Jones}, {Kewley}, {Kumari}, {Leitherer}, {Llerena},
  {Maseda}, {Mingozzi}, {Nanayakkara}, {Ouchi}, {Plat}, {Pogge},
  {Ravindranath}, {Rigby}, {Sanders}, {Scarlata}, {Senchyna}, {Skillman},
  {Steidel}, {Strom}, {Sugahara}, {Wilkins}, {Wofford}, {Xu}, \& {Classy
  Team}}]{Berg2022}
{Berg}, D.~A., {James}, B.~L., {King}, T., {et~al.} 2022, \apjs, 261, 31,
  \dodoi{10.3847/1538-4365/ac6c03}

\bibitem[{{Bertin} \& {Arnouts}(1996)}]{Bertin1996}
{Bertin}, E., \& {Arnouts}, S. 1996, \aaps, 117, 393,
  \dodoi{10.1051/aas:1996164}

\bibitem[{{Bian} {et~al.}(2016){Bian}, {Kewley}, {Dopita}, \&
  {Juneau}}]{Bian2016}
{Bian}, F., {Kewley}, L.~J., {Dopita}, M.~A., \& {Juneau}, S. 2016, \apj, 822,
  62, \dodoi{10.3847/0004-637X/822/2/62}

\bibitem[{{Blanc} {et~al.}(2015){Blanc}, {Kewley}, {Vogt}, \&
  {Dopita}}]{Blanc2015}
{Blanc}, G.~A., {Kewley}, L., {Vogt}, F. P.~A., \& {Dopita}, M.~A. 2015, \apj,
  798, 99, \dodoi{10.1088/0004-637X/798/2/99}

\bibitem[{{Boquien} {et~al.}(2019){Boquien}, {Burgarella}, {Roehlly}, {Buat},
  {Ciesla}, {Corre}, {Inoue}, \& {Salas}}]{Boquien2019}
{Boquien}, M., {Burgarella}, D., {Roehlly}, Y., {et~al.} 2019, \aap, 622, A103,
  \dodoi{10.1051/0004-6361/201834156}

\bibitem[{{Brammer} \& {Matharu}(2021)}]{Brammer2021}
{Brammer}, G., \& {Matharu}, J. 2021, {gbrammer/grizli: Release 2021}, 1.3.2,
  Zenodo, \dodoi{10.5281/zenodo.5012699}

\bibitem[{{Brinchmann} {et~al.}(2004){Brinchmann}, {Charlot}, {White},
  {Tremonti}, {Kauffmann}, {Heckman}, \& {Brinkmann}}]{Brinchmann2004}
{Brinchmann}, J., {Charlot}, S., {White}, S.~D.~M., {et~al.} 2004, \mnras, 351,
  1151, \dodoi{10.1111/j.1365-2966.2004.07881.x}

\bibitem[{{Bruzual} \& {Charlot}(2003)}]{Bruzual2003}
{Bruzual}, G., \& {Charlot}, S. 2003, \mnras, 344, 1000,
  \dodoi{10.1046/j.1365-8711.2003.06897.x}

\bibitem[{{Calzetti} {et~al.}(2000){Calzetti}, {Armus}, {Bohlin}, {Kinney},
  {Koornneef}, \& {Storchi-Bergmann}}]{Calzetti2000}
{Calzetti}, D., {Armus}, L., {Bohlin}, R.~C., {et~al.} 2000, \apj, 533, 682,
  \dodoi{10.1086/308692}

\bibitem[{{Cardelli} {et~al.}(1989){Cardelli}, {Clayton}, \&
  {Mathis}}]{Cardelli1989}
{Cardelli}, J.~A., {Clayton}, G.~C., \& {Mathis}, J.~S. 1989, \apj, 345, 245,
  \dodoi{10.1086/167900}

\bibitem[{{Chabrier}(2003)}]{Chabrier2003}
{Chabrier}, G. 2003, \pasp, 115, 763, \dodoi{10.1086/376392}

\bibitem[{{Chisholm} {et~al.}(2018){Chisholm}, {Gazagnes}, {Schaerer},
  {Verhamme}, {Rigby}, {Bayliss}, {Sharon}, {Gladders}, \&
  {Dahle}}]{Chisholm2018}
{Chisholm}, J., {Gazagnes}, S., {Schaerer}, D., {et~al.} 2018, \aap, 616, A30,
  \dodoi{10.1051/0004-6361/201832758}

\bibitem[{{Choustikov} {et~al.}(2024){Choustikov}, {Katz}, {Saxena}, {Cameron},
  {Devriendt}, {Slyz}, {Rosdahl}, {Blaizot}, \&
  {Michel-Dansac}}]{Choustikov2024}
{Choustikov}, N., {Katz}, H., {Saxena}, A., {et~al.} 2024, \mnras, 529, 3751,
  \dodoi{10.1093/mnras/stae776}

\bibitem[{{Conroy} \& {Gunn}(2010)}]{Conroy2010}
{Conroy}, C., \& {Gunn}, J.~E. 2010, {FSPS: Flexible Stellar Population
  Synthesis}, Astrophysics Source Code Library, record ascl:1010.043

\bibitem[{{Cullen} {et~al.}(2014){Cullen}, {Cirasuolo}, {McLure}, {Dunlop}, \&
  {Bowler}}]{Cullen2014}
{Cullen}, F., {Cirasuolo}, M., {McLure}, R.~J., {Dunlop}, J.~S., \& {Bowler},
  R.~A.~A. 2014, \mnras, 440, 2300, \dodoi{10.1093/mnras/stu443}

\bibitem[{{Curti} {et~al.}(2020){Curti}, {Mannucci}, {Cresci}, \&
  {Maiolino}}]{Curti2020}
{Curti}, M., {Mannucci}, F., {Cresci}, G., \& {Maiolino}, R. 2020, \mnras, 491,
  944, \dodoi{10.1093/mnras/stz2910}

\bibitem[{{Curti} {et~al.}(2024){Curti}, {Maiolino}, {Curtis-Lake},
  {Chevallard}, {Carniani}, {D'Eugenio}, {Looser}, {Scholtz}, {Charlot},
  {Cameron}, {{\"U}bler}, {Witstok}, {Boyett}, {Laseter}, {Sandles}, {Arribas},
  {Bunker}, {Giardino}, {Maseda}, {Rawle}, {Rodr{\'\i}guez Del Pino}, {Smit},
  {Willott}, {Eisenstein}, {Hausen}, {Johnson}, {Rieke}, {Robertson},
  {Tacchella}, {Williams}, {Willmer}, {Baker}, {Bhatawdekar}, {Egami},
  {Helton}, {Ji}, {Kumari}, {Perna}, {Shivaei}, \& {Sun}}]{Curti2024}
{Curti}, M., {Maiolino}, R., {Curtis-Lake}, E., {et~al.} 2024, \aap, 684, A75,
  \dodoi{10.1051/0004-6361/202346698}

\bibitem[{{Davies} {et~al.}(2021){Davies}, {F{\"o}rster Schreiber}, {Genzel},
  {Shimizu}, {Davies}, {Schruba}, {Tacconi}, {{\"U}bler}, {Wisnioski}, {Wuyts},
  {Fossati}, {Herrera-Camus}, {Lutz}, {Mendel}, {Naab}, {Price}, {Renzini},
  {Wilman}, {Beifiori}, {Belli}, {Burkert}, {Chan}, {Contursi}, {Fabricius},
  {Lee}, {Saglia}, \& {Sternberg}}]{Davies2021}
{Davies}, R.~L., {F{\"o}rster Schreiber}, N.~M., {Genzel}, R., {et~al.} 2021,
  \apj, 909, 78, \dodoi{10.3847/1538-4357/abd551}

\bibitem[{{de los Reyes} \& {Kennicutt}(2019)}]{delosReyes2019}
{de los Reyes}, M. A.~C., \& {Kennicutt}, Robert~C., J. 2019, \apj, 872, 16,
  \dodoi{10.3847/1538-4357/aafa82}

\bibitem[{{Dickinson} {et~al.}(2024){Dickinson}, {Amorin}, {Arrabal Haro},
  {Bagley}, {Barro}, {Buat}, {Burgarella}, {Calabro'}, {Carnall}, {Casey},
  {Chworowsky}, {Cleri}, {Cole}, {Cooper}, {Cullen}, {Daddi}, {Donnan},
  {Dunlop}, {Elbaz}, {Ferguson}, {Fernandez}, {Finkelstein}, {Fontana},
  {Fujimoto}, {Giavalisco}, {Hamilton}, {Hathi}, {Hirschmann}, {Hutchison},
  {Juneau}, {Jung}, {Kartaltepe}, {Kocevski}, {Koekemoer}, {Larson}, {Long},
  {Lucas}, {Mascia}, {McGrath}, {McLeod}, {McLure}, {Napolitano}, {Papovich},
  {Pentericci}, {Perez Gonzalez}, {Simons}, {Somerville}, {Trump}, {Wang},
  {Weiner}, {Wilkins}, {Yung}, \& {Zavala}}]{Dickinson2024}
{Dickinson}, M., {Amorin}, R., {Arrabal Haro}, P., {et~al.} 2024, {The
  CANDELS-Area Prism Epoch of Reionization Survey (CAPERS)}, JWST Proposal.
  Cycle 3, ID. \#6368

\bibitem[{{Dopita} \& {Sutherland}(1996)}]{Dopita1996}
{Dopita}, M.~A., \& {Sutherland}, R.~S. 1996, \apjs, 102, 161,
  \dodoi{10.1086/192255}

\bibitem[{{Dopita} \& {Sutherland}(2003)}]{Dopita2003}
---. 2003, {Astrophysics of the diffuse universe},
  \dodoi{10.1007/978-3-662-05866-4}

\bibitem[{{Eisenstein} {et~al.}(2023){Eisenstein}, {Willott}, {Alberts},
  {Arribas}, {Bonaventura}, {Bunker}, {Cameron}, {Carniani}, {Charlot},
  {Curtis-Lake}, {D'Eugenio}, {Endsley}, {Ferruit}, {Giardino}, {Hainline},
  {Hausen}, {Jakobsen}, {Johnson}, {Maiolino}, {Rieke}, {Rieke}, {Rix},
  {Robertson}, {Stark}, {Tacchella}, {Williams}, {Willmer}, {Baker}, {Baum},
  {Bhatawdekar}, {Boyett}, {Chen}, {Chevallard}, {Circosta}, {Curti},
  {Danhaive}, {DeCoursey}, {de Graaff}, {Dressler}, {Egami}, {Helton},
  {Hviding}, {Ji}, {Jones}, {Kumari}, {L{\"u}tzgendorf}, {Laseter}, {Looser},
  {Lyu}, {Maseda}, {Nelson}, {Parlanti}, {Perna}, {Pusk{\'a}s}, {Rawle},
  {Rodr{\'\i}guez Del Pino}, {Sandles}, {Saxena}, {Scholtz}, {Sharpe},
  {Shivaei}, {Silcock}, {Simmonds}, {Skarbinski}, {Smit}, {Stone}, {Suess},
  {Sun}, {Tang}, {Topping}, {{\"U}bler}, {Villanueva}, {Wallace}, {Whitler},
  {Witstok}, \& {Woodrum}}]{Eisenstein2023}
{Eisenstein}, D.~J., {Willott}, C., {Alberts}, S., {et~al.} 2023, arXiv
  e-prints, arXiv:2306.02465, \dodoi{10.48550/arXiv.2306.02465}

\bibitem[{{Emami} {et~al.}(2020){Emami}, {Siana}, {Alavi}, {Gburek}, {Freeman},
  {Richard}, {Weisz}, \& {Stark}}]{Emami2020}
{Emami}, N., {Siana}, B., {Alavi}, A., {et~al.} 2020, \apj, 895, 116,
  \dodoi{10.3847/1538-4357/ab8f97}

\bibitem[{{Emami} {et~al.}(2019){Emami}, {Siana}, {Weisz}, {Johnson}, {Ma}, \&
  {El-Badry}}]{Emami2019}
{Emami}, N., {Siana}, B., {Weisz}, D.~R., {et~al.} 2019, \apj, 881, 71,
  \dodoi{10.3847/1538-4357/ab211a}

\bibitem[{{Endsley} {et~al.}(2021){Endsley}, {Stark}, {Chevallard}, \&
  {Charlot}}]{Endsley2021}
{Endsley}, R., {Stark}, D.~P., {Chevallard}, J., \& {Charlot}, S. 2021, \mnras,
  500, 5229, \dodoi{10.1093/mnras/staa3370}

\bibitem[{{Erb} {et~al.}(2006){Erb}, {Steidel}, {Shapley}, {Pettini}, {Reddy},
  \& {Adelberger}}]{Erb2006}
{Erb}, D.~K., {Steidel}, C.~C., {Shapley}, A.~E., {et~al.} 2006, \apj, 647,
  128, \dodoi{10.1086/505341}

\bibitem[{{Estrada-Carpenter} {et~al.}(2019){Estrada-Carpenter}, {Papovich},
  {Momcheva}, {Brammer}, {Long}, {Quadri}, {Bridge}, {Dickinson}, {Ferguson},
  {Finkelstein}, {Giavalisco}, {Gosmeyer}, {Lotz}, {Salmon}, {Skelton},
  {Trump}, \& {Weiner}}]{Estrada-Carpenter2019}
{Estrada-Carpenter}, V., {Papovich}, C., {Momcheva}, I., {et~al.} 2019, \apj,
  870, 133, \dodoi{10.3847/1538-4357/aaf22e}

\bibitem[{{Faisst} {et~al.}(2019){Faisst}, {Capak}, {Emami}, {Tacchella}, \&
  {Larson}}]{Faisst2019}
{Faisst}, A.~L., {Capak}, P.~L., {Emami}, N., {Tacchella}, S., \& {Larson},
  K.~L. 2019, \apj, 884, 133, \dodoi{10.3847/1538-4357/ab425b}

\bibitem[{{Faisst} {et~al.}(2018){Faisst}, {Masters}, {Wang}, {Merson},
  {Capak}, {Malhotra}, \& {Rhoads}}]{Faisst2018}
{Faisst}, A.~L., {Masters}, D., {Wang}, Y., {et~al.} 2018, \apj, 855, 132,
  \dodoi{10.3847/1538-4357/aab1fc}

\bibitem[{{Finkelstein} {et~al.}(2022){Finkelstein}, {Bagley}, {Song},
  {Larson}, {Papovich}, {Dickinson}, {Finkelstein}, {Koekemoer}, {Pirzkal},
  {Somerville}, {Yung}, {Behroozi}, {Ferguson}, {Giavalisco}, {Grogin},
  {Hathi}, {Hutchison}, {Jung}, {Kocevski}, {Kawinwanichakij}, {Rojas-Ruiz},
  {Ryan}, {Snyder}, \& {Tacchella}}]{Finkelstein2022}
{Finkelstein}, S.~L., {Bagley}, M., {Song}, M., {et~al.} 2022, \apj, 928, 52,
  \dodoi{10.3847/1538-4357/ac3aed}

\bibitem[{{Finkelstein} {et~al.}(2024){Finkelstein}, {Leung}, {Bagley},
  {Dickinson}, {Ferguson}, {Papovich}, {Akins}, {Arrabal Haro}, {Dav{\'e}},
  {Dekel}, {Kartaltepe}, {Kocevski}, {Koekemoer}, {Pirzkal}, {Somerville},
  {Yung}, {Amor{\'\i}n}, {Backhaus}, {Behroozi}, {Bisigello}, {Bromm}, {Casey},
  {Ch{\'a}vez Ortiz}, {Cheng}, {Chworowsky}, {Cleri}, {Cooper}, {Davis}, {de la
  Vega}, {Elbaz}, {Franco}, {Fontana}, {Fujimoto}, {Giavalisco}, {Grogin},
  {Holwerda}, {Huertas-Company}, {Hirschmann}, {Iyer}, {Jogee}, {Jung},
  {Larson}, {Lucas}, {Mobasher}, {Morales}, {Morley}, {Mukherjee},
  {P{\'e}rez-Gonz{\'a}lez}, {Ravindranath}, {Rodighiero}, {Rowland},
  {Tacchella}, {Taylor}, {Trump}, \& {Wilkins}}]{Finkelstein2024}
{Finkelstein}, S.~L., {Leung}, G. C.~K., {Bagley}, M.~B., {et~al.} 2024, \apjl,
  969, L2, \dodoi{10.3847/2041-8213/ad4495}

\bibitem[{{Fukugita} {et~al.}(1996){Fukugita}, {Ichikawa}, {Gunn}, {Doi},
  {Shimasaku}, \& {Schneider}}]{Fukugita1996}
{Fukugita}, M., {Ichikawa}, T., {Gunn}, J.~E., {et~al.} 1996, \aj, 111, 1748,
  \dodoi{10.1086/117915}

\bibitem[{{Grogin} {et~al.}(2011){Grogin}, {Kocevski}, {Faber}, {Ferguson},
  {Koekemoer}, {Riess}, {Acquaviva}, {Alexander}, {Almaini}, {Ashby}, {Barden},
  {Bell}, {Bournaud}, {Brown}, {Caputi}, {Casertano}, {Cassata}, {Castellano},
  {Challis}, {Chary}, {Cheung}, {Cirasuolo}, {Conselice}, {Roshan Cooray},
  {Croton}, {Daddi}, {Dahlen}, {Dav{\'e}}, {de Mello}, {Dekel}, {Dickinson},
  {Dolch}, {Donley}, {Dunlop}, {Dutton}, {Elbaz}, {Fazio}, {Filippenko},
  {Finkelstein}, {Fontana}, {Gardner}, {Garnavich}, {Gawiser}, {Giavalisco},
  {Grazian}, {Guo}, {Hathi}, {H{\"a}ussler}, {Hopkins}, {Huang}, {Huang},
  {Jha}, {Kartaltepe}, {Kirshner}, {Koo}, {Lai}, {Lee}, {Li}, {Lotz}, {Lucas},
  {Madau}, {McCarthy}, {McGrath}, {McIntosh}, {McLure}, {Mobasher},
  {Moustakas}, {Mozena}, {Nandra}, {Newman}, {Niemi}, {Noeske}, {Papovich},
  {Pentericci}, {Pope}, {Primack}, {Rajan}, {Ravindranath}, {Reddy}, {Renzini},
  {Rix}, {Robaina}, {Rodney}, {Rosario}, {Rosati}, {Salimbeni}, {Scarlata},
  {Siana}, {Simard}, {Smidt}, {Somerville}, {Spinrad}, {Straughn}, {Strolger},
  {Telford}, {Teplitz}, {Trump}, {van der Wel}, {Villforth}, {Wechsler},
  {Weiner}, {Wiklind}, {Wild}, {Wilson}, {Wuyts}, {Yan}, \& {Yun}}]{Grogin2011}
{Grogin}, N.~A., {Kocevski}, D.~D., {Faber}, S.~M., {et~al.} 2011, \apjs, 197,
  35, \dodoi{10.1088/0067-0049/197/2/35}

\bibitem[{{Guo} {et~al.}(2016){Guo}, {Rafelski}, {Faber}, {Koo}, {Krumholz},
  {Trump}, {Willner}, {Amor{\'\i}n}, {Barro}, {Bell}, {Gardner}, {Gawiser},
  {Hathi}, {Koekemoer}, {Pacifici}, {P{\'e}rez-Gonz{\'a}lez}, {Ravindranath},
  {Reddy}, {Teplitz}, \& {Yesuf}}]{Guo2016}
{Guo}, Y., {Rafelski}, M., {Faber}, S.~M., {et~al.} 2016, \apj, 833, 37,
  \dodoi{10.3847/1538-4357/833/1/37}

\bibitem[{{Hu} {et~al.}(2024){Hu}, {Papovich}, {Dickinson}, {Kennicutt},
  {Shen}, {Amor{\'\i}n}, {Arrabal Haro}, {Bagley}, {Bhatawdekar}, {Cleri},
  {Cole}, {Dekel}, {de la Vega}, {Finkelstein}, {Grogin}, {Hathi},
  {Hirschmann}, {Holwerda}, {Hutchison}, {Jung}, {Koekemoer}, {Kartaltepe},
  {Lucas}, {Llerena}, {Mascia}, {Mobasher}, {Napolitano}, {Newman},
  {Pentericci}, {P{\'e}rez-Gonz{\'a}lez}, {Trump}, {Wilkins}, \&
  {Yung}}]{Hu2024}
{Hu}, W., {Papovich}, C., {Dickinson}, M., {et~al.} 2024, \apj, 971, 21,
  \dodoi{10.3847/1538-4357/ad5015}

\bibitem[{{Hunter}(2007)}]{Hunter2007}
{Hunter}, J.~D. 2007, Computing in Science and Engineering, 9, 90,
  \dodoi{10.1109/MCSE.2007.55}

\bibitem[{{Inoue}(2011)}]{Inoue2011}
{Inoue}, A.~K. 2011, \mnras, 415, 2920,
  \dodoi{10.1111/j.1365-2966.2011.18906.x}

\bibitem[{{Izotov} {et~al.}(2021){Izotov}, {Worseck}, {Schaerer}, {Guseva},
  {Chisholm}, {Thuan}, {Fricke}, \& {Verhamme}}]{Izotov2021}
{Izotov}, Y.~I., {Worseck}, G., {Schaerer}, D., {et~al.} 2021, \mnras, 503,
  1734, \dodoi{10.1093/mnras/stab612}

\bibitem[{{Izotov} {et~al.}(2018){Izotov}, {Worseck}, {Schaerer}, {Guseva},
  {Thuan}, {Fricke}, \& {Orlitov{\'a}}}]{Izotov2018}
---. 2018, \mnras, 478, 4851, \dodoi{10.1093/mnras/sty1378}

\bibitem[{{Jones} {et~al.}(2015){Jones}, {Martin}, \& {Cooper}}]{Jones2015}
{Jones}, T., {Martin}, C., \& {Cooper}, M.~C. 2015, \apj, 813, 126,
  \dodoi{10.1088/0004-637X/813/2/126}

\bibitem[{{Jung} {et~al.}(2024){Jung}, {Finkelstein}, {Arrabal Haro},
  {Dickinson}, {Ferguson}, {Hutchison}, {Kartaltepe}, {Larson}, {Simons},
  {Papovich}, {Park}, {Pentericci}, {Trump}, {Amor{\'\i}n}, {Backhaus},
  {Bagley}, {Casey}, {Cheng}, {Cleri}, {Cooper}, {Cooper}, {Gardner},
  {Gawiser}, {Grazian}, {Hathi}, {Hirschmann}, {Koekemoer}, {Lucas},
  {Mobasher}, {Pirzkal}, {Ravindranath}, {Straughn}, {Yung}, \& {de la
  Vega}}]{Jung2024}
{Jung}, I., {Finkelstein}, S.~L., {Arrabal Haro}, P., {et~al.} 2024, \apj, 967,
  73, \dodoi{10.3847/1538-4357/ad3913}

\bibitem[{{Kaasinen} {et~al.}(2018){Kaasinen}, {Kewley}, {Bian}, {Groves},
  {Kashino}, {Silverman}, \& {Kartaltepe}}]{Kaasinen2018}
{Kaasinen}, M., {Kewley}, L., {Bian}, F., {et~al.} 2018, \mnras, 477, 5568,
  \dodoi{10.1093/mnras/sty1012}

\bibitem[{{Kashino} {et~al.}(2019){Kashino}, {Silverman}, {Sanders},
  {Kartaltepe}, {Daddi}, {Renzini}, {Rodighiero}, {Puglisi}, {Valentino},
  {Juneau}, {Arimoto}, {Nagao}, {Ilbert}, {Le F{\`e}vre}, \&
  {Koekemoer}}]{Kashino2019}
{Kashino}, D., {Silverman}, J.~D., {Sanders}, D., {et~al.} 2019, \apjs, 241,
  10, \dodoi{10.3847/1538-4365/ab06c4}

\bibitem[{{Kauffmann} {et~al.}(2003){Kauffmann}, {Heckman}, {Tremonti},
  {Brinchmann}, {Charlot}, {White}, {Ridgway}, {Brinkmann}, {Fukugita}, {Hall},
  {Ivezi{\'c}}, {Richards}, \& {Schneider}}]{Kauffmann2003}
{Kauffmann}, G., {Heckman}, T.~M., {Tremonti}, C., {et~al.} 2003, \mnras, 346,
  1055, \dodoi{10.1111/j.1365-2966.2003.07154.x}

\bibitem[{{Kelly}(2007)}]{Kelly2007}
{Kelly}, B.~C. 2007, \apj, 665, 1489, \dodoi{10.1086/519947}

\bibitem[{{Kennicutt} {et~al.}(2007){Kennicutt}, {Calzetti}, {Walter}, {Helou},
  {Hollenbach}, {Armus}, {Bendo}, {Dale}, {Draine}, {Engelbracht}, {Gordon},
  {Prescott}, {Regan}, {Thornley}, {Bot}, {Brinks}, {de Blok}, {de Mello},
  {Meyer}, {Moustakas}, {Murphy}, {Sheth}, \& {Smith}}]{Kennicutt2007}
{Kennicutt}, Robert~C., J., {Calzetti}, D., {Walter}, F., {et~al.} 2007, \apj,
  671, 333, \dodoi{10.1086/522300}

\bibitem[{{Kennicutt} \& {Evans}(2012)}]{Kennicutt2012}
{Kennicutt}, R.~C., \& {Evans}, N.~J. 2012, \araa, 50, 531,
  \dodoi{10.1146/annurev-astro-081811-125610}

\bibitem[{{Kewley} \& {Dopita}(2002)}]{Kewley2002}
{Kewley}, L.~J., \& {Dopita}, M.~A. 2002, \apjs, 142, 35,
  \dodoi{10.1086/341326}

\bibitem[{{Kewley} {et~al.}(2013){Kewley}, {Dopita}, {Leitherer}, {Dav{\'e}},
  {Yuan}, {Allen}, {Groves}, \& {Sutherland}}]{Kewley2013}
{Kewley}, L.~J., {Dopita}, M.~A., {Leitherer}, C., {et~al.} 2013, \apj, 774,
  100, \dodoi{10.1088/0004-637X/774/2/100}

\bibitem[{{Kewley} {et~al.}(2019{\natexlab{a}}){Kewley}, {Nicholls},
  {Sutherland}, {Rigby}, {Acharya}, {Dopita}, \& {Bayliss}}]{Kewley2019b}
{Kewley}, L.~J., {Nicholls}, D.~C., {Sutherland}, R., {et~al.}
  2019{\natexlab{a}}, \apj, 880, 16, \dodoi{10.3847/1538-4357/ab16ed}

\bibitem[{{Kewley} {et~al.}(2019{\natexlab{b}}){Kewley}, {Nicholls}, \&
  {Sutherland}}]{Kewley2019}
{Kewley}, L.~J., {Nicholls}, D.~C., \& {Sutherland}, R.~S. 2019{\natexlab{b}},
  \araa, 57, 511, \dodoi{10.1146/annurev-astro-081817-051832}

\bibitem[{{Kewley} {et~al.}(2015){Kewley}, {Zahid}, {Geller}, {Dopita},
  {Hwang}, \& {Fabricant}}]{Kewley2015}
{Kewley}, L.~J., {Zahid}, H.~J., {Geller}, M.~J., {et~al.} 2015, \apjl, 812,
  L20, \dodoi{10.1088/2041-8205/812/2/L20}

\bibitem[{{Koekemoer} {et~al.}(2011){Koekemoer}, {Faber}, {Ferguson}, {Grogin},
  {Kocevski}, {Koo}, {Lai}, {Lotz}, {Lucas}, {McGrath}, {Ogaz}, {Rajan},
  {Riess}, {Rodney}, {Strolger}, {Casertano}, {Castellano}, {Dahlen},
  {Dickinson}, {Dolch}, {Fontana}, {Giavalisco}, {Grazian}, {Guo}, {Hathi},
  {Huang}, {van der Wel}, {Yan}, {Acquaviva}, {Alexander}, {Almaini}, {Ashby},
  {Barden}, {Bell}, {Bournaud}, {Brown}, {Caputi}, {Cassata}, {Challis},
  {Chary}, {Cheung}, {Cirasuolo}, {Conselice}, {Roshan Cooray}, {Croton},
  {Daddi}, {Dav{\'e}}, {de Mello}, {de Ravel}, {Dekel}, {Donley}, {Dunlop},
  {Dutton}, {Elbaz}, {Fazio}, {Filippenko}, {Finkelstein}, {Frazer}, {Gardner},
  {Garnavich}, {Gawiser}, {Gruetzbauch}, {Hartley}, {H{\"a}ussler},
  {Herrington}, {Hopkins}, {Huang}, {Jha}, {Johnson}, {Kartaltepe},
  {Khostovan}, {Kirshner}, {Lani}, {Lee}, {Li}, {Madau}, {McCarthy},
  {McIntosh}, {McLure}, {McPartland}, {Mobasher}, {Moreira}, {Mortlock},
  {Moustakas}, {Mozena}, {Nandra}, {Newman}, {Nielsen}, {Niemi}, {Noeske},
  {Papovich}, {Pentericci}, {Pope}, {Primack}, {Ravindranath}, {Reddy},
  {Renzini}, {Rix}, {Robaina}, {Rosario}, {Rosati}, {Salimbeni}, {Scarlata},
  {Siana}, {Simard}, {Smidt}, {Snyder}, {Somerville}, {Spinrad}, {Straughn},
  {Telford}, {Teplitz}, {Trump}, {Vargas}, {Villforth}, {Wagner}, {Wandro},
  {Wechsler}, {Weiner}, {Wiklind}, {Wild}, {Wilson}, {Wuyts}, \&
  {Yun}}]{Koekemoer2011}
{Koekemoer}, A.~M., {Faber}, S.~M., {Ferguson}, H.~C., {et~al.} 2011, \apjs,
  197, 36, \dodoi{10.1088/0067-0049/197/2/36}

\bibitem[{{Kriek} {et~al.}(2015){Kriek}, {Shapley}, {Reddy}, {Siana}, {Coil},
  {Mobasher}, {Freeman}, {de Groot}, {Price}, {Sanders}, {Shivaei}, {Brammer},
  {Momcheva}, {Skelton}, {van Dokkum}, {Whitaker}, {Aird}, {Azadi}, {Kassis},
  {Bullock}, {Conroy}, {Dav{\'e}}, {Kere{\v{s}}}, \& {Krumholz}}]{Kriek2015}
{Kriek}, M., {Shapley}, A.~E., {Reddy}, N.~A., {et~al.} 2015, \apjs, 218, 15,
  \dodoi{10.1088/0067-0049/218/2/15}

\bibitem[{{Kroupa}(2001)}]{Kroupa2001}
{Kroupa}, P. 2001, \mnras, 322, 231, \dodoi{10.1046/j.1365-8711.2001.04022.x}

\bibitem[{{Leitherer} \& {Heckman}(1995)}]{Leitherer1995}
{Leitherer}, C., \& {Heckman}, T.~M. 1995, \apjs, 96, 9, \dodoi{10.1086/192112}

\bibitem[{{Leung, Gene} {et~al.}(2023){Leung, Gene}, {Finkelstein, Steven},
  {Papovich, Casey}, \& {Pirzkal,
  Norbert}}]{https://doi.org/10.17909/3s7h-8k54}
{Leung, Gene}, {Finkelstein, Steven}, {Papovich, Casey}, \& {Pirzkal, Norbert}.
  2023, NGDEEP Epoch 1 NIRCam imaging data,  STScI/MAST,
  \dodoi{10.17909/3S7H-8K54}

\bibitem[{{Lyu} {et~al.}(2022){Lyu}, {Alberts}, {Rieke}, \&
  {Rujopakarn}}]{Lyu2022}
{Lyu}, J., {Alberts}, S., {Rieke}, G.~H., \& {Rujopakarn}, W. 2022, \apj, 941,
  191, \dodoi{10.3847/1538-4357/ac9e5d}

\bibitem[{{Madau} \& {Dickinson}(2014)}]{Madau2014}
{Madau}, P., \& {Dickinson}, M. 2014, \araa, 52, 415,
  \dodoi{10.1146/annurev-astro-081811-125615}

\bibitem[{{Maiolino} {et~al.}(2008){Maiolino}, {Nagao}, {Grazian}, {Cocchia},
  {Marconi}, {Mannucci}, {Cimatti}, {Pipino}, {Ballero}, {Calura}, {Chiappini},
  {Fontana}, {Granato}, {Matteucci}, {Pastorini}, {Pentericci}, {Risaliti},
  {Salvati}, \& {Silva}}]{Maiolino2008}
{Maiolino}, R., {Nagao}, T., {Grazian}, A., {et~al.} 2008, \aap, 488, 463,
  \dodoi{10.1051/0004-6361:200809678}

\bibitem[{{Matharu} {et~al.}(2021){Matharu}, {Muzzin}, {Brammer}, {Nelson},
  {Auger}, {Hewett}, {van der Burg}, {Balogh}, {Demarco}, {Marchesini},
  {Noble}, {Rudnick}, {van der Wel}, {Wilson}, \& {Yee}}]{Matharu2021}
{Matharu}, J., {Muzzin}, A., {Brammer}, G.~B., {et~al.} 2021, \apj, 923, 222,
  \dodoi{10.3847/1538-4357/ac26c3}

\bibitem[{{Matharu} {et~al.}(2023){Matharu}, {Muzzin}, {Sarrouh}, {Brammer},
  {Abraham}, {Asada}, {Brada{\v{c}}}, {Desprez}, {Martis}, {Mowla}, {Noirot},
  {Sawicki}, {Strait}, {Willott}, {Gould}, {Grindlay}, \&
  {Harshan}}]{Matharu2023}
{Matharu}, J., {Muzzin}, A., {Sarrouh}, G. T.~E., {et~al.} 2023, \apjl, 949,
  L11, \dodoi{10.3847/2041-8213/acd1db}

\bibitem[{{Mingozzi} {et~al.}(2020){Mingozzi}, {Belfiore}, {Cresci}, {Bundy},
  {Bershady}, {Bizyaev}, {Blanc}, {Boquien}, {Drory}, {Fu}, {Maiolino},
  {Riffel}, {Schaefer}, {Storchi-Bergmann}, {Telles}, {Tremonti}, {Zakamska},
  \& {Zhang}}]{Mingozzi2020}
{Mingozzi}, M., {Belfiore}, F., {Cresci}, G., {et~al.} 2020, \aap, 636, A42,
  \dodoi{10.1051/0004-6361/201937203}

\bibitem[{{Nakajima} {et~al.}(2020){Nakajima}, {Ellis}, {Robertson}, {Tang}, \&
  {Stark}}]{Nakajima2020}
{Nakajima}, K., {Ellis}, R.~S., {Robertson}, B.~E., {Tang}, M., \& {Stark},
  D.~P. 2020, \apj, 889, 161, \dodoi{10.3847/1538-4357/ab6604}

\bibitem[{{Nakajima} \& {Ouchi}(2014)}]{Nakajima2014}
{Nakajima}, K., \& {Ouchi}, M. 2014, \mnras, 442, 900,
  \dodoi{10.1093/mnras/stu902}

\bibitem[{{Noirot} {et~al.}(2023){Noirot}, {Desprez}, {Asada}, {Sawicki},
  {Estrada-Carpenter}, {Martis}, {Sarrouh}, {Strait}, {Abraham},
  {Brada{\v{c}}}, {Brammer}, {Iyer}, {MacFarland}, {Matharu}, {Mowla},
  {Muzzin}, {Pacifici}, {Ravindranath}, {Willott}, {Albert}, {Doyon},
  {Hutchings}, \& {Rowlands}}]{Noirot2023}
{Noirot}, G., {Desprez}, G., {Asada}, Y., {et~al.} 2023, \mnras, 525, 1867,
  \dodoi{10.1093/mnras/stad1019}

\bibitem[{{Oke} \& {Gunn}(1983)}]{Oke1983}
{Oke}, J.~B., \& {Gunn}, J.~E. 1983, \apj, 266, 713, \dodoi{10.1086/160817}

\bibitem[{{Osterbrock} \& {Ferland}(2006)}]{Osterbrock2006}
{Osterbrock}, D.~E., \& {Ferland}, G.~J. 2006, {Astrophysics of gaseous nebulae
  and active galactic nuclei}

\bibitem[{{Papovich} {et~al.}(2022){Papovich}, {Simons}, {Estrada-Carpenter},
  {Matharu}, {Momcheva}, {Trump}, {Backhaus}, {Brammer}, {Cleri},
  {Finkelstein}, {Giavalisco}, {Ji}, {Jung}, {Kewley}, {Nicholls}, {Pirzkal},
  {Rafelski}, \& {Weiner}}]{Papovich2022}
{Papovich}, C., {Simons}, R.~C., {Estrada-Carpenter}, V., {et~al.} 2022, \apj,
  937, 22, \dodoi{10.3847/1538-4357/ac8058}

\bibitem[{Pasha \& Miller(2023)}]{Pasha2023}
Pasha, I., \& Miller, T.~B. 2023, Journal of Open Source Software, 8, 5703,
  \dodoi{10.21105/joss.05703}

\bibitem[{{Pessa} {et~al.}(2021){Pessa}, {Schinnerer}, {Belfiore}, {Emsellem},
  {Leroy}, {Schruba}, {Kruijssen}, {Pan}, {Blanc}, {Sanchez-Blazquez},
  {Bigiel}, {Chevance}, {Congiu}, {Dale}, {Faesi}, {Glover}, {Grasha},
  {Groves}, {Ho}, {Jim{\'e}nez-Donaire}, {Klessen}, {Kreckel}, {Koch}, {Liu},
  {Meidt}, {Pety}, {Querejeta}, {Rosolowsky}, {Saito}, {Santoro}, {Sun},
  {Usero}, {Watkins}, \& {Williams}}]{Pessa2021}
{Pessa}, I., {Schinnerer}, E., {Belfiore}, F., {et~al.} 2021, \aap, 650, A134,
  \dodoi{10.1051/0004-6361/202140733}

\bibitem[{{Pirzkal} {et~al.}(2023){Pirzkal}, {Rothberg}, {Papovich}, {Shen},
  {Leung}, {Bagley}, {Finkelstein}, {Lotz}, {Koekemoer}, {Hathi}, {Cheng},
  {Cleri}, {Y.}, {Yung}, {Backhaus}, {Gardner}, {P{\'e}rez-Gonz{\'a}lez},
  {Ferguson}, {Grogin}, {Matharu}, {Ravindranath}, {Ryan}, {Berg}, {Casey},
  {Castellano}, {Ch{\'a}vez Ortiz}, {Chworowsky}, {Dickinson}, {Somerville},
  {Cox}, {Dav{\'e}}, {Davis}, {Estrada-Carpenter}, {Fontana}, {Fujimoto},
  {Giavalisco}, {Grazian}, {Hutchison}, {Jaskot}, {Jung}, {Kartaltepe},
  {Kewley}, {Kirkpatrick}, {Kocevski}, {Larson}, {Natarajan}, {Pentericci},
  {Simons}, {Snyder}, {Trump}, {Vanderhoof}, \& {Wilkins}}]{Pirzkal2023}
{Pirzkal}, N., {Rothberg}, B., {Papovich}, C., {et~al.} 2023, arXiv e-prints,
  arXiv:2312.09972, \dodoi{10.48550/arXiv.2312.09972}

\bibitem[{{Planck Collaboration} {et~al.}(2016){Planck Collaboration}, {Ade},
  {Aghanim}, {Arnaud}, {Ashdown}, {Aumont}, {Baccigalupi}, {Banday},
  {Barreiro}, {Bartlett}, {Bartolo}, {Battaner}, {Battye}, {Benabed},
  {Beno{\^\i}t}, {Benoit-L{\'e}vy}, {Bernard}, {Bersanelli}, {Bielewicz},
  {Bock}, {Bonaldi}, {Bonavera}, {Bond}, {Borrill}, {Bouchet}, {Boulanger},
  {Bucher}, {Burigana}, {Butler}, {Calabrese}, {Cardoso}, {Catalano},
  {Challinor}, {Chamballu}, {Chary}, {Chiang}, {Chluba}, {Christensen},
  {Church}, {Clements}, {Colombi}, {Colombo}, {Combet}, {Coulais}, {Crill},
  {Curto}, {Cuttaia}, {Danese}, {Davies}, {Davis}, {de Bernardis}, {de Rosa},
  {de Zotti}, {Delabrouille}, {D{\'e}sert}, {Di Valentino}, {Dickinson},
  {Diego}, {Dolag}, {Dole}, {Donzelli}, {Dor{\'e}}, {Douspis}, {Ducout},
  {Dunkley}, {Dupac}, {Efstathiou}, {Elsner}, {En{\ss}lin}, {Eriksen},
  {Farhang}, {Fergusson}, {Finelli}, {Forni}, {Frailis}, {Fraisse},
  {Franceschi}, {Frejsel}, {Galeotta}, {Galli}, {Ganga}, {Gauthier}, {Gerbino},
  {Ghosh}, {Giard}, {Giraud-H{\'e}raud}, {Giusarma}, {Gjerl{\o}w},
  {Gonz{\'a}lez-Nuevo}, {G{\'o}rski}, {Gratton}, {Gregorio}, {Gruppuso},
  {Gudmundsson}, {Hamann}, {Hansen}, {Hanson}, {Harrison}, {Helou},
  {Henrot-Versill{\'e}}, {Hern{\'a}ndez-Monteagudo}, {Herranz}, {Hildebrandt},
  {Hivon}, {Hobson}, {Holmes}, {Hornstrup}, {Hovest}, {Huang}, {Huffenberger},
  {Hurier}, {Jaffe}, {Jaffe}, {Jones}, {Juvela}, {Keih{\"a}nen}, {Keskitalo},
  {Kisner}, {Kneissl}, {Knoche}, {Knox}, {Kunz}, {Kurki-Suonio}, {Lagache},
  {L{\"a}hteenm{\"a}ki}, {Lamarre}, {Lasenby}, {Lattanzi}, {Lawrence}, {Leahy},
  {Leonardi}, {Lesgourgues}, {Levrier}, {Lewis}, {Liguori}, {Lilje},
  {Linden-V{\o}rnle}, {L{\'o}pez-Caniego}, {Lubin}, {Mac{\'\i}as-P{\'e}rez},
  {Maggio}, {Maino}, {Mandolesi}, {Mangilli}, {Marchini}, {Maris}, {Martin},
  {Martinelli}, {Mart{\'\i}nez-Gonz{\'a}lez}, {Masi}, {Matarrese}, {McGehee},
  {Meinhold}, {Melchiorri}, {Melin}, {Mendes}, {Mennella}, {Migliaccio},
  {Millea}, {Mitra}, {Miville-Desch{\^e}nes}, {Moneti}, {Montier}, {Morgante},
  {Mortlock}, {Moss}, {Munshi}, {Murphy}, {Naselsky}, {Nati}, {Natoli},
  {Netterfield}, {N{\o}rgaard-Nielsen}, {Noviello}, {Novikov}, {Novikov},
  {Oxborrow}, {Paci}, {Pagano}, {Pajot}, {Paladini}, {Paoletti}, {Partridge},
  {Pasian}, {Patanchon}, {Pearson}, {Perdereau}, {Perotto}, {Perrotta},
  {Pettorino}, {Piacentini}, {Piat}, {Pierpaoli}, {Pietrobon}, {Plaszczynski},
  {Pointecouteau}, {Polenta}, {Popa}, {Pratt}, {Pr{\'e}zeau}, {Prunet},
  {Puget}, {Rachen}, {Reach}, {Rebolo}, {Reinecke}, {Remazeilles}, {Renault},
  {Renzi}, {Ristorcelli}, {Rocha}, {Rosset}, {Rossetti}, {Roudier},
  {Rouill{\'e} d'Orfeuil}, {Rowan-Robinson}, {Rubi{\~n}o-Mart{\'\i}n},
  {Rusholme}, {Said}, {Salvatelli}, {Salvati}, {Sandri}, {Santos},
  {Savelainen}, {Savini}, {Scott}, {Seiffert}, {Serra}, {Shellard}, {Spencer},
  {Spinelli}, {Stolyarov}, {Stompor}, {Sudiwala}, {Sunyaev}, {Sutton},
  {Suur-Uski}, {Sygnet}, {Tauber}, {Terenzi}, {Toffolatti}, {Tomasi},
  {Tristram}, {Trombetti}, {Tucci}, {Tuovinen}, {T{\"u}rler}, {Umana},
  {Valenziano}, {Valiviita}, {Van Tent}, {Vielva}, {Villa}, {Wade}, {Wandelt},
  {Wehus}, {White}, {White}, {Wilkinson}, {Yvon}, {Zacchei}, \&
  {Zonca}}]{Planck2016}
{Planck Collaboration}, {Ade}, P.~A.~R., {Aghanim}, N., {et~al.} 2016, \aap,
  594, A13, \dodoi{10.1051/0004-6361/201525830}

\bibitem[{{Prieto-Lyon} {et~al.}(2023){Prieto-Lyon}, {Strait}, {Mason},
  {Brammer}, {Caminha}, {Mercurio}, {Acebron}, {Bergamini}, {Grillo}, {Rosati},
  {Vanzella}, {Castellano}, {Merlin}, {Paris}, {Boyett}, {Calabr{\`o}},
  {Morishita}, {Mascia}, {Pentericci}, {Roberts-Borsani}, {Roy}, {Treu}, \&
  {Vulcani}}]{Prieto-Lyon2023}
{Prieto-Lyon}, G., {Strait}, V., {Mason}, C.~A., {et~al.} 2023, \aap, 672,
  A186, \dodoi{10.1051/0004-6361/202245532}

\bibitem[{{Rafelski} {et~al.}(2015){Rafelski}, {Teplitz}, {Gardner}, {Coe},
  {Bond}, {Koekemoer}, {Grogin}, {Kurczynski}, {McGrath}, {Bourque}, {Atek},
  {Brown}, {Colbert}, {Codoreanu}, {Ferguson}, {Finkelstein}, {Gawiser},
  {Giavalisco}, {Gronwall}, {Hanish}, {Lee}, {Mehta}, {de Mello},
  {Ravindranath}, {Ryan}, {Scarlata}, {Siana}, {Soto}, \&
  {Voyer}}]{Rafelski2015}
{Rafelski}, M., {Teplitz}, H.~I., {Gardner}, J.~P., {et~al.} 2015, \aj, 150,
  31, \dodoi{10.1088/0004-6256/150/1/31}

\bibitem[{{Reddy} {et~al.}(2023{\natexlab{a}}){Reddy}, {Topping}, {Sanders},
  {Shapley}, \& {Brammer}}]{Reddy2023b}
{Reddy}, N.~A., {Topping}, M.~W., {Sanders}, R.~L., {Shapley}, A.~E., \&
  {Brammer}, G. 2023{\natexlab{a}}, \apj, 952, 167,
  \dodoi{10.3847/1538-4357/acd754}

\bibitem[{{Reddy} {et~al.}(2015){Reddy}, {Kriek}, {Shapley}, {Freeman},
  {Siana}, {Coil}, {Mobasher}, {Price}, {Sanders}, \& {Shivaei}}]{Reddy2015}
{Reddy}, N.~A., {Kriek}, M., {Shapley}, A.~E., {et~al.} 2015, \apj, 806, 259,
  \dodoi{10.1088/0004-637X/806/2/259}

\bibitem[{{Reddy} {et~al.}(2018){Reddy}, {Shapley}, {Sanders}, {Kriek}, {Coil},
  {Shivaei}, {Freeman}, {Mobasher}, {Siana}, {Azadi}, {Fetherolf}, {Fornasini},
  {Leung}, {Price}, {Zick}, \& {Barro}}]{Reddy2018}
{Reddy}, N.~A., {Shapley}, A.~E., {Sanders}, R.~L., {et~al.} 2018, \apj, 869,
  92, \dodoi{10.3847/1538-4357/aaed1e}

\bibitem[{{Reddy} {et~al.}(2023{\natexlab{b}}){Reddy}, {Sanders}, {Shapley},
  {Topping}, {Kriek}, {Coil}, {Mobasher}, {Siana}, \& {Rezaee}}]{Reddy2023a}
{Reddy}, N.~A., {Sanders}, R.~L., {Shapley}, A.~E., {et~al.}
  2023{\natexlab{b}}, \apj, 951, 56, \dodoi{10.3847/1538-4357/acd0b1}

\bibitem[{{Rieke} \& {the JADES Collaboration}(2023)}]{Rieke2023}
{Rieke}, M., \& {the JADES Collaboration}. 2023, arXiv e-prints,
  arXiv:2306.02466, \dodoi{10.48550/arXiv.2306.02466}

\bibitem[{{Rieke, Marcia} {et~al.}(2023){Rieke, Marcia}, {Robertson, Brant},
  {Tacchella, Sandro}, {Willmer, Christopher}, {Johnson, Ben}, {Carniani,
  Stefano}, {Bunker, Andy}, \& {Willott,
  Chris}}]{https://doi.org/10.17909/8tdj-8n28}
{Rieke, Marcia}, {Robertson, Brant}, {Tacchella, Sandro}, {et~al.} 2023, Data
  from the JWST Advanced Deep Extragalactic Survey (JADES),  STScI/MAST,
  \dodoi{10.17909/8TDJ-8N28}

\bibitem[{{Sanders} {et~al.}(2024){Sanders}, {Shapley}, {Topping}, {Reddy}, \&
  {Brammer}}]{Sanders2024}
{Sanders}, R.~L., {Shapley}, A.~E., {Topping}, M.~W., {Reddy}, N.~A., \&
  {Brammer}, G.~B. 2024, \apj, 962, 24, \dodoi{10.3847/1538-4357/ad15fc}

\bibitem[{{Sanders} {et~al.}(2016){Sanders}, {Shapley}, {Kriek}, {Reddy},
  {Freeman}, {Coil}, {Siana}, {Mobasher}, {Shivaei}, {Price}, \& {de
  Groot}}]{Sanders2016}
{Sanders}, R.~L., {Shapley}, A.~E., {Kriek}, M., {et~al.} 2016, \apj, 816, 23,
  \dodoi{10.3847/0004-637X/816/1/23}

\bibitem[{{Sanders} {et~al.}(2018){Sanders}, {Shapley}, {Kriek}, {Freeman},
  {Reddy}, {Siana}, {Coil}, {Mobasher}, {Dav{\'e}}, {Shivaei}, {Azadi},
  {Price}, {Leung}, {Fetherolf}, {de Groot}, {Zick}, {Fornasini}, \&
  {Barro}}]{Sanders2018}
---. 2018, \apj, 858, 99, \dodoi{10.3847/1538-4357/aabcbd}

\bibitem[{{Sanders} {et~al.}(2020){Sanders}, {Shapley}, {Reddy}, {Kriek},
  {Siana}, {Coil}, {Mobasher}, {Shivaei}, {Freeman}, {Azadi}, {Price}, {Leung},
  {Fetherolf}, {de Groot}, {Zick}, {Fornasini}, \& {Barro}}]{Sanders2020}
{Sanders}, R.~L., {Shapley}, A.~E., {Reddy}, N.~A., {et~al.} 2020, \mnras, 491,
  1427, \dodoi{10.1093/mnras/stz3032}

\bibitem[{{Sanders} {et~al.}(2021){Sanders}, {Shapley}, {Jones}, {Reddy},
  {Kriek}, {Siana}, {Coil}, {Mobasher}, {Shivaei}, {Dav{\'e}}, {Azadi},
  {Price}, {Leung}, {Freeman}, {Fetherolf}, {de Groot}, {Zick}, \&
  {Barro}}]{Sanders2021}
{Sanders}, R.~L., {Shapley}, A.~E., {Jones}, T., {et~al.} 2021, \apj, 914, 19,
  \dodoi{10.3847/1538-4357/abf4c1}

\bibitem[{{Scarlata} {et~al.}(2024){Scarlata}, {Hayes}, {Panagia}, {Mehta},
  {Haardt}, \& {Bagley}}]{Scarlata2024}
{Scarlata}, C., {Hayes}, M., {Panagia}, N., {et~al.} 2024, arXiv e-prints,
  arXiv:2404.09015, \dodoi{10.48550/arXiv.2404.09015}

\bibitem[{{Shen} {et~al.}(2023){Shen}, {Papovich}, {Yang}, {Matharu}, {Wang},
  {Magnelli}, {Elbaz}, {Jogee}, {Alavi}, {Arrabal Haro}, {Backhaus}, {Bagley},
  {Bell}, {Bisigello}, {Calabr{\`o}}, {Cooper}, {Costantin}, {Daddi},
  {Dickinson}, {Finkelstein}, {Fujimoto}, {Giavalisco}, {Grogin}, {Guo},
  {Holwerda}, {Kartaltepe}, {Koekemoer}, {Kurczynski}, {Lucas},
  {P{\'e}rez-Gonz{\'a}lez}, {Pirzkal}, {Prichard}, {Rafelski}, {Ronayne},
  {Simons}, {Sunnquist}, {Teplitz}, {Trump}, {Weiner}, {Windhorst}, \&
  {Yung}}]{Shen2023}
{Shen}, L., {Papovich}, C., {Yang}, G., {et~al.} 2023, \apj, 950, 7,
  \dodoi{10.3847/1538-4357/acc944}

\bibitem[{{Shen} {et~al.}(2024){Shen}, {Papovich}, {Matharu}, {Pirzkal}, {Hu},
  {Backhaus}, {Bagley}, {Cheng}, {Cleri}, {Finkelstein}, {Huertas-Company},
  {Giavalisco}, {Grogin}, {Jung}, {Kartaltepe}, {Koekemoer}, {Lotz}, {Maseda},
  {P{\'e}rez-Gonz{\'a}lez}, {Rothberg}, {Simons}, {Tacchella}, {Williams}, \&
  {Yung}}]{Shen2024}
{Shen}, L., {Papovich}, C., {Matharu}, J., {et~al.} 2024, \apjl, 963, L49,
  \dodoi{10.3847/2041-8213/ad28bd}

\bibitem[{{Shimakawa} {et~al.}(2015){Shimakawa}, {Kodama}, {Steidel}, {Tadaki},
  {Tanaka}, {Strom}, {Hayashi}, {Koyama}, {Suzuki}, \&
  {Yamamoto}}]{Shimakawa2015}
{Shimakawa}, R., {Kodama}, T., {Steidel}, C.~C., {et~al.} 2015, \mnras, 451,
  1284, \dodoi{10.1093/mnras/stv915}

\bibitem[{{Simons} {et~al.}(2021){Simons}, {Papovich}, {Momcheva}, {Trump},
  {Brammer}, {Estrada-Carpenter}, {Backhaus}, {Cleri}, {Finkelstein},
  {Giavalisco}, {Ji}, {Jung}, {Matharu}, \& {Weiner}}]{Simons2021}
{Simons}, R.~C., {Papovich}, C., {Momcheva}, I., {et~al.} 2021, \apj, 923, 203,
  \dodoi{10.3847/1538-4357/ac28f4}

\bibitem[{{Simons} {et~al.}(2023){Simons}, {Papovich}, {Momcheva}, {Brammer},
  {Estrada-Carpenter}, {Finkelstein}, {Gosmeyer}, {Matharu}, {Trump},
  {Backhaus}, {Cheng}, {Cleri}, {Ferguson}, {Finlator}, {Giavalisco}, {Ji},
  {Jung}, {Lotz}, {O'Brien}, {Skelton}, {Tilvi}, \& {Weiner}}]{Simons2023}
{Simons}, R.~C., {Papovich}, C., {Momcheva}, I.~G., {et~al.} 2023, \apjs, 266,
  13, \dodoi{10.3847/1538-4365/acc517}

\bibitem[{{Somerville} \& {Dav{\'e}}(2015)}]{Somerville2015}
{Somerville}, R.~S., \& {Dav{\'e}}, R. 2015, \araa, 53, 51,
  \dodoi{10.1146/annurev-astro-082812-140951}

\bibitem[{{Steidel} {et~al.}(2014){Steidel}, {Rudie}, {Strom}, {Pettini},
  {Reddy}, {Shapley}, {Trainor}, {Erb}, {Turner}, {Konidaris}, {Kulas}, {Mace},
  {Matthews}, \& {McLean}}]{Steidel2014}
{Steidel}, C.~C., {Rudie}, G.~C., {Strom}, A.~L., {et~al.} 2014, \apj, 795,
  165, \dodoi{10.1088/0004-637X/795/2/165}

\bibitem[{{Strom} {et~al.}(2018){Strom}, {Steidel}, {Rudie}, {Trainor}, \&
  {Pettini}}]{Strom2018}
{Strom}, A.~L., {Steidel}, C.~C., {Rudie}, G.~C., {Trainor}, R.~F., \&
  {Pettini}, M. 2018, \apj, 868, 117, \dodoi{10.3847/1538-4357/aae1a5}

\bibitem[{{Strom} {et~al.}(2017){Strom}, {Steidel}, {Rudie}, {Trainor},
  {Pettini}, \& {Reddy}}]{Strom2017}
{Strom}, A.~L., {Steidel}, C.~C., {Rudie}, G.~C., {et~al.} 2017, \apj, 836,
  164, \dodoi{10.3847/1538-4357/836/2/164}

\bibitem[{{Sutherland} \& {Dopita}(1993)}]{Sutherland1993}
{Sutherland}, R.~S., \& {Dopita}, M.~A. 1993, \apjs, 88, 253,
  \dodoi{10.1086/191823}

\bibitem[{{Tang} {et~al.}(2019){Tang}, {Stark}, {Chevallard}, \&
  {Charlot}}]{Tang2019}
{Tang}, M., {Stark}, D.~P., {Chevallard}, J., \& {Charlot}, S. 2019, \mnras,
  489, 2572, \dodoi{10.1093/mnras/stz2236}

\bibitem[{{Teplitz} {et~al.}(2013){Teplitz}, {Rafelski}, {Kurczynski}, {Bond},
  {Grogin}, {Koekemoer}, {Atek}, {Brown}, {Coe}, {Colbert}, {Ferguson},
  {Finkelstein}, {Gardner}, {Gawiser}, {Giavalisco}, {Gronwall}, {Hanish},
  {Lee}, {de Mello}, {Ravindranath}, {Ryan}, {Siana}, {Scarlata}, {Soto},
  {Voyer}, \& {Wolfe}}]{Teplitz2013}
{Teplitz}, H.~I., {Rafelski}, M., {Kurczynski}, P., {et~al.} 2013, \aj, 146,
  159, \dodoi{10.1088/0004-6256/146/6/159}

\bibitem[{{Tomczak} {et~al.}(2014){Tomczak}, {Quadri}, {Tran}, {Labb{\'e}},
  {Straatman}, {Papovich}, {Glazebrook}, {Allen}, {Brammer}, {Kacprzak},
  {Kawinwanichakij}, {Kelson}, {McCarthy}, {Mehrtens}, {Monson}, {Persson},
  {Spitler}, {Tilvi}, \& {van Dokkum}}]{Tomczak2014}
{Tomczak}, A.~R., {Quadri}, R.~F., {Tran}, K.-V.~H., {et~al.} 2014, \apj, 783,
  85, \dodoi{10.1088/0004-637X/783/2/85}

\bibitem[{{Tomczak} {et~al.}(2016){Tomczak}, {Quadri}, {Tran}, {Labb{\'e}},
  {Straatman}, {Papovich}, {Glazebrook}, {Allen}, {Brammer}, {Cowley},
  {Dickinson}, {Elbaz}, {Inami}, {Kacprzak}, {Morrison}, {Nanayakkara},
  {Persson}, {Rees}, {Salmon}, {Schreiber}, {Spitler}, \&
  {Whitaker}}]{Tomczak2016}
---. 2016, \apj, 817, 118, \dodoi{10.3847/0004-637X/817/2/118}

\bibitem[{{Tremonti} {et~al.}(2004){Tremonti}, {Heckman}, {Kauffmann},
  {Brinchmann}, {Charlot}, {White}, {Seibert}, {Peng}, {Schlegel}, {Uomoto},
  {Fukugita}, \& {Brinkmann}}]{Tremonti2004}
{Tremonti}, C.~A., {Heckman}, T.~M., {Kauffmann}, G., {et~al.} 2004, \apj, 613,
  898, \dodoi{10.1086/423264}

\bibitem[{{van der Walt} {et~al.}(2011){van der Walt}, {Colbert}, \&
  {Varoquaux}}]{vanderWalt2011}
{van der Walt}, S., {Colbert}, S.~C., \& {Varoquaux}, G. 2011, Computing in
  Science and Engineering, 13, 22, \dodoi{10.1109/MCSE.2011.37}

\bibitem[{Virtanen {et~al.}(2020)Virtanen, Gommers, Oliphant, Haberland, Reddy,
  Cournapeau, Burovski, Peterson, Weckesser, Bright, {van der Walt}, Brett,
  Wilson, Millman, Mayorov, Nelson, Jones, Kern, Larson, Carey, Polat, Feng,
  Moore, {VanderPlas}, Laxalde, Perktold, Cimrman, Henriksen, Quintero, Harris,
  Archibald, Ribeiro, Pedregosa, {van Mulbregt}, \& {SciPy 1.0
  Contributors}}]{2020SciPy-NMeth}
Virtanen, P., Gommers, R., Oliphant, T.~E., {et~al.} 2020, Nature Methods, 17,
  261, \dodoi{10.1038/s41592-019-0686-2}

\bibitem[{{Wang} {et~al.}(2022){Wang}, {Jones}, {Vulcani}, {Treu}, {Morishita},
  {Roberts-Borsani}, {Malkan}, {Henry}, {Brammer}, {Strait}, {Brada{\v{c}}},
  {Boyett}, {Calabr{\`o}}, {Castellano}, {Fontana}, {Glazebrook}, {Kelly},
  {Leethochawalit}, {Marchesini}, {Santini}, {Trenti}, \& {Yang}}]{Wang2022}
{Wang}, X., {Jones}, T., {Vulcani}, B., {et~al.} 2022, \apjl, 938, L16,
  \dodoi{10.3847/2041-8213/ac959e}

\bibitem[{{Whitaker} {et~al.}(2011){Whitaker}, {Labb{\'e}}, {van Dokkum},
  {Brammer}, {Kriek}, {Marchesini}, {Quadri}, {Franx}, {Muzzin}, {Williams},
  {Bezanson}, {Illingworth}, {Lee}, {Lundgren}, {Nelson}, {Rudnick}, {Tal}, \&
  {Wake}}]{Whitaker2011}
{Whitaker}, K.~E., {Labb{\'e}}, I., {van Dokkum}, P.~G., {et~al.} 2011, \apj,
  735, 86, \dodoi{10.1088/0004-637X/735/2/86}

\bibitem[{{Whitaker} {et~al.}(2014){Whitaker}, {Franx}, {Leja}, {van Dokkum},
  {Henry}, {Skelton}, {Fumagalli}, {Momcheva}, {Brammer}, {Labb{\'e}},
  {Nelson}, \& {Rigby}}]{Whitaker2014}
{Whitaker}, K.~E., {Franx}, M., {Leja}, J., {et~al.} 2014, \apj, 795, 104,
  \dodoi{10.1088/0004-637X/795/2/104}

\bibitem[{{Williams} {et~al.}(2023){Williams}, {Tacchella}, {Maseda},
  {Robertson}, {Johnson}, {Willott}, {Eisenstein}, {Willmer}, {Ji}, {Hainline},
  {Helton}, {Alberts}, {Baum}, {Bhatawdekar}, {Boyett}, {Bunker}, {Carniani},
  {Charlot}, {Chevallard}, {Curtis-Lake}, {de Graaf}, {Egami}, {Franx},
  {Kumari}, {Maiolino}, {Nelson}, {Rieke}, {Sandles}, {Shivaei}, {Simmonds},
  {Smit}, {Suess}, {Sun}, {Ubler}, \& {Witstok}}]{Williams2023}
{Williams}, C.~C., {Tacchella}, S., {Maseda}, M.~V., {et~al.} 2023, arXiv
  e-prints, arXiv:2301.09780, \dodoi{10.48550/arXiv.2301.09780}

\bibitem[{{Williams, Christina} {et~al.}(2023){Williams, Christina},
  {Tacchella, Sandro}, \& {Maseda,
  Michael}}]{https://doi.org/10.17909/fsc4-dt61}
{Williams, Christina}, {Tacchella, Sandro}, \& {Maseda, Michael}. 2023, Data
  from the JWST Extragalactic Medium-band Survey (JEMS),  STScI/MAST,
  \dodoi{10.17909/FSC4-DT61}

\bibitem[{{Yang} {et~al.}(2020){Yang}, {Boquien}, {Buat}, {Burgarella},
  {Ciesla}, {Duras}, {Stalevski}, {Brandt}, \& {Papovich}}]{Yang2020}
{Yang}, G., {Boquien}, M., {Buat}, V., {et~al.} 2020, \mnras, 491, 740,
  \dodoi{10.1093/mnras/stz3001}

\bibitem[{{Yang} {et~al.}(2017{\natexlab{a}}){Yang}, {Malhotra}, {Rhoads}, \&
  {Wang}}]{Yang2017b}
{Yang}, H., {Malhotra}, S., {Rhoads}, J.~E., \& {Wang}, J. 2017{\natexlab{a}},
  \apj, 847, 38, \dodoi{10.3847/1538-4357/aa8809}

\bibitem[{{Yang} {et~al.}(2017{\natexlab{b}}){Yang}, {Malhotra}, {Gronke},
  {Rhoads}, {Leitherer}, {Wofford}, {Jiang}, {Dijkstra}, {Tilvi}, \&
  {Wang}}]{Yang2017a}
{Yang}, H., {Malhotra}, S., {Gronke}, M., {et~al.} 2017{\natexlab{b}}, \apj,
  844, 171, \dodoi{10.3847/1538-4357/aa7d4d}

\bibitem[{{Yates} {et~al.}(2020){Yates}, {Schady}, {Chen}, {Schweyer}, \&
  {Wiseman}}]{Yates2020}
{Yates}, R.~M., {Schady}, P., {Chen}, T.~W., {Schweyer}, T., \& {Wiseman}, P.
  2020, \aap, 634, A107, \dodoi{10.1051/0004-6361/201936506}

\bibitem[{{Zahid} {et~al.}(2014){Zahid}, {Dima}, {Kudritzki}, {Kewley},
  {Geller}, {Hwang}, {Silverman}, \& {Kashino}}]{Zahid2014}
{Zahid}, H.~J., {Dima}, G.~I., {Kudritzki}, R.-P., {et~al.} 2014, \apj, 791,
  130, \dodoi{10.1088/0004-637X/791/2/130}

\bibitem[{{Zahid} {et~al.}(2011){Zahid}, {Kewley}, \& {Bresolin}}]{Zahid2011}
{Zahid}, H.~J., {Kewley}, L.~J., \& {Bresolin}, F. 2011, \apj, 730, 137,
  \dodoi{10.1088/0004-637X/730/2/137}

\end{thebibliography}
\bibliographystyle{aasjournal}



\end{document}